\documentclass[twocolumn,superscriptaddress,nofootinbib,showkeys,aps,pra,10pt]{revtex4-2}

\usepackage[utf8]{inputenc}
\usepackage[T1]{fontenc}
\usepackage{amsmath,amssymb,amsthm,mathtools}
\usepackage{bm}
\usepackage{graphicx}
\usepackage{booktabs}
\usepackage{enumitem}
\usepackage{algpseudocode}
\usepackage{xcolor}
\usepackage{appendix}
\usepackage[colorlinks=true,linkcolor=blue!55!black,citecolor=blue!55!black,
            urlcolor=blue!55!black]{hyperref}
\usepackage{cleveref}

\graphicspath{{figures/}}

\theoremstyle{plain}
\newtheorem{theorem}{Theorem}
\newtheorem{proposition}[theorem]{Proposition}
\newtheorem{lemma}[theorem]{Lemma}
\newtheorem{corollary}[theorem]{Corollary}
\theoremstyle{definition}
\newtheorem{definition}[theorem]{Definition}

\theoremstyle{remark}
\newtheorem{remark}[theorem]{Remark}

\DeclareMathOperator{\rank}{rank}
\DeclareMathOperator{\supp}{supp}
\DeclareMathOperator{\row}{row}
\DeclareMathOperator{\Tr}{Tr}

\newcommand{\Ftwo}{\mathbb{F}_2}
\newcommand{\Gphys}{G_{\mathrm{phys}}}
\newcommand{\Pauth}{P_{\mathrm{auth}}}
\newcommand{\Pforb}{P_{\mathrm{forb}}}
\newcommand{\Pint}{P_{\mathrm{int}}}
\newcommand{\Wtot}{W_{\mathrm{tot}}}
\newcommand{\Wrec}{W_{\mathrm{rec}}}
\newcommand{\Crec}{C_{\mathrm{rec}}}
\newcommand{\CSt}{C_{\mathrm{St}}}
\newcommand{\Dsched}{D_{\mathrm{sched}}}
\newcommand{\tsec}{t_{\mathrm{sec}}}
\newcommand{\trec}{t_{\mathrm{rec}}}
\newcommand{\Aramp}{A_{\mathrm{ramp}}}

\newcounter{algcnt}
\renewcommand{\thealgcnt}{\arabic{algcnt}}
\makeatletter
\newenvironment{algorithm}[1][]{%
  \par\addvspace{\medskipamount}%
  \hrule\smallskip
  \refstepcounter{algcnt}%
  \renewcommand{\caption}[1]{%
    \noindent\textbf{Algorithm \thealgcnt:} ##1\par\smallskip\hrule\smallskip}%
}{%
  \smallskip\hrule\par\addvspace{\medskipamount}%
}
\makeatother

\algrenewcommand\algorithmicrequire{\textbf{Input:}}
\algrenewcommand\algorithmicensure{\textbf{Output:}}

\begin{document}

\title{Quantum Secret Sharing from CSS qLDPC Codes:
Computable Access Structures, Accessible Logical Dimension,
and Topology-Aware Network Cost}

\author{Malik~Mouaji}
\author{Saif~Al-Kuwari}

\affiliation{%
    Qatar Center for Quantum Computing,
    College of Science and Engineering,
    Hamad Bin Khalifa University,
    Doha, Qatar
}




\begin{abstract}
Quantum secret sharing (QSS) schemes can be constructed from quantum error-correcting codes, but characterizing their access structures and implementation costs becomes challenging for finite-length, high-rate quantum low-density parity-check (qLDPC) codes. We develop an exact computational framework for QSS induced by CSS qLDPC codes. The framework characterizes authorized, forbidden, and intermediate coalitions using restricted parity-check matrices and binary linear algebra, and introduces the accessible logical dimension \(\kappa(A)\), defined through the symplectic rank of locally supported logical classes, to refine the characterization of intermediate coalitions. We further incorporate physical network constraints through separate Tanner-incidence, Steiner-routing, and scheduling-depth costs, and treat the partitioning of physical qubits into participants as an explicit design variable. Numerical experiments on four CSS qLDPC codes verify the predicted complement duality and monotonicity properties and show substantial variation in access transitions and network costs. For the codes studied, the bivariate bicycle instance exhibits the sharpest transition, while network embedding costs vary by more than an order of magnitude across physical topologies. We also show that participant partitioning can substantially improve block-level security and reconstruction thresholds. These results provide a practical bridge between the algebraic structure of CSS qLDPC codes, the graded accessibility of quantum secrets, and the physical constraints of networked QSS implementations.

\end{abstract}


\maketitle

\section{Introduction}
\label{sec:introduction}

Quantum secret sharing (QSS) is a central primitive of quantum
cryptography. It enables a dealer to distribute a classical or quantum
secret among $n$ participants such that authorized coalitions can reconstruct the entire secret, forbidden coalitions obtain no information about it, while intermediate coalitions may access only part of the encoded logical information. Since the seminal works of Hillery, Bu\v{z}ek and Berthiaume \cite{HilleryBuzekBerthiaume1999} and of Cleve, Gottesman and Lo \cite{CleveGottesmanLo1999}, QSS has been regarded as the quantum analogue of classical secret sharing. However, unlike the classical setting, quantum secret sharing is constrained by the no-cloning theorem, the monogamy of quantum correlations, and the cost of generating and distributing multipartite entanglement~\cite{MarkhamSanders2008,Gheorghiu2012}.

A fundamental connection relates QSS to quantum error correction
\cite{CleveGottesmanLo1999,Gottesman2000}. An encoding that corrects
erasures in a set $E$ allows the complement $E^c$ to reconstruct the
logical secret; conversely, if an erasure in a set $A$ is correctable, the reduced state held by $A$ is independent of the encoded secret. For
stabilizer codes this correspondence admits a direct logical-operator
formulation: an erasure set is correctable precisely when it supports no nontrivial logical operator\cite{BravyiTerhal2009,YoshidaChuang2010}.
Hence an error-correcting code induces a QSS access structure whose
authorized, forbidden and intermediate coalitions are determined by the
supports of logical operators. When the code encodes more than one logical qubit, the induced scheme is generally a ramp scheme rather than a perfect threshold scheme~\cite{OgawaEtAl2005,SarvepalliKlappenecker2009,Matsumoto2019}.

Quantum low-density parity-check (qLDPC) codes are an appealing candidate substrate for scalable QSS. Their sparse parity-check matrices yield bounded-weight stabilizer generators and bounded incidence degree in the associated Tanner graph~\cite{BreuckmannEberhardt2021}.
Hypergraph-product \cite{TillichZemor2014} and bivariate bicycle
\cite{BravyiEtAl2024} codes are especially relevant finite-length CSS
families, since they combine nonzero encoding rate with sparse checks and practical decoding; lifted-product and quantum Tanner constructions
further establish the existence of asymptotically good qLDPC families
\cite{PanteleevKalachev2022,LeverrierZemor2022}. Sparsity alone, however, does not guarantee a geometrically local implementation: qubits participating in a low-weight check may still be far apart in a physical network, and geometric locality provably constrains what stabilizer measurement circuits can achieve
\cite{DelfosseBeverlandTremblay2021,BerthusenEtAl2025}. Since QSS is
inherently a network primitive~\cite{PirkerDur2019,AzumaEtAl2020}, this distinction motivates a joint study of access structures and topology-aware implementation costs.

The link between erasure correction and QSS access structures dates back to the work of Cleve, Gottesman, and Lo \cite{CleveGottesmanLo1999} and Gottesman \cite{Gottesman2000}. The use of logical operators to characterize correctability is also well established \cite{BravyiTerhal2009,YoshidaChuang2010}. Stabilizer-based QSS schemes are generally ramp schemes. In this setting, an intermediate set may hold only part of the secret. Gheorghiu showed that this partial information can be described by a subgroup of the logical Pauli group, called the \emph{information group} \cite{Gheorghiu2012}. This work builds on the earlier analysis of quantum ramp schemes by Ogawa \emph{et al.} \cite{OgawaEtAl2005} and the matroid and code-based results of Sarvepalli and Klappenecker \cite{SarvepalliKlappenecker2009}. Matsumoto later expressed these ideas directly in terms of the stabilizer check matrix \cite{Matsumoto2019,Matsumoto2020}. Strong security for classical secrets is a related but distinct problem. In quantum ramp schemes, it generally requires additional randomization \cite{ZhangMatsumoto2014,Matsumoto2019}. Other recent directions include communication-efficient QSS \cite{SenthoorSarvepalli2021,LiEtAl2026,LiLi2026}, advance sharing of quantum shares \cite{ShibataMatsumoto2023}, and measurement-free reconstruction circuits \cite{ChiwakiMatsumoto2025}. On the coding side,
hypergraph-product \cite{TillichZemor2014} and bivariate bicycle
\cite{BravyiEtAl2024} codes are the finite-length CSS families we use.
The cost of realizing their nonlocal checks on constrained hardware has
been studied for reconfigurable atom arrays \cite{XuEtAl2024}, modular and networked architectures \cite{ChandraKaurSeshadreesan2025,ChandraEtAl2026,ShawRengaswamy2026}, surgery-based schemes \cite{BaspinBerentCohen2025}, and connectivity reduction by routing \cite{ZhaoYanNi2025,BerthusenEtAl2025}. The rate bounds
for the sparse stabilizer codes in the erasure channel are due to Delfosse and Z{\'e}mor \cite{DelfosseZemor2013}, with finite-rate sparse constructions in \cite{TremblayEtAl2023} and erasure decoding in \cite{KuoOuyang2026}.
Against this background, the present paper does not claim the QSS--QEC correspondence, the existence of ramp behaviour, or the existence of an information group. It contributes a computational specialization of these facts to CSS qLDPC codes, and a network-cost layer that the information-theoretic literature does not address.
Table~\ref{tab:novelty-map} explicitly states this division.

\begin{table*}[t]
\centering
\caption{Positioning of the present work relative to established QEC--QSS concepts. The contribution of this work lies primarily in the computational specialization and integration of these ingredients for CSS qLDPC codes and topology-aware participant design.}
\label{tab:novelty-map}
\begin{tabular}{p{0.28\textwidth}
                p{0.30\textwidth}
                p{0.33\textwidth}}
\toprule
\textbf{Aspect}
&
\textbf{Established foundation}
&
\textbf{Role in the present work}
\\
\midrule

QEC--QSS correspondence
&
Quantum codes can realize secret-sharing schemes, with erasure correctability determining recovery properties.
&
We take this correspondence as a starting point and develop explicit computational tests for CSS qLDPC codes.
\\

Logical-operator support
&
Logical-operator supports characterize which subsystems can access encoded information.
&
We express the relevant support conditions through restricted ranks of $H_X$ and $H_Z$, enabling direct coalition classification.
\\

Information accessible to a subsystem
&
Stabilizer and logical-information formalisms describe the encoded information accessible to a subsystem.
&
We implement this structure through binary symplectic linear algebra and compute the accessible logical dimension $\kappa(A)$ for arbitrary coalitions.
\\

Share/participant assignment
&
Previous QSS work considers assignment of shares to participants and optimization for prescribed access structures.
&
For a fixed CSS code, we treat the partition of physical qubits among participants as a design variable and evaluate the induced security and reconstruction thresholds.
\\

Network-aware implementation
&
Graph routing and scheduling methods quantify the cost of implementing distributed interactions.
&
We combine CSS Tanner incidence, approximate Steiner routing, and scheduling depth with QSS access analysis under a common physical placement.
\\

Joint design
&
The above ingredients have largely been studied in separate contexts.
&
We integrate code-level access, logical-information profiling, participant partitioning, physical placement, and network cost in a common computational pipeline.
\\

\bottomrule
\end{tabular}
\end{table*}

Concretely, this work contributes the following.
\begin{enumerate}[label=\textup{(\arabic*)}]
    \item \emph{An exact computational characterization of CSS-induced
    access structures.} A rank criterion
    (Theorem~\ref{thm:exact-css-qss-access}) classifies any coalition from
    $\ell_X(A)$ and $\ell_Z(A)$ by $\Ftwo$ Gaussian elimination on
    restricted parity-check matrices, and the CSS complement duality
    (Lemma~\ref{lem:css-complement-duality}) yields exact profile
    symmetries that double as an implementation test.
    \item \emph{A symplectic accessibility profile for intermediate
    coalitions.} The accessible logical dimension $\kappa(A)$ (Definition~\ref{def:accessible-logical-dimension}) is the $\mathbb F_2$-rank of the symplectic pairing between locally supported logical classes. It is basis-independent and monotone (Proposition~\ref{prop:kappa-properties}) and counts the number of quantum logical qubits in the noncommuting matrix-factor part of the local logical algebra (Proposition~\ref{prop:accessible-subsystem}). This gives a CSS-specific computable realization of the symplectic component of the information-group structure known for stabilizer QSS.
    \item \emph{A network-aware design layer.} We separate the Tanner
    incidence count $\Wtot$, the Steiner routing proxy $\CSt$, and the
    scheduling-depth proxy $\Dsched$ of Definition~\ref{def:dsched}, and
    give a constructive recovery procedure based on cleaning logical
    operators onto authorized coalitions
    (Section~\ref{sec:topology-cost}).
    \item \emph{The participant partition as a design variable.} Treating
    the triple $(\mathcal{C},\mathcal{P},\Gphys)$ of code, partition, and
    physical graph as the object to be optimized, we show that block-level
    thresholds $\tsec,\trec$ are properties of the partition and not of the
    code alone, and quantify this on four codes
    (Sections~\ref{sec:numerical-results} and~\ref{sec:block-level-qss}).
\end{enumerate}

The remainder of the paper is organized as follows.
Section~\ref{sec:preliminaries} reviews CSS codes, quantum erasures and
the QEC--QSS correspondence.
Section~\ref{sec:access-structure} derives the algebraic
access-structure criterion and its duality consequences.
Section~\ref{sec:topology-cost} introduces the topology-aware embedding
cost and the constructive recovery procedure.
Section~\ref{sec:numerical-results} presents the numerical results.
Section~\ref{sec:block-level-qss} extends the framework to block-level
participant partitions and compares it with GHZ- and graph-state QSS.
Section~\ref{sec:discussion} discusses noisy channels, approximate QSS and
the limitations of the model, and
Section~\ref{sec:conclusion} concludes.

\section{Preliminaries: Quantum Secret Sharing, Erasures, and CSS qLDPC Codes}
\label{sec:preliminaries}

This section introduces the notation and concepts used throughout the
paper. We first review quantum secret-sharing access structures and their
connection to quantum error correction against erasures. We then recall
the stabilizer and CSS formalisms, with emphasis on the logical-operator
criterion for erasure correctability. Finally, we introduce the qLDPC
code families considered in our numerical study.

\subsection{Quantum secret sharing and access structures}
\label{subsec:qss-access}

Let
\[
    [n] := \{1,2,\ldots,n\}
\]
denote a set of $n$ participants. A quantum secret-sharing (QSS) scheme
for a secret system $S$ is specified by an encoding isometry
\begin{equation}
    V:\mathcal{H}_S
    \longrightarrow
    \mathcal{H}_{1}\otimes\mathcal{H}_{2}
    \otimes \cdots \otimes \mathcal{H}_{n},
    \label{eq:qss-encoding-isometry}
\end{equation}
where subsystem $i$ is distributed to participant $i$. In the present
work, the secret consists of $k_L$ logical qubits and every physical
share is a qubit, such that
\[
    \mathcal{H}_S \cong (\mathbb{C}^2)^{\otimes k_L},
    \qquad
    \mathcal{H}_{i} \cong \mathbb{C}^2.
\]

For a coalition $A\subseteq[n]$, let $\bar{A}:=[n]\setminus A$ denote
its complement. The channel from the secret to the shares held by $A$ is
\begin{equation}
    \mathcal{N}_{A}(\rho)
    :=
    \Tr_{\bar{A}}\!\left[V\rho V^\dagger\right].
    \label{eq:reduced-qss-channel}
\end{equation}
The complementary channel is correspondingly
\begin{equation}
    \mathcal{N}_{\bar{A}}(\rho)
    :=
    \Tr_{A}\!\left[V\rho V^\dagger\right].
    \label{eq:complementary-qss-channel}
\end{equation}

\begin{definition}[Authorized coalition]
\label{def:authorized-coalition}
A coalition $A\subseteq[n]$ is called \emph{authorized} if there exists
a recovery channel
\[
    \mathcal{R}_{A}:
    \mathcal{L}\!\left(\bigotimes_{i\in A}\mathcal{H}_{i}\right)
    \longrightarrow
    \mathcal{L}(\mathcal{H}_S)
\]
such that
\begin{equation}
    \mathcal{R}_{A}\circ\mathcal{N}_{A}
    =
    \operatorname{id}_{S}.
    \label{eq:authorized-reconstruction}
\end{equation}
Thus, every encoded quantum state can be recovered perfectly using only
the shares held by $A$.
\end{definition}

\begin{definition}[Forbidden coalition]
\label{def:forbidden-coalition}
A coalition $A\subseteq[n]$ is called \emph{forbidden} if its reduced
state is independent of the secret. Equivalently, there exists a density
operator $\sigma_A$ such that
\begin{equation}
    \mathcal{N}_{A}(\rho)=\sigma_A
    \qquad
    \text{for every density operator } \rho \text{ on } \mathcal{H}_S.
    \label{eq:forbidden-no-information}
\end{equation}
A forbidden coalition therefore obtains no information, classical or
quantum, about the encoded secret.
\end{definition}

\begin{definition}[Intermediate coalition]
\label{def:intermediate-coalition}
A coalition $A\subseteq[n]$ is \emph{intermediate} if it is neither
authorized nor forbidden.
\end{definition}

The corresponding families of coalitions are denoted by
\[
    \mathcal{A}_{\mathrm{auth}},
    \qquad
    \mathcal{A}_{\mathrm{forb}},
    \qquad
    \mathcal{A}_{\mathrm{int}},
\]
respectively. Together, they partition the power set $2^{[n]}$.

A perfect quantum threshold scheme $((t,n))$ for an arbitrary
quantum secret is a special case in which every coalition of
cardinality at least $t$ is authorized, while every coalition of
cardinality smaller than $t$ is forbidden. More generally, a ramp
QSS scheme admits intermediate coalitions. Such coalitions may retain
partial information about the secret while lacking the ability to
reconstruct it perfectly. Ramp schemes arise naturally from general
stabilizer and qLDPC codes, whose access structures are typically not
determined solely by coalition cardinality.

The no-cloning theorem imposes a fundamental constraint on a quantum
access structure: two disjoint coalitions cannot both be authorized.
Indeed, if $A\cap B=\varnothing$ and both coalitions could independently
recover an arbitrary unknown secret, the encoding would generate two
independent copies of that secret. For a pure-state perfect quantum threshold scheme sharing an arbitrary quantum secret, the no-cloning constraint requires $n<2t$. Under the standard assumptions of pure-state perfect QSS, authorized and
forbidden coalitions are complementary, which further gives
\[
n=2t-1.
\]
\cite{CleveGottesmanLo1999,Gottesman2000}.

\subsection{Quantum erasures and the QEC--QSS correspondence}
\label{subsec:qec-qss-correspondence}

A quantum erasure channel removes a known set of physical subsystems.
Unlike a general quantum error, the location of every erased qubit is
assumed to be available to the decoder. Let $E\subseteq[n]$ be an
erasure set and let $\bar{E}$ denote the surviving qubits. The erasure
of $E$ is correctable when there exists a recovery channel acting only
on $\bar{E}$ that restores the encoded logical state.

For an encoding isometry $V$, erasure correctability can be stated as
\begin{equation}
    \exists\,\mathcal{R}_{\bar{E}}
    \quad \text{such that} \quad
    \mathcal{R}_{\bar{E}}
    \circ
    \Tr_E\!\left[V(\cdot)V^\dagger\right]
    =
    \operatorname{id}_{S}.
    \label{eq:erasure-correction-definition}
\end{equation}
The QEC--QSS correspondence follows immediately from comparing
Eq.~\eqref{eq:erasure-correction-definition} with
Eq.~\eqref{eq:authorized-reconstruction}: a coalition $A$ is authorized
if and only if the erasure of $\bar{A}$ is correctable.

The complementary statement is equally important. If the erasure of
$A$ is correctable, then the channel to $A$ is constant and $A$ is
forbidden. This is a consequence of information--disturbance
complementarity: if the complement $\bar{A}$ can perfectly restore an
arbitrary secret, the subsystem $A$ cannot retain any information about
that secret. Therefore,
\begin{equation}
    A\in\mathcal{A}_{\mathrm{auth}}
    \iff
    \bar{A}\text{ is a correctable erasure set},
    \label{eq:authorized-erasure-equivalence}
\end{equation}
and
\begin{equation}
    A\in\mathcal{A}_{\mathrm{forb}}
    \iff
    A\text{ is a correctable erasure set}.
    \label{eq:forbidden-erasure-equivalence}
\end{equation}

These identities explain why quantum secret sharing is naturally
described in the language of erasure correction. They also show that,
for a general code, a coalition may be neither authorized nor forbidden.
Such a coalition belongs to the ramp region:
\begin{equation}
\begin{split}
    A\in\mathcal{A}_{\mathrm{int}}
    \iff
    &\;A \text{ and } \bar{A} \text{ are both}\\
    &\;\text{uncorrectable erasure sets}.
\end{split}
\label{eq:intermediate-erasure-equivalence}
\end{equation}

The distance $d$ of an $[[n,k_L,d]]$ quantum code gives an immediate
cardinality-based sufficient condition. Since all erasures of cardinality
at most $d-1$ are correctable,
\begin{equation}
    |E|\leq d-1
    \quad\Longrightarrow\quad
    E\text{ is correctable}.
    \label{eq:distance-erasure-guarantee}
\end{equation}
Accordingly, every coalition $A$ satisfying
\begin{equation}
    |A|\geq n-d+1
    \label{eq:distance-authorized-guarantee}
\end{equation}
is authorized, whereas every coalition satisfying
\begin{equation}
    |A|\leq d-1
    \label{eq:distance-forbidden-guarantee}
\end{equation}
is forbidden. Between these two guaranteed regions, code distance alone
does not determine the access structure. The exact classification depends
on the support geometry of logical operators, which is the central object
of our analysis.

\subsection{Stabilizer codes and logical operators}
\label{subsec:stabilizer-logicals}

Let $\mathcal{P}_n$ denote the $n$-qubit Pauli group, modulo global
phases. An $[[n,k_L,d]]$ stabilizer code is specified by an abelian
subgroup
\[
    \mathcal{S}\subseteq\mathcal{P}_n,
    \qquad
    -I\notin\mathcal{S},
\]
called the stabilizer group. The code space $\mathcal{C}$ is the
simultaneous $+1$ eigenspace of all elements of $\mathcal{S}$. If
$\mathcal{S}$ has $n-k_L$ independent generators, then
\[
    \dim(\mathcal{C})=2^{k_L}.
\]

The normalizer of the stabilizer group is
\[
    \mathcal{N}(\mathcal{S})
    :=
    \left\{
    P\in\mathcal{P}_n:
    PS=SP \text{ for all } S\in\mathcal{S}
    \right\}.
\]
Elements of $\mathcal{N}(\mathcal{S})\setminus\mathcal{S}$ preserve the
code space but act nontrivially on logical degrees of freedom. Logical
Pauli operators are therefore equivalence classes in
\[
    \mathcal{N}(\mathcal{S})/\mathcal{S}.
\]

For a Pauli operator $P$, let $\supp(P)\subseteq[n]$ be its support and
let $\mathrm{wt}(P):=|\supp(P)|$ be its weight. The code distance is
\begin{equation}
    d
    =
    \min_{P\in\mathcal{N}(\mathcal{S})\setminus\mathcal{S}}
    \mathrm{wt}(P).
    \label{eq:stabilizer-distance}
\end{equation}
Thus, the distance is the minimum number of qubits on which a nontrivial
logical Pauli can be supported.

For erasure correction, the relevant criterion is stronger than the
global distance condition because it concerns a specified set of qubits.

\begin{proposition}[Logical-operator criterion for erasure correction]
\label{prop:logical-erasure-criterion}
Let $E\subseteq[n]$. The erasure of $E$ is correctable if and only if
\begin{equation}
    \mathcal{N}(\mathcal{S})\cap\mathcal{P}(E)
    =
    \mathcal{S}\cap\mathcal{P}(E),
    \label{eq:logical-erasure-criterion}
\end{equation}
where $\mathcal{P}(E)$ is the subgroup of Pauli operators supported
entirely on $E$.
Equivalently, $E$ is correctable if and only if it supports no nontrivial
logical Pauli operator.
\end{proposition}

\emph{Proof.} See Appendix~\ref{app:secII-proofs}.

Proposition~\ref{prop:logical-erasure-criterion} directly links logical
operator supports to QSS access structures. In particular, a coalition
$A$ is authorized whenever no nontrivial logical operator can be
supported on $\bar{A}$; it is forbidden whenever no nontrivial logical
operator can be supported on $A$.

\subsection{CSS codes}
\label{subsec:css-codes}

CSS codes form a particularly convenient subclass of stabilizer codes.
They admit a binary representation in terms of two parity-check matrices
\[
    H_X\in\Ftwo^{m_X\times n},
    \qquad
    H_Z\in\Ftwo^{m_Z\times n},
\]
satisfying the commutation condition
\begin{equation}
    H_XH_Z^{T}=0
    \qquad
    \text{over } \Ftwo.
    \label{eq:css-commutation}
\end{equation}
Each row of $H_X$ specifies an $X$-type stabilizer generator, while each
row of $H_Z$ specifies a $Z$-type stabilizer generator. The number of
encoded logical qubits is
\begin{equation}
    k_L
    =
    n-\rank(H_X)-\rank(H_Z).
    \label{eq:css-logical-dimension}
\end{equation}

For a binary vector $x\in\Ftwo^n$, define
\[
    X(x):=\bigotimes_{j=1}^{n}X_j^{x_j},
    \qquad
    Z(x):=\bigotimes_{j=1}^{n}Z_j^{x_j}.
\]
The $X$-type logical-operator space is
\begin{equation}
    \mathcal{L}_{X}
    =
    \ker(H_Z)\big/\operatorname{row}(H_X),
    \label{eq:x-logical-space}
\end{equation}
while the $Z$-type logical-operator space is
\begin{equation}
    \mathcal{L}_{Z}
    =
    \ker(H_X)\big/\operatorname{row}(H_Z).
    \label{eq:z-logical-space}
\end{equation}
Their dimensions are both $k_L$.

Let $E\subseteq[n]$, and let $H_X|_E$ and $H_Z|_E$ denote the
restrictions of $H_X$ and $H_Z$ to the columns indexed by $E$. The
number of independent $X$-type logical operators whose support can be
contained in $E$ is denoted by $\ell_X(E)$; similarly,
$\ell_Z(E)$ denotes the corresponding number of $Z$-type logical
operators. Section~\ref{sec:access-structure} derives explicit
rank-based expressions for these quantities and uses them to classify
arbitrary coalitions.

For a CSS code, the logical-operator criterion of
Proposition~\ref{prop:logical-erasure-criterion} becomes
\begin{equation}
    E\text{ is correctable}
    \iff
    \ell_X(E)=0
    \quad\text{and}\quad
    \ell_Z(E)=0.
    \label{eq:css-erasure-correctability}
\end{equation}
This decomposition is computationally useful because all required
operations reduce to binary linear algebra.

\subsection{qLDPC code families considered in this work}
\label{subsec:qldpc-families}

A CSS code is called a quantum low-density parity-check (qLDPC) code if
the weights of the rows and columns of $H_X$ and $H_Z$ remain bounded, or
at least grow slowly, with the block length. Equivalently, every
stabilizer generator has bounded weight and every physical qubit
participates in a bounded number of parity checks. The associated Tanner
graph is sparse.

The sparsity of a qLDPC code should be interpreted carefully. It bounds
the local combinatorial degree of the code, but it does not automatically
imply a geometrically local physical implementation. In particular, a
weight-$w$ stabilizer may involve $w$ qubits that are widely separated in
a physical network. The mismatch between the Tanner graph and the
physical communication graph is therefore a central source of
entanglement-routing overhead. This motivates the topology-aware model
introduced in Section~\ref{sec:topology-cost}.

Our analysis considers the following representative CSS families.

\paragraph{Repetition- and Hamming-derived CSS constructions.}
These small and highly structured codes provide analytically
interpretable baselines. They are used to validate the access-structure
classification and to compare numerical observations with known
finite-size behavior.

\paragraph{Hypergraph-product codes.}
Given two binary parity-check matrices
\[
    H_1\in\Ftwo^{r_1\times n_1},
    \qquad
    H_2\in\Ftwo^{r_2\times n_2},
\]
the hypergraph-product (HGP) construction defines a CSS code with
\begin{equation}
    H_X
    =
    \left[
        H_1\otimes I_{n_2}
        \;\middle|\;
        I_{r_1}\otimes H_2^T
    \right],
    \label{eq:hgp-hx}
\end{equation}
and
\begin{equation}
    H_Z
    =
    \left[
        I_{n_1}\otimes H_2
        \;\middle|\;
        H_1^T\otimes I_{r_2}
    \right].
    \label{eq:hgp-hz}
\end{equation}
The relation $H_XH_Z^T=0$ follows directly from the binary tensor-product
identities. When $H_1$ and $H_2$ are sparse, the resulting CSS code is
qLDPC. In this work, we consider HGP codes derived from repetition,
Hamming, and random sparse parity-check matrices.

\paragraph{Random sparse HGP codes.}
To probe how the access transition depends on the combinatorial
structure of the underlying classical matrices, we also construct HGP
codes from sparse random binary matrices with prescribed column weights.
These instances allow controlled numerical comparisons between regular
or algebraically structured constructions and unstructured sparse
ensembles.

\paragraph{Bivariate bicycle codes.}
Bivariate bicycle (BB) codes are quasi-cyclic CSS qLDPC codes defined
from two sparse polynomials over a bivariate quotient ring. They provide
finite-length codes with favorable rate, sparse checks, and practical
decoding properties. Our numerical study includes a representative
\([[72,12,6]]\) BB code, which provides a high-rate finite-size benchmark
for the access-structure analysis.

The remainder of the paper develops the exact access-structure test for
these codes, studies its finite-size numerical behavior, and quantifies
the topology-dependent overhead required to implement the corresponding
QSS protocols.

\section{Exact Algebraic Characterization of CSS-Induced Access Structures}
\label{sec:access-structure}

In this section, we derive an exact algebraic characterization of the
quantum secret-sharing access structure induced by an arbitrary CSS code.
The central observation is that the classification of a coalition reduces
to determining whether its support, or the support of its complement,
contains a nontrivial logical operator. For CSS codes, this problem can be
solved entirely through binary linear algebra.

Throughout this section, let
\[
    \mathcal{C}
    =
    [[n,k_L,d]]
\]
be a CSS stabilizer code specified by binary parity-check matrices
\[
    H_X\in\mathbb{F}_2^{m_X\times n},
    \qquad
    H_Z\in\mathbb{F}_2^{m_Z\times n},
\]
satisfying
\[
    H_XH_Z^T=0.
\]
We write
\[
    r_X:=\rank(H_X),
    \qquad
    r_Z:=\rank(H_Z),
\]
so that the number of encoded logical qubits is
\begin{equation}
    k_L=n-r_X-r_Z.
    \label{eq:section3-klogical}
\end{equation}
Proofs of the technical results in this section are collected
in Appendix~\ref{app:secIII-proofs} to keep the main
development focused on the computational framework.

\subsection{Restricted supports and binary subspaces}
\label{subsec:restricted-supports}

Let $E\subseteq[n]$ be a subset of physical qubits. We denote its
complement by
\[
    E^c=[n]\setminus E.
\]
For any binary matrix $M$ with $n$ columns, $M|_E$ denotes the
restriction of $M$ to the columns indexed by $E$. Likewise, $M|_{E^c}$
is the restriction to the complementary set of columns.

We introduce the coordinate subspace
\begin{equation}
    \mathbb{F}_2^E
    :=
    \left\{
        x\in\mathbb{F}_2^n:
        \supp(x)\subseteq E
    \right\}.
    \label{eq:coordinate-subspace}
\end{equation}
Thus, $\mathbb{F}_2^E$ is naturally isomorphic to
$\mathbb{F}_2^{|E|}$.

Recall from Section~\ref{subsec:css-codes} that the CSS logical-operator
spaces are
\begin{equation}
    \mathcal{L}_X
    =
    \ker(H_Z)/\operatorname{row}(H_X),
    \label{eq:section3-lx-global}
\end{equation}
and
\begin{equation}
    \mathcal{L}_Z
    =
    \ker(H_X)/\operatorname{row}(H_Z).
    \label{eq:section3-lz-global}
\end{equation}
An element of $\ker(H_Z)$ represents an $X$-type Pauli operator commuting
with every $Z$-type stabilizer. It is a nontrivial logical $X$ operator
precisely when it is not contained in $\operatorname{row}(H_X)$.
Similarly, vectors in $\ker(H_X)\setminus\operatorname{row}(H_Z)$
represent nontrivial logical $Z$ operators.

To determine whether a logical operator can be supported entirely on
$E$, one must restrict both the normalizer condition and the stabilizer
equivalence relation to this subset.

\subsection{Logical operators supported on a prescribed subset}
\label{subsec:supported-logicals}

\begin{definition}[Supported logical dimensions]
\label{def:supported-logical-dimensions}
For $E\subseteq[n]$, define
\begin{equation}
    \ell_X(E)
    :=
    \dim
    \left[
    \frac{\ker(H_Z)\cap\mathbb{F}_2^E}
    {\operatorname{row}(H_X)\cap\mathbb{F}_2^E}
    \right],
    \label{eq:ellx-definition}
\end{equation}
and
\begin{equation}
    \ell_Z(E)
    :=
    \dim
    \left[
    \frac{\ker(H_X)\cap\mathbb{F}_2^E}
    {\operatorname{row}(H_Z)\cap\mathbb{F}_2^E}
    \right].
    \label{eq:ellz-definition}
\end{equation}
The quantities $\ell_X(E)$ and $\ell_Z(E)$ count independent
nontrivial $X$-type and $Z$-type logical operators, respectively, that
admit representatives supported entirely on $E$.
\end{definition}

The following result transforms
Definitions~\ref{def:supported-logical-dimensions} into formulas involving
only ranks of restricted parity-check matrices.

\begin{lemma}[Dimension of locally supported normalizer spaces]
\label{lem:local-normalizer-dimensions}
For every $E\subseteq[n]$,
\begin{equation}
    \dim\!\left(\ker(H_Z)\cap\mathbb{F}_2^E\right)
    =
    |E|-\rank(H_Z|_E),
    \label{eq:kernel-hz-restricted}
\end{equation}
and
\begin{equation}
    \dim\!\left(\ker(H_X)\cap\mathbb{F}_2^E\right)
    =
    |E|-\rank(H_X|_E).
    \label{eq:kernel-hx-restricted}
\end{equation}
\end{lemma}


The denominator in Definition~\ref{def:supported-logical-dimensions}
requires more care. A stabilizer vector belongs to
$\operatorname{row}(H_X)\cap\mathbb{F}_2^E$ when its restriction to
$E^c$ vanishes. The dimension of this subspace is determined by the
rank loss caused by restricting $H_X$ to $E^c$.

\begin{lemma}[Dimension of locally supported stabilizer spaces]
\label{lem:local-stabilizer-dimensions}
For every $E\subseteq[n]$,
\begin{equation}
    \dim\!\left(
    \operatorname{row}(H_X)\cap\mathbb{F}_2^E
    \right)
    =
    \rank(H_X)-\rank(H_X|_{E^c}),
    \label{eq:local-x-stabilizer-dim}
\end{equation}
and
\begin{equation}
    \dim\!\left(
    \operatorname{row}(H_Z)\cap\mathbb{F}_2^E
    \right)
    =
    \rank(H_Z)-\rank(H_Z|_{E^c}).
    \label{eq:local-z-stabilizer-dim}
\end{equation}
\end{lemma}


Combining Lemmas~\ref{lem:local-normalizer-dimensions} and
\ref{lem:local-stabilizer-dimensions} gives the desired rank formulas.

\begin{proposition}[Rank formula for supported CSS logical operators]
\label{prop:rank-formula-supported-logicals}
For every subset $E\subseteq[n]$,
\begin{equation}
    \ell_X(E)
    =
    |E|
    -\rank(H_Z|_E)
    -\rank(H_X)
    +\rank(H_X|_{E^c}),
    \label{eq:ellx-rank-formula}
\end{equation}
and
\begin{equation}
    \ell_Z(E)
    =
    |E|
    -\rank(H_X|_E)
    -\rank(H_Z)
    +\rank(H_Z|_{E^c}).
    \label{eq:ellz-rank-formula}
\end{equation}
\end{proposition}

\emph{Proof.} See Appendix~\ref{app:secIII-proofs-subsection-3}.

The quantities in Proposition~\ref{prop:rank-formula-supported-logicals}
are nonnegative integers. In practice, this provides a direct
computational test based only on Gaussian elimination over
$\mathbb{F}_2$.

\subsection{Exact classification of coalitions}
\label{subsec:exact-coalition-classification}

We now translate the supported-logical dimensions into a classification
of QSS coalitions.

\begin{theorem}[Restricted-rank characterization of CSS-induced QSS access structures]
\label{thm:exact-css-qss-access}
Let $\mathcal{C}$ be a CSS code and let $A\subseteq[n]$ be a coalition
of participants. Then:

\begin{enumerate}[label=\textup{(\roman*)}]
    \item The coalition $A$ is authorized if and only if
    \begin{equation}
        \ell_X(A^c)=0
        \qquad\text{and}\qquad
        \ell_Z(A^c)=0.
        \label{eq:authorized-ell-condition}
    \end{equation}

    \item The coalition $A$ is forbidden if and only if
    \begin{equation}
        \ell_X(A)=0
        \qquad\text{and}\qquad
        \ell_Z(A)=0.
        \label{eq:forbidden-ell-condition}
    \end{equation}

    \item The coalition $A$ is intermediate if and only if both $A$ and
    $A^c$ support at least one nontrivial logical operator:
    \begin{equation}
        \ell_X(A)+\ell_Z(A)>0
        \quad\text{and}\quad
        \ell_X(A^c)+\ell_Z(A^c)>0.
        \label{eq:intermediate-ell-condition}
    \end{equation}
\end{enumerate}
\end{theorem}

\emph{Proof.} See Appendix~\ref{app:secIII-proofs-subsection-4}.

Theorem~\ref{thm:exact-css-qss-access} gives an exact directly computable restricted-parity-check characterization of the QSS access structure induced by a CSS code. In particular, it
shows that the access structure is determined not only by the code
distance but by the detailed distribution of logical-operator supports.

\begin{corollary}[Distance-based guaranteed regions]
\label{cor:distance-based-regions}
Let $\mathcal{C}$ be an $[[n,k_L,d]]$ CSS code. Then:
\begin{equation}
    |A|\leq d-1
    \quad\Longrightarrow\quad
    A\in\mathcal{A}_{\mathrm{forb}},
    \label{eq:guaranteed-forbidden-region}
\end{equation}
and
\begin{equation}
    |A|\geq n-d+1
    \quad\Longrightarrow\quad
    A\in\mathcal{A}_{\mathrm{auth}}.
    \label{eq:guaranteed-authorized-region}
\end{equation}
\end{corollary}


Corollary~\ref{cor:distance-based-regions} provides only extremal guarantees. In particular, for codes with $k_L>1$, the region
\[
    d \leq |A| \leq n-d
\]
can contain a nontrivial mixture of authorized, forbidden, and
intermediate coalitions. This is precisely the regime in which numerical
sampling and exact rank tests are required.

\subsection{Monotonicity and complementarity properties}
\label{subsec:monotonicity-complementarity}

The access structures induced by quantum codes obey several general
properties.

\begin{proposition}[Monotonicity of authorization]
\label{prop:authorization-monotonicity}
Let $A,B\subseteq[n]$ satisfy $A\subseteq B$. If $A$ is authorized,
then $B$ is authorized.
\end{proposition}


\begin{proposition}[Monotonicity of forbidden coalitions]
\label{prop:forbidden-monotonicity}
Let $A,B\subseteq[n]$ satisfy $B\subseteq A$. If $A$ is forbidden,
then $B$ is forbidden.
\end{proposition}


For pure-state QSS schemes, authorized and forbidden structures are
dual under complementation. This fact follows from the no-cloning theorem
and from the reversibility of the encoding isometry
\cite{CleveGottesmanLo1999,Gottesman2000}.

\begin{corollary}[Complementarity for pure-state QSS]
\label{cor:pure-qss-complementarity}
For a pure-state QSS scheme induced by an isometric encoding, a coalition
$A$ is authorized if and only if $A^c$ is forbidden.
\end{corollary}


The complementarity relation does not imply that every subset is either
authorized or forbidden. For general CSS qLDPC codes with multiple
logical qubits, intermediate coalitions can occur. What complementarity
guarantees is the duality
\begin{equation}
    A\in\mathcal{A}_{\mathrm{auth}}
    \iff
    A^c\in\mathcal{A}_{\mathrm{forb}}.
    \label{eq:authorized-forbidden-duality}
\end{equation}
The intermediate family is closed under complementation:
\begin{equation}
    A\in\mathcal{A}_{\mathrm{int}}
    \iff
    A^c\in\mathcal{A}_{\mathrm{int}}.
    \label{eq:intermediate-complement-duality}
\end{equation}

The information-theoretic duality~\eqref{eq:authorized-forbidden-duality}
has an exact algebraic counterpart at the level of supported logical
dimensions, which we now state and which will be used as a numerical
consistency check in Section~\ref{sec:numerical-results}.

\begin{lemma}[CSS complement duality]
\label{lem:css-complement-duality}
Let $\mathcal{C}$ be a CSS code with parity-check matrices $H_X,H_Z$
satisfying $H_XH_Z^{T}=0$ and $k_L=n-\rank H_X-\rank H_Z$. Then for every
$A\subseteq[n]$,
\begin{equation}
    \ell_X(A)=k_L-\ell_Z(A^c),
    \qquad
    \ell_Z(A)=k_L-\ell_X(A^c).
    \label{eq:css-complement-duality}
\end{equation}
\end{lemma}

\emph{Proof.} See Appendix~\ref{app:secIII-proofs-subsection-8}.

\begin{corollary}[Exact symmetry of the access profile]
\label{cor:access-profile-symmetry}
For a pure CSS-induced QSS scheme and every $0\le a\le n$,
\begin{equation}
    \Pauth(a)=\Pforb(n-a),
    \qquad
    \Pint(a)=\Pint(n-a).
    \label{eq:access-profile-symmetry}
\end{equation}
\end{corollary}


\subsection{Access-transition observables}
\label{subsec:access-transition-observables}

The exact characterization above can be used to enumerate the complete
access structure for small block lengths. For qLDPC codes of moderate or
large length, exhaustive enumeration over $2^n$ coalitions is
infeasible. We therefore characterize access structures statistically as
a function of coalition size.

For each $a\in\{0,1,\ldots,n\}$, define
\begin{equation}
    P_{\mathrm{auth}}(a)
    :=
    \Pr_{A\subseteq[n],\,|A|=a}
    \left[
        A\in\mathcal{A}_{\mathrm{auth}}
    \right],
    \label{eq:pauth-definition}
\end{equation}
\begin{equation}
    P_{\mathrm{forb}}(a)
    :=
    \Pr_{A\subseteq[n],\,|A|=a}
    \left[
        A\in\mathcal{A}_{\mathrm{forb}}
    \right],
    \label{eq:pforb-definition}
\end{equation}
and
\begin{equation}
    P_{\mathrm{int}}(a)
    :=
    \Pr_{A\subseteq[n],\,|A|=a}
    \left[
        A\in\mathcal{A}_{\mathrm{int}}
    \right].
    \label{eq:pint-definition}
\end{equation}
By construction,
\begin{equation}
    P_{\mathrm{auth}}(a)
    +
    P_{\mathrm{forb}}(a)
    +
    P_{\mathrm{int}}(a)
    =
    1.
    \label{eq:probability-normalization}
\end{equation}

For pure-state QSS encodings, complementarity implies the finite-size
symmetries
\begin{equation}
    P_{\mathrm{auth}}(a)
    =
    P_{\mathrm{forb}}(n-a),
    \label{eq:pauth-pforb-symmetry}
\end{equation}
and
\begin{equation}
    P_{\mathrm{int}}(a)
    =
    P_{\mathrm{int}}(n-a).
    \label{eq:pint-symmetry}
\end{equation}
These identities provide useful consistency checks for numerical
implementations.

To summarize the authorization transition, we introduce the normalized
coalition fraction
\begin{equation}
    q:=\frac{a}{n}.
    \label{eq:normalized-coalition-size}
\end{equation}
For $p\in(0,1)$, the empirical $p$-authorization threshold is defined as
\begin{equation}
    q_p
    :=
    \inf\left\{
        q\in[0,1]:
        P_{\mathrm{auth}}(q)\geq p
    \right\}.
    \label{eq:qp-definition}
\end{equation}
In numerical experiments, we use in particular
\[
    q_{10},
    \qquad
    q_{50},
    \qquad
    q_{90},
\] where $q_{50}$ is an empirical statistical midpoint and define the transition width
\begin{equation}
    \Delta q
    :=
    q_{90}-q_{10}.
    \label{eq:transition-width}
\end{equation}
A small value of $\Delta q$ indicates a sharp authorization
transition, whereas a large value indicates a broad authorization
transition. Although a broad authorization transition is often
associated with a substantial intermediate-coalition regime,
$\Delta q$ does not directly measure the size of the ramp region,
because it is defined solely from $P_{\mathrm{auth}}(q)$. We therefore
use $A_{\mathrm{ramp}}$, defined from $P_{\mathrm{int}}(q)$, as the
complementary observable that directly quantifies the prevalence of
intermediate coalitions.

Finally, we define the normalized ramp area
\begin{equation}
    A_{\mathrm{ramp}}
    :=
    \int_0^1 P_{\mathrm{int}}(q)\,dq,
    \label{eq:ramp-area}
\end{equation}
which is evaluated numerically using the discrete set of sampled
coalition sizes. This quantity measures the average fraction of
coalitions that belong to the intermediate structure across all relative
coalition sizes.

\subsection{Algorithmic implementation}
\label{subsec:access-algorithm}

The rank formulas of
Proposition~\ref{prop:rank-formula-supported-logicals} lead to a direct
algorithm for coalition classification.

\begin{algorithm}[t]
\caption{Exact CSS-QSS coalition classification}
\label{alg:css-qss-classification}
\begin{algorithmic}[1]
\Require CSS parity-check matrices $H_X,H_Z$ and coalition $A\subseteq[n]$
\Ensure \texttt{authorized}, \texttt{forbidden}, or \texttt{intermediate}
\State Compute $A^c=[n]\setminus A$
\State Compute $\ell_X(A),\ell_Z(A)$ using
Eqs.~\eqref{eq:ellx-rank-formula}--\eqref{eq:ellz-rank-formula}
\State Compute $\ell_X(A^c),\ell_Z(A^c)$ using
Eqs.~\eqref{eq:ellx-rank-formula}--\eqref{eq:ellz-rank-formula}
\If{$\ell_X(A^c)=0$ and $\ell_Z(A^c)=0$}
    \State \Return \texttt{authorized}
\ElsIf{$\ell_X(A)=0$ and $\ell_Z(A)=0$}
    \State \Return \texttt{forbidden}
\Else
    \State \Return \texttt{intermediate}
\EndIf
\end{algorithmic}
\end{algorithm}

For each coalition, the dominant operations are rank computations over
$\mathbb{F}_2$ for the restricted matrices
\[
    H_X|_A,\quad H_Z|_A,\quad
    H_X|_{A^c},\quad H_Z|_{A^c}.
\]
The global ranks $r_X$ and $r_Z$ are computed once per code instance and
then reused for every coalition. This is the procedure implemented in
our numerical framework using finite-field linear algebra in SageMath.

For small codes, all coalitions of a given size may be enumerated
exactly. For larger qLDPC codes, we estimate
$P_{\mathrm{auth}}$, $P_{\mathrm{forb}}$, and $P_{\mathrm{int}}$ by
uniform Monte Carlo sampling of coalitions of fixed cardinality. The
resulting curves, threshold parameters, and ramp-area observables are
reported in Section~\ref{sec:numerical-results}.

\section{Topology-Aware Embedding Cost and Constructive Recovery}
\label{sec:topology-cost}

The algebraic analysis of Section~\ref{sec:access-structure} determines
which coalitions can reconstruct the encoded secret and which coalitions
obtain no information about it. This characterization is independent of
the physical layout of the shares. In a distributed quantum network,
however, the physical cost of preparing, maintaining, and measuring a
CSS code depends strongly on the connectivity between the network nodes.

In this section, we introduce a topology-aware cost model for CSS qLDPC
codes. The model explicitly distinguishes the sparse combinatorial
structure of a Tanner graph from its physical realization on a quantum
communication network. We use weighted Steiner trees to quantify the
cost of connecting the physical qubits participating in a stabilizer
measurement or preparation operation. The present model abstracts any distributed stabilizer-measurement primitive whose communication support is a connected subgraph spanning the check qubits; architecture-specific constants relating tree edges to Bell-pair consumption or gate count are intentionally left outside the present model. We then describe a constructive procedure for recovering logical operators on authorized coalitions by solving binary linear systems.

\subsection{Physical communication graph}
\label{subsec:physical-network}

Let
\begin{equation}
    G_{\mathrm{phys}}
    =
    (V_{\mathrm{phys}},E_{\mathrm{phys}},w)
    \label{eq:physical-graph}
\end{equation}
be a connected, undirected weighted graph representing a quantum
communication network. The vertex set $V_{\mathrm{phys}}$ represents
quantum processors, repeater stations, or storage nodes. An edge
$(u,v)\in E_{\mathrm{phys}}$ represents an available physical
entanglement-generation link. Its nonnegative weight
\[
    w(u,v)\geq 0
\]
quantifies the cost of using this link. Depending on the application,
$w(u,v)$ may represent physical distance, inverse Bell-pair generation
rate, expected Bell-pair consumption, latency, loss, infidelity, or a
weighted combination of such quantities.

Let the $n$ physical data qubits of a CSS code be mapped to network
vertices by a placement map
\begin{equation}
    \pi:[n]\longrightarrow V_{\mathrm{phys}}.
    \label{eq:qubit-placement-map}
\end{equation}
The simplest setting assigns one data qubit to each participant:
\[
    V_{\mathrm{phys}}=[n],
    \qquad
    \pi(i)=i.
\]
More generally, several physical qubits may be located at the same
network node. This setting is essential when a participant holds a block
of shares, as considered in Section~\ref{sec:block-level-qss}.

The code structure is described by the $X$- and $Z$-Tanner graphs
\[
    T_X=(V_Q\sqcup V_X,E_X),
    \qquad
    T_Z=(V_Q\sqcup V_Z,E_Z),
\]
where:
\begin{itemize}
    \item $V_Q=[n]$ is the set of data-qubit vertices;
    \item $V_X$ is the set of $X$-check vertices, one for each row of
    $H_X$;
    \item $V_Z$ is the set of $Z$-check vertices, one for each row of
    $H_Z$;
    \item $(i,s)\in E_X$ if the $X$-type check $s$ acts nontrivially on
    qubit $i$;
    \item $(i,s)\in E_Z$ if the $Z$-type check $s$ acts nontrivially on
    qubit $i$.
\end{itemize}

For a qLDPC code, the degrees of the Tanner graphs are bounded or slowly
growing with the block length. Equivalently, each stabilizer generator
has low weight and each data qubit participates in a bounded number of
checks. This sparsity is advantageous, but it is not sufficient to
guarantee a local physical implementation. A stabilizer of low weight
may act on qubits that are separated by long paths in
$G_{\mathrm{phys}}$. This distinction between Tanner-graph sparsity and
geometric locality is central to qLDPC implementation constraints
\cite{DelfosseBeverlandTremblay2021,BerthusenEtAl2025}.

\subsection{Stabilizer supports and Steiner-tree routing}
\label{subsec:steiner-cost}

For a stabilizer generator $s$, let
\[
    \supp(s)\subseteq[n]
\]
be the set of data qubits on which $s$ acts nontrivially. Its physical
terminal set under the placement $\pi$ is
\begin{equation}
    \Pi(s)
    :=
    \{\pi(i):i\in\supp(s)\}
    \subseteq V_{\mathrm{phys}}.
    \label{eq:stabilizer-terminal-set}
\end{equation}

A distributed measurement of $s$, or a distributed entangling operation
that prepares correlations associated with $s$, requires the nodes in
$\Pi(s)$ to be connected by entanglement-routing resources. A natural
graph-theoretic model is the minimum-weight Steiner tree spanning these
terminals.

\begin{definition}[Steiner routing cost]
\label{def:steiner-cost}
For a terminal set $U\subseteq V_{\mathrm{phys}}$, define
\begin{equation}
    \operatorname{ST}_{G_{\mathrm{phys}}}(U)
    :=
    \min_{\substack{T\subseteq G_{\mathrm{phys}}\\
                    T\text{ connected}\\
                    U\subseteq V(T)}}
    \sum_{e\in E(T)} w(e).
    \label{eq:steiner-tree-cost}
\end{equation}
The minimization is over connected subgraphs $T$ containing all
terminals in $U$. Since all edge weights are nonnegative, an optimal
connected subgraph can be chosen to be a tree.
\end{definition}

The quantity
$\operatorname{ST}_{G_{\mathrm{phys}}}(\Pi(s))$ measures the minimum
routing cost needed to connect the qubits in the support of the
stabilizer $s$. It is a proxy for the Bell-pair resources and
entanglement-swapping operations required to realize a nonlocal
multipartite interaction. Steiner trees have previously appeared in
quantum-network protocols for multipartite entanglement and graph-state
distribution, where a tree connects a chosen set of terminal users and
local fusion or measurement operations convert pairwise entanglement into
a multipartite resource \cite{PirkerDur2019,AzumaEtAl2020}.

For the CSS code, let $\mathcal{S}_X$ and $\mathcal{S}_Z$ denote chosen
independent generating sets of $X$- and $Z$-type stabilizers. We define
the total topology-aware stabilizer cost by
\begin{equation}
    \CSt(\pi;\Gphys)
    :=
    \sum_{s\in\mathcal{S}_X}
    \operatorname{ST}_{\Gphys}(\Pi(s))
    \;+\!\!
    \sum_{s\in\mathcal{S}_Z}
    \operatorname{ST}_{\Gphys}(\Pi(s)).
    \label{eq:total-stabilizer-cost}
\end{equation}

When $G_{\mathrm{phys}}$ is unweighted, we set $w(e)=1$ and
$C_{\mathrm{St}}$ equals the total number of physical network edges used
by the minimum Steiner trees, counted with multiplicity across
stabilizer generators. If a network link must be consumed repeatedly
during distinct stabilizer measurements, this multiplicity has a direct
resource interpretation.

\begin{remark}[Cost model interpretation]
\label{rem:cost-interpretation}
The cost in Eq.~\eqref{eq:total-stabilizer-cost} is not claimed to equal
the exact minimum number of Bell pairs for every hardware platform or
measurement protocol. Rather, it is a code- and topology-dependent
routing proxy that captures the basic obstruction caused by nonlocal
stabilizer supports. More detailed architectures may associate different
weights with elementary links, swaps, purification rounds, memories, or
teleported gates.
\end{remark}

\subsection{Preparation and syndrome-extraction proxies}
\label{subsec:preparation-cost}

The Steiner cost captures nonlocal routing requirements but does not
alone account for local gate operations. We therefore introduce
additional circuit-level proxies.

Let
\[
    W_X
    :=
    \sum_{s\in\mathcal{S}_X}|\supp(s)|,
    \qquad
    W_Z
    :=
    \sum_{s\in\mathcal{S}_Z}|\supp(s)|.
\]
Equivalently,
\begin{equation}
    W_X=\|H_X\|_0,
    \qquad
    W_Z=\|H_Z\|_0,
    \label{eq:incidence-counts}
\end{equation}
where $\|H\|_0$ denotes the number of nonzero binary entries in $H$.
The total Tanner incidence count is
\begin{equation}
    W_{\mathrm{tot}}
    :=
    W_X+W_Z.
    \label{eq:total-incidence-count}
\end{equation}

Under a standard ancilla-assisted circuit, measuring a weight-$w_s$
Pauli stabilizer requires $O(w_s)$ two-qubit data--ancilla interactions.
Thus, $W_{\mathrm{tot}}$ provides a lower-level proxy for the total
number of local entangling gates in one full syndrome-extraction round.

For a qLDPC family with bounded row weights, one has
\begin{equation}
    W_{\mathrm{tot}}=O(n).
    \label{eq:qldpc-linear-incidence}
\end{equation}
This linear scaling is a direct consequence of check sparsity. However,
the circuit depth depends on whether interactions corresponding to
different Tanner-graph edges can be scheduled in parallel.

Let $\Delta_X$ and $\Delta_Z$ denote the maximum degrees of the Tanner
graphs $T_X$ and $T_Z$, respectively. In an all-to-all connectivity model
with one ancilla per check, edge coloring of each Tanner graph yields a
syndrome-extraction schedule of depth bounded by
\begin{equation}
    D_{\mathrm{syn}}
    \lesssim
    \Delta_X+\Delta_Z+O(1).
    \label{eq:all-to-all-depth-bound}
\end{equation}
For a bounded-degree qLDPC family, this is $O(1)$. In contrast, on a
restricted geometric network, long-range routing and edge congestion can
increase the depth substantially. General lower bounds on local
stabilizer-measurement circuits show that nonlocal logical and
stabilizer structure can force a depth cost controlled by cuts of the
interaction graph \cite{DelfosseBeverlandTremblay2021}.

We now make the scheduling proxy $\Dsched$ explicit, since it is the one
quantity in Eq.~\eqref{eq:embedding-score} that couples the combinatorics
of the code to the congestion of the network.

\begin{definition}[Scheduling-depth proxy]
\label{def:dsched}
For a Tanner graph $T\in\{T_X,T_Z\}$ let $\chi'(T)$ be its edge chromatic
number, i.e.\ the least number of rounds in which all data--ancilla
interactions of $T$ can be performed subject to the constraint that no
qubit and no ancilla participates in two interactions in the same round.
For a placement $\pi$ on $\Gphys$, let $T_s$ denote the Steiner tree
routing check $s$ and define the load of a physical link $e$ as
\begin{equation}
    \mathrm{load}(e)
    :=
    \bigl|\{\,s\in\mathcal{S}_X\cup\mathcal{S}_Z:\;
    e\in E(T_s)\,\}\bigr| .
    \label{eq:link-load}
\end{equation}
The \emph{scheduling-depth proxy} is
\begin{equation}
    \Dsched(\pi;\Gphys)
    :=
    \max\bigl\{
        \chi'(T_X)+\chi'(T_Z),\,
        \textstyle\max_{e}\mathrm{load}(e)
    \bigr\}.
    \label{eq:dsched-definition}
\end{equation}
\end{definition}

The two arguments of the maximum are two independent bottlenecks. The
first is intrinsic to the code: with one ancilla per check, a round of
syndrome extraction is a proper edge coloring of the Tanner graph, and the
$X$- and $Z$-rounds are executed sequentially, giving
$\chi'(T_X)+\chi'(T_Z)$. The second is intrinsic to the embedding: a
physical link carrying $\mathrm{load}(e)$ distinct check-routing trees can
serve at most one of them per time step if entanglement is generated
link-by-link, so its load is a lower bound on the number of rounds. Taking
the maximum rather than the sum treats the two constraints as
simultaneously satisfiable, which is optimistic; $\Dsched$ is therefore a
proxy and not a circuit-depth guarantee.

The two components of $D_{\rm sched}$ are efficiently computable for the sparse Tanner graphs considered here. Implementation details, approximation guarantees, and the relation to K\"onig's edge-coloring theorem are given in Appendix~\ref{app:secIV-details-subsection-1}.

To combine routing and local-operation costs, we define the weighted
embedding score
\begin{equation}
\begin{split}
    C_{\mathrm{emb}}(\pi;G_{\mathrm{phys}})
    :=\;
    &\alpha\, C_{\mathrm{St}}(\pi;G_{\mathrm{phys}})
    +
    \beta\, W_{\mathrm{tot}}\\
    &+
    \gamma\, D_{\mathrm{sched}}(\pi;G_{\mathrm{phys}}),
\end{split}
    \label{eq:embedding-score}
\end{equation}
where:
\begin{itemize}
    \item $C_{\mathrm{St}}$ measures nonlocal entanglement-routing cost;
    \item $W_{\mathrm{tot}}$ measures the number of stabilizer--qubit
    incidences and hence local two-qubit interactions;
    \item $D_{\mathrm{sched}}$ is an estimated scheduling depth in the
    physical network;
    \item $\alpha,\beta,\gamma\geq0$ are application-dependent weights.
\end{itemize}

In the numerical results below, we report $C_{\mathrm{St}}$,
$W_{\mathrm{tot}}$, and $\Dsched$ separately. This avoids imposing a
hardware-specific choice of the coefficients $\alpha,\beta,\gamma$ while
retaining a transparent decomposition of the physical overhead into a
routing term, a gate-count term, and a depth term.

We consider a modular setting in which stabilizer checks spanning several network nodes are implemented using distributed entanglement. In this interpretation, $W_{\rm tot}$ counts local data--ancilla interactions, $C_{\rm St}$ captures the network resources required to connect the modules participating in each check, and $D_{\rm sched}$ captures the resulting parallelization constraints. The model is deliberately architecture agnostic; connections with GHZ-mediated parity measurements, lattice-surgery implementations, and routing-based constructions are discussed in Appendix~\ref{app:secIV-details-subsection-2}.

\begin{remark}[$\CSt$ is a routing proxy]
\label{rem:cst-is-proxy}
We will refer to $\CSt$ throughout as a \emph{topology-aware routing
proxy}. It is a sum of approximate Steiner-tree weights, hence (i) an
upper bound on the optimal routing cost of each individual check, (ii)
agnostic to entanglement fidelity, purification, and rate mismatch, and
(iii) additive across checks, thus ignoring reuse of already-distributed
entanglement between checks with overlapping supports. It orders
placements and topologies; it does not predict wall-clock or Bell-pair
budgets.
\end{remark}

\subsection{Placement optimization}
\label{subsec:placement-optimization}

The placement map $\pi$ can be optimized to reduce the cost of embedding
a qLDPC code into a constrained quantum network. Given a fixed physical
graph $G_{\mathrm{phys}}$, we consider the optimization problem
\begin{equation}
    \pi^\star
    \in
    \arg\min_{\pi\in\mathfrak{P}}
    C_{\mathrm{St}}(\pi;G_{\mathrm{phys}}),
    \label{eq:placement-optimization}
\end{equation}
where $\mathfrak{P}$ is a set of feasible placements. 
For a fixed placement, the Steiner cost of each stabilizer support is approximated using a metric-closure minimum-spanning-tree construction. We then optimize the placement by local search over feasible qubit assignments. The complete routing approximation and implementation details are given in Appendix~\ref{app:secIV-details-subsection-3}.

\begin{algorithm}[t]
\caption{Topology-aware placement heuristic}
\label{alg:topology-placement}
\begin{algorithmic}[1]
\Require CSS matrices $H_X,H_Z$, physical graph $G_{\mathrm{phys}}$,
         feasible placement set $\mathfrak{P}$
\Ensure A low-cost placement $\pi$
\State Initialize $\pi\in\mathfrak{P}$
\State Compute $C_{\mathrm{St}}(\pi;G_{\mathrm{phys}})$
\For{$r=1$ to $N_{\mathrm{iter}}$}
    \State Generate a feasible neighboring placement $\pi'$
    \State Estimate $C_{\mathrm{St}}(\pi';G_{\mathrm{phys}})$
    \If{$C_{\mathrm{St}}(\pi';G_{\mathrm{phys}})
       <C_{\mathrm{St}}(\pi;G_{\mathrm{phys}})$}
        \State $\pi\leftarrow\pi'$
    \Else
        \State Optionally accept $\pi'$ with an annealing probability
    \EndIf
\EndFor
\State \Return $\pi$
\end{algorithmic}
\end{algorithm}

The objective in Eq.~\eqref{eq:placement-optimization} expresses a
precise version of the network-aware qLDPC design problem: a Tanner graph
is desirable not only when it is sparse, but also when it admits a
low-cost embedding into the physical network graph.

\subsection{Linear-cost condition for matched network topologies}
\label{subsec:linear-cost-condition}

We now state a simple sufficient condition under which sparse CSS codes
admit linear total routing cost.

\begin{proposition}[Linear routing cost under bounded-distortion embedding]
\label{prop:linear-steiner-cost}
Let $\{\mathcal{C}_n\}$ be a family of CSS qLDPC codes with a bounded
number of stabilizer generators per physical qubit and bounded
stabilizer weights. Suppose there exists a family of placements
$\pi_n$ into physical graphs $G_{\mathrm{phys},n}$ such that, for every
stabilizer generator $s$,
\begin{equation}
    \operatorname{ST}_{G_{\mathrm{phys},n}}(\Pi_n(s))
    \leq c_{\mathrm{emb}}\left|\supp(s)\right|
    \label{eq:bounded-distortion-condition}
\end{equation}
for a constant $c_{\mathrm{emb}}$ independent of $n$. Then
\begin{equation}
    C_{\mathrm{St}}(\pi_n;G_{\mathrm{phys},n})
    =
    O(n).
    \label{eq:linear-routing-cost}
\end{equation}
\end{proposition}

\emph{Proof.} See Appendix~\ref{app:secIV-details-subsection-5}.

Proposition~\ref{prop:linear-steiner-cost} does not state that every
qLDPC code is linearly embeddable into every network topology. Rather,
it identifies the extra condition required beyond sparsity: the physical
network must support a bounded-distortion realization of the relevant
stabilizer supports. This is plausible for modular or expander-like
networks designed around the code Tanner graph, but it can fail on
strictly local low-dimensional geometries. The gap between these cases
will be investigated numerically for complete, ring, grid, modular, and
expander physical graphs.

\subsection{Constructive recovery on authorized coalitions}
\label{subsec:constructive-recovery}

The access-structure criterion of Section~\ref{sec:access-structure}
certifies that an authorized coalition can reconstruct the logical
secret. We now describe how to construct explicit representatives of the
logical Pauli operators whose support lies entirely within such a
coalition.

Let $A\subseteq[n]$ be authorized. By
Theorem~\ref{thm:exact-css-qss-access}, the complement $A^c$ supports no
nontrivial logical operator. Consequently, every global logical operator
has an equivalent representative, modulo stabilizers, supported on $A$.
This is the stabilizer ``cleaning'' property.

Consider a global $X$-type logical representative
\[
    x\in\ker(H_Z)
\]
whose equivalence class is nontrivial modulo
$\operatorname{row}(H_X)$. We seek a stabilizer vector
\[
    s_X\in\operatorname{row}(H_X)
\]
such that
\begin{equation}
    (x+s_X)|_{A^c}=0.
    \label{eq:x-cleaning-condition}
\end{equation}
Writing $s_X=uH_X$ for a binary row vector $u$, the cleaning condition
becomes the linear system
\begin{equation}
    u(H_X|_{A^c})
    =
    x|_{A^c}.
    \label{eq:x-cleaning-linear-system}
\end{equation}
A solution exists whenever the logical class of $x$ can be cleaned from
$A^c$, which is guaranteed for an authorized set. The cleaned
representative is
\begin{equation}
    x_A
    :=
    x+uH_X,
    \qquad
    \supp(x_A)\subseteq A.
    \label{eq:cleaned-x-logical}
\end{equation}

Similarly, for a $Z$-type logical representative
\[
    z\in\ker(H_X),
\]
one solves
\begin{equation}
    v(H_Z|_{A^c})
    =
    z|_{A^c}
    \label{eq:z-cleaning-linear-system}
\end{equation}
and defines
\begin{equation}
    z_A
    :=
    z+vH_Z,
    \qquad
    \supp(z_A)\subseteq A.
    \label{eq:cleaned-z-logical}
\end{equation}

\begin{proposition}[Cleaning of logical operators onto authorized sets]
\label{prop:logical-cleaning}
Let $A$ be an authorized coalition for a CSS code. Every logical
$X$-type or $Z$-type Pauli operator admits an equivalent representative
whose support is contained entirely in $A$.
\end{proposition}

\emph{Proof.} See Appendix~\ref{app:secIV-details-subsection-6}.

The supports of the cleaned representatives determine the physical
resources needed for reconstruction by coalition $A$. We define the
recovery weight of a chosen logical Pauli basis as
\begin{equation}
    W_{\mathrm{rec}}(A)
    :=
    \sum_{j=1}^{k_L}
    \left[
        \mathrm{wt}(\bar{X}_{j,A})
        +
        \mathrm{wt}(\bar{Z}_{j,A})
    \right],
    \label{eq:recovery-weight}
\end{equation}
where $\bar{X}_{j,A}$ and $\bar{Z}_{j,A}$ are cleaned representatives
supported on $A$. A topology-aware recovery-routing cost is defined by
\begin{multline}
    C_{\mathrm{rec}}(A;\pi,G_{\mathrm{phys}})
    :=
    \sum_{j=1}^{k_L}
    \Bigl[
        \operatorname{ST}_{G_{\mathrm{phys}}}
        \bigl(\pi(\supp(\bar{X}_{j,A}))\bigr)
    \\
        +\;
        \operatorname{ST}_{G_{\mathrm{phys}}}
        \bigl(\pi(\supp(\bar{Z}_{j,A}))\bigr)
    \Bigr].
    \label{eq:recovery-routing-cost}
\end{multline}

The quantities $W_{\mathrm{rec}}(A)$ and
$C_{\mathrm{rec}}(A;\pi,G_{\mathrm{phys}})$ distinguish two authorized
coalitions that are equivalent from an information-theoretic viewpoint
but have different physical recovery costs. They will be evaluated for
sampled authorized coalitions in Section~\ref{sec:numerical-results}.

\begin{remark}[$\Wrec$ depends on a basis and on a heuristic]
\label{rem:wrec-basis-dependence}
$W_{\rm rec}(A)$ is an achievable recovery cost rather than an invariant of $(\mathcal C,A)$: it depends on the chosen logical basis and cleaning heuristic. Throughout the numerical study these choices are held fixed, so that comparisons across coalition sizes remain consistent. Implementation details are given in Appendix~\ref{app:secIV-details-subsection-7}.
\end{remark}

The cleaned representatives are obtained by solving Eqs.~\eqref{eq:x-cleaning-linear-system}, \eqref{eq:cleaned-x-logical}, \eqref{eq:z-cleaning-linear-system}, and \eqref{eq:cleaned-z-logical} over $\mathbb F_2$, followed by a greedy stabilizer-based weight reduction. The complete algorithm, its failure criterion, and its polynomial-time complexity are given in Appendix~\ref{app:secIV-details-subsection-8}.

\subsection{Summary of topology-aware observables}
\label{subsec:topology-observables}

For each code instance, placement, and physical network topology, we
evaluate the following observables:
\begin{align}
    W_{\mathrm{tot}}
    &=
    \|H_X\|_0+\|H_Z\|_0,
    \label{eq:summary-wtot}\\
    C_{\mathrm{St}}(\pi;G_{\mathrm{phys}})
    &=
    \sum_{s\in\mathcal{S}_X\cup\mathcal{S}_Z}
    \operatorname{ST}_{G_{\mathrm{phys}}}(\Pi(s)),
    \label{eq:summary-cst}\\
    D_{\mathrm{sched}}(\pi;G_{\mathrm{phys}})
    &=
    \text{stabilizer-measurement depth},
    \label{eq:summary-depth}\\
    W_{\mathrm{rec}}(A)
    &=
    \sum_{j=1}^{k_L}
    \left[
        \mathrm{wt}(\bar{X}_{j,A})
        +
        \mathrm{wt}(\bar{Z}_{j,A})
    \right],
    \label{eq:summary-wrec}\\
    C_{\mathrm{rec}}(A;\pi,G_{\mathrm{phys}})
    &=
    \sum_{j=1}^{k_L}
    \bigl[
        \sigma(\bar{X}_{j,A})
        +
        \sigma(\bar{Z}_{j,A})
    \bigr],
    \label{eq:summary-crec}
\end{align}

where we abbreviate
$\sigma(P):=\operatorname{ST}_{G_{\mathrm{phys}}}(\pi(\supp(P)))$ for the
Steiner cost of the routed support of a Pauli operator $P$.

Together with the access-structure observables introduced in
Section~\ref{subsec:access-transition-observables}, these quantities
provide a joint characterization of:
\begin{center}
secret rate \;vs.\; access structure \;vs.\;
network embedding cost.
\end{center}

The next section applies this framework to repetition- and
Hamming-derived constructions, random sparse hypergraph-product codes,
and the \([[72,12,6]]\) bivariate bicycle code.

\section{Numerical Study of qLDPC-Induced Access Structures and Network Costs}
\label{sec:numerical-results}

This section applies the analytical framework of
Sections~\ref{sec:access-structure} and~\ref{sec:topology-cost} to
representative CSS and qLDPC code families. Our objectives are threefold:
to characterize the access structures induced by each code, to quantify
finite-size authorization transitions and ramp regions, and to evaluate
the topology-aware embedding and recovery costs of
Section~\ref{sec:topology-cost}.

The workflow uses exact binary linear algebra over $\Ftwo$. For a
coalition $A\subseteq[n]$ we compute $\ell_X(A)$, $\ell_Z(A)$,
$\ell_X(A^c)$ and $\ell_Z(A^c)$ from the rank formulas of
Proposition~\ref{prop:rank-formula-supported-logicals}, classify $A$
according to Theorem~\ref{thm:exact-css-qss-access}, and evaluate the
accessible logical dimension $\kappa(A)$ of
Definition~\ref{def:accessible-logical-dimension}. All linear algebra is
performed with exact Gauss--Jordan elimination on bit-packed matrices, so
no floating-point tolerance enters the classification.

\subsection{Code ensembles and numerical protocol}
\label{subsec:numerical-protocol}

We evaluate four representative CSS instances: two structured HGP baselines derived from repetition and Hamming codes, one random sparse HGP instance, and the bivariate bicycle code $[[72,12,6]]$. Their principal parameters are summarized in Table~\ref{tab:code-parameters}.

Coalitions are classified using the exact rank tests of Proposition~\ref{prop:rank-formula-supported-logicals} and Theorem~\ref{thm:exact-css-qss-access}. For each code, we sample 21 coalition sizes and use exact enumeration whenever feasible and uniform Monte Carlo sampling otherwise. Additional construction details,
sparsity statistics, sampling parameters, and implementation information are given in
Appendix~\ref{app:numerical-details-subsection-1}.

\begin{table}[t]
    \centering
    \caption{Code instances used in the numerical study. Here
    $\Wtot=\|H_X\|_0+\|H_Z\|_0$ is the total Tanner
    incidence count, and $R_s=k_L/n$ is the finite-size secret rate.}
    \label{tab:code-parameters}
    \small
    \begin{tabular}{lrrrrrr}
        \toprule
        Code & $n$ & $k_L$ & $\rank H_X$ & $\rank H_Z$ & $\Wtot$ & $R_s$ \\
        \midrule
        HGP(rep$_5$,rep$_5$)          & $41$  & $1$  & $20$ & $20$ & $144$  & $0.0244$ \\
        HGP(Ham$_{7,4}$,Ham$_{7,4}$) & $58$  & $16$ & $21$ & $21$ & $240$  & $0.2759$ \\
        Random sparse HGP             & $208$ & $20$ & $96$ & $92$ & $1440$ & $0.0962$ \\
        BB $[[72,12,6]]$              & $72$  & $12$ & $30$ & $30$ & $432$  & $0.1667$ \\
        \bottomrule
    \end{tabular}
\end{table}
The random HGP instance is irregular, whereas the BB instance is $(6,3)$-regular; detailed row- and column-weight statistics are reported in Appendix~\ref{app:numerical-details-subsection-1}.


Before collecting numerical statistics, we verified the CSS commutation and dimension identities together with the duality, complementarity, and monotonicity predictions of Section~\ref{sec:access-structure}. All $280$ validation tests passed without violation; complete validation details are reported in Appendix~\ref{app:numerical-details}.

\subsection{Access transitions}
\label{subsec:access-results}

\begin{figure*}[t]
    \centering
    \includegraphics[width=0.92\linewidth]{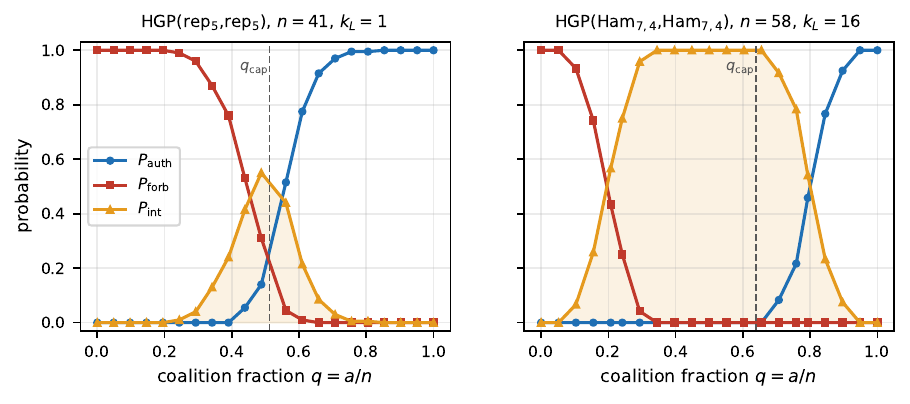}
    \caption{Access probabilities for the two structured HGP baselines,
    $[[41,1]]$ (left) and $[[58,16]]$ (right), as functions of the
    normalized coalition size $q=a/n$. Curves show $\Pauth$, $\Pforb$ and
    $\Pint$. The $k_L=1$ baseline is nearly a perfect threshold scheme,
    while the $k_L=16$ baseline is dominated by intermediate coalitions.}
    \label{fig:baseline-access}
\end{figure*}

\begin{figure}[t]
    \centering
    \includegraphics[width=0.98\linewidth]{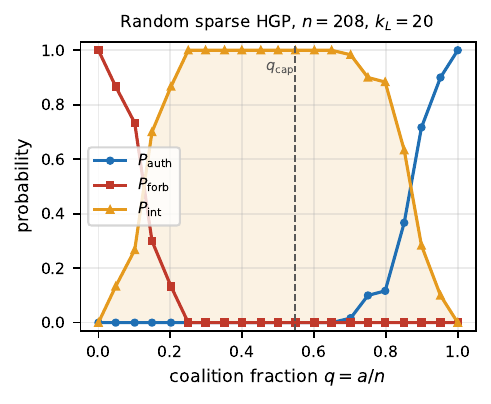}
    \caption{Estimated access probabilities for the random sparse HGP code
    $[[208,20]]$. The dashed vertical line marks the rate-dependent
    asymptotic erasure benchmark $q_{\mathrm{cap}}=(1+R_s)/2\approx0.548$
    of Eq.~\eqref{eq:qcap-benchmark}. The comparison is a finite-size
    diagnostic and carries no asymptotic capacity claim.}
    \label{fig:random-hgp-access}
\end{figure}

\begin{figure}[t]
    \centering
    \includegraphics[width=0.98\linewidth]{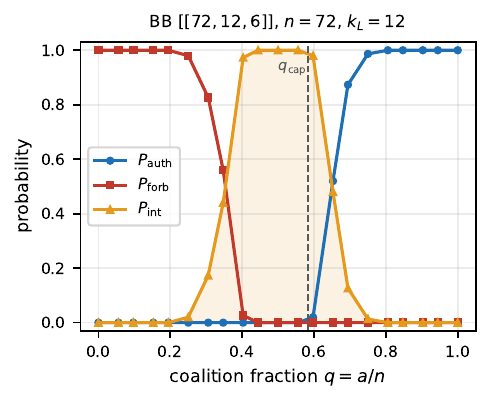}
    \caption{Access probabilities for the bivariate bicycle code
    $[[72,12,6]]$, obtained by exact enumeration where feasible and Monte
    Carlo sampling otherwise. The dashed line marks
    $q_{\mathrm{cap}}=(1+k_L/n)/2=7/12\approx0.583$. The shaded band indicates the
    distance-guaranteed forbidden region $a\le d-1=5$.}
    \label{fig:bb72-access}
\end{figure}

Figures~\ref{fig:baseline-access}, \ref{fig:random-hgp-access}
and~\ref{fig:bb72-access} show the measured profiles $\Pauth(q)$,
$\Pforb(q)$ and $\Pint(q)$ as functions of the normalized coalition size
$q=a/n$; Table~\ref{tab:access-summary} collects the derived observables.

\begin{table*}[t]
    \centering
    \caption{Access-structure summary. Threshold quantities are normalized coalition fractions $q=a/n$.}
    \label{tab:access-summary}
    \begin{tabular}{lrrrrrrrrr}
        \toprule
        Code & $n$ & $k_L$ & $R_s$ & $d$ & $q_{10}$ & $q_{50}$ & $q_{90}$ & $\Delta q$ & $A_{\mathrm{ramp}}$ \\
        \midrule
        HGP(rep$_5$,rep$_5$) & 41 & 1 & 0.0244 & --- & 0.4648 & 0.5580 & 0.6533 & 0.1885 & 0.1177 \\
        HGP(Ham$_{7,4}$,Ham$_{7,4}$) & 58 & 16 & 0.2759 & --- & 0.7134 & 0.8001 & 0.8884 & 0.1750 & 0.6057 \\
        Random sparse HGP & 208 & 20 & 0.0962 & --- & 0.7500 & 0.8693 & 0.9519 & 0.2019 & 0.7376 \\
        BB $[[72,12,6]]$ & 72 & 12 & 0.1667 & --- & 0.6061 & 0.6506 & 0.7075 & 0.1014 & 0.3089 \\
        \bottomrule
    \end{tabular}
\end{table*}

Several observations follow.

\paragraph{Rate matters but does not determine the ramp.}
The ramp area $\Aramp=\int_0^1\Pint(q)\,dq$ grows broadly with $R_s$
across the four instances. As Table~\ref{tab:access-summary} shows, $A_{\rm ramp}$ does not vary monotonically with the secret rate. In particular, the random HGP instance has the largest ramp area despite its intermediate rate, excluding a rate-only explanation within this sample. The random instance breaks monotonicity in $R_s$ outright, and does so by a wide margin, so among the four instances studied the rate is at best a partial predictor: the random code combines a moderate rate with a highly irregular distribution of logical supports, and here irregularity, not rate, appears to widen the intermediate family. With four instances we can exclude a monotone rate-only law but cannot isolate the responsible mechanism; the controlled ensemble of Section~\ref{subsec:ensemble-regularity} addresses that question directly. The $[[41,1]]$ baseline, with a single logical qubit, has the smallest ramp area and comes closest to a perfect threshold scheme, as expected from Corollary~\ref{cor:pure-qss-complementarity} when $k_L=1$.

\paragraph{The BB instance has the sharpest transition among the four.}
The BB code has the narrowest transition of the four instances studied,
$\Delta q=q_{90}-q_{10}=0.101$, against $0.175$ for $[[58,16]]$,
$0.188$ for $[[41,1]]$ and $0.202$ for the random HGP code. As reported in
Appendix~\ref{app:numerical-details-subsection-3}, the bootstrap confidence interval for
the BB value, $[0.085,0.116]$, does not overlap those of the other three,
whereas the intervals of the remaining three overlap heavily; the
separation of BB from the rest is therefore statistically supported, while
any ordering among $[[41,1]]$, $[[58,16]]$ and the random instance is not.
One plausible mechanism is that quasi-cyclic symmetry homogenizes coalition
statistics: in a $(6,3)$-regular code all qubits are structurally
equivalent, so the classification of a coalition depends mainly on its
size, whereas in a random instance qubits of degree one coexist with
qubits of degree eight and coalitions of equal cardinality behave very
differently. This suggests, but does not establish, a regularity effect,
since the four instances also differ in $n$, $k_L$, and check weight; the
controlled ensemble of Section~\ref{subsec:ensemble-regularity} tests the
hypothesis at matched parameters. For distributed cryptography a narrow
$\Delta q$ is desirable, since it makes the authorization boundary
predictable from coalition size alone.

\paragraph{Finite-size thresholds sit above the asymptotic erasure-capacity benchmark.}
The rate-dependent benchmark against which $q_{50}$ should be read is not
$1-R_s$ but the erasure-channel bound for sparse stabilizer codes. Delfosse and Z{\'e}mor showed that a family of low-density stabilizer codes of rate
$R$ correcting a random erasure fraction $p$ must satisfy $R\le 1-2p$
\cite{DelfosseZemor2013}; writing $q=1-p$ for the surviving fraction, an
asymptotically achieving family would recover from any
\begin{equation}
    q\;\ge\;q_{\mathrm{cap}}^{\mathrm{erasure}}(R_s):=\frac{1+R_s}{2}.
    \label{eq:qcap-benchmark}
\end{equation}
Comparable finite-rate sparse constructions and their erasure performance
are discussed in \cite{TremblayEtAl2023,KuoOuyang2026}.

Every instance we study lies \emph{above} this benchmark. The
half-authorization points are $q_{50}=0.558$ against
$q_{\mathrm{cap}}^{\mathrm{erasure}}=0.512$ for $[[41,1]]$, $0.800$ against $0.638$ for
$[[58,16]]$, $0.869$ against $0.548$ for the random $[[208,20]]$ instance,
and $0.651$ against $0.583$ for the BB code. Finite-size coalitions
therefore need to be \emph{larger} than the asymptotic bound suggests
before reconstruction becomes typical, with the gap ranging from $0.05$ for
$[[41,1]]$ to $0.32$ for the random instance. This ordering is the
expected direction: $q_{\mathrm{cap}}^{\mathrm{erasure}}$ is an asymptotic bound on what an
optimal family could achieve, not a finite-size access threshold for a
particular code, and finite length, sublinear distance, and support
irregularity all push the empirical transition to the right. The two codes
closest to the benchmark are the two with the most regular logical support
structure, and the random instance is the furthest from it. No asymptotic
claim can be extracted from a single code length at each rate, and none is
made here.

\begin{figure}[t]
    \centering
    \includegraphics[width=0.98\linewidth]{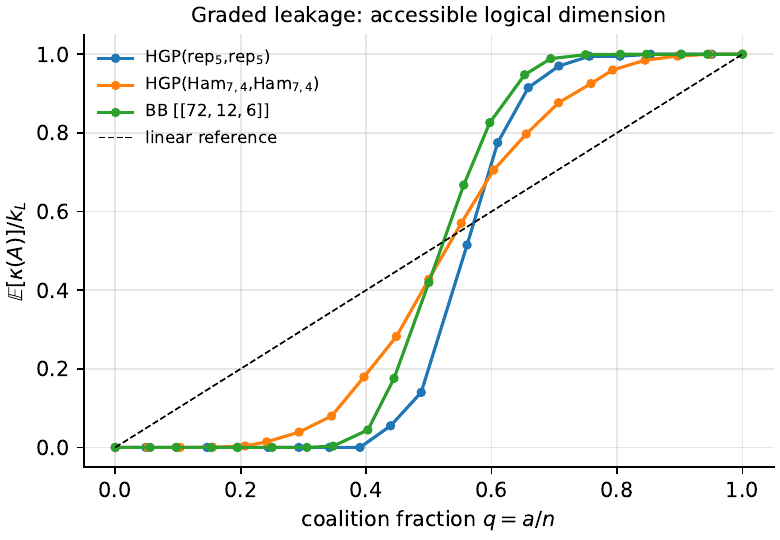}
    \caption{Mean normalized accessible logical dimension
    $\overline{\kappa}(a)/k_L$ against normalized coalition size for the
    codes with $k_L>1$. The quantity interpolates continuously between the
    forbidden regime ($\kappa=0$) and the authorized regime
    ($\kappa=k_L$), grading the intermediate family that the ternary
    classification of Theorem~\ref{thm:exact-css-qss-access} leaves undifferentiated.}
    \label{fig:kappa-leakage}
\end{figure}

\paragraph{Accessible logical dimension in the ramp region.}
Figure~\ref{fig:kappa-leakage} shows the mean normalized accessible
logical dimension $\overline{\kappa}(a)/k_L$ from
Definition~\ref{def:accessible-logical-dimension}. For the BB code,
$\overline{\kappa}$ remains exactly zero up to $a=25$ ($q=0.35$), then
rises through $\overline{\kappa}/k_L=0.176$ at $a=32$, $0.419$ at $a=36$,
$0.667$ at $a=40$ and $0.826$ at $a=43$, and saturates at $1$ for
$a\ge58$. This is the central quantitative content of the graded
description: at $a=36$, $\kappa(A)$ is exact for each sampled coalition; the plotted mean is estimated from the sampled ensemble. As stressed in Remark~\ref{rem:kappa-vs-information-group}, this is a
statement about which logical operators are jointly addressable on $A$, not
a claim that $42\%$ of the secret's information has leaked. A binary authorized/forbidden description of the same regime discards this structure, whereas $\kappa$ quantifies it exactly.

Bootstrap confidence intervals confirm that the BB transition width is separated from those of the other three instances; the complete uncertainty analysis is reported in Appendix~\ref{app:numerical-details-subsection-3}.

\subsection{Comparison of Controlled Quasi-Cyclic and Random Biregular HGP Ensembles}
\label{subsec:ensemble-regularity}

The four headline codes differ simultaneously in block length, rate and
check weight, so they cannot attribute the sharp BB transition to algebraic
regularity rather than to length or sparsity. We therefore construct a
controlled ensemble in which the confounders are held fixed by
construction. Both families are hypergraph products of $9\times12$ seed
matrices of column weight $3$ and row weight $4$; the quasi-cyclic family
imposes in addition a $3\times4$ block-circulant structure of lifting size
$L=3$, while the random family draws biregular seeds with the same weight
profile. Every instance in both families has
\[
    n=225,
    \qquad
    \|H_X\|_0=\|H_Z\|_0=756,
    \qquad
    \Wtot=1512,
\]
so block length, total Tanner incidence and both weight profiles are
identical across all $16$ instances. We draw $8$ instances per family and
compare means by a two-sided permutation test with $20\,000$ permutations.
Although the quasi-cyclic and random biregular HGP ensembles have identical physical block lengths and check-weight distributions, their encoded dimensions differ substantially. This difference can be explained by the ranks of the underlying classical parity-check matrices. For a symetric hypergraph-product construction based on a binary matrix $H\in\mathbb{F_2^{r\times n_0}}$, let 
\[
\rho = \operatorname{rank}_{\mathbb{F}_2}(H).
\]
The number of encoded logical qubits is
\begin{equation}
    \label{eq:hpg_encoded_dimension}
    k_L = (n_0 - \rho)^2+(r - \rho)^2.
\end{equation}
In our numerical experiments, $n_0 = 12$ and $r = 9$, giving 
\begin{equation}
    k_L = (12 - \rho)^2+(9 - \rho)^2.
\end{equation}
The observed logical dimensions $k_L\in\{9,17,29,65\}$ correspond to seed-matrix ranks $\rho\in\{9,8,7,5\}$, respectively.

Thus, matching matrix dimensions and check-weight distributions does not guarantee matching encoded dimensions. In the present sample, the quasi-cyclic instances exhibit lower seed-matrix ranks and consequently higher encoding rates than the random biregular instances. The additional algebraic structure imposed by quasi-cyclic constructions can introduce linear dependencies among parity checks, which provides a mechanism for the observed rank differences. However, this behavior is instance-dependent and should not be interpreted as a universal property of quasi-cyclic HGP codes.

Figure~\ref{fig:ensemble-regularity} and Table~\ref{tab:ensemble-regularity} (in appendix ~\ref{app:numerical-details-subsection-4}) report the outcome. The numerical results indicate that the quasi-cyclic instances considered in this study exhibit sharper transitions compared to the random biregular instances,
\begin{align}
    \Delta q_{\mathrm{QC}}   &= 0.1379\pm0.0232,\notag\\
    \Delta q_{\mathrm{rand}} &= 0.1874\pm0.0268,\notag
\end{align}
with $p\approx0.001$, and simultaneously the larger ramp area,
$\Aramp=0.733\pm0.134$ against $0.389\pm0.173$ ($p\approx0.001$), and the larger midpoint, $q_{50}=0.871\pm0.070$ against $0.688\pm0.089$. The mean normalized accessible dimension is the one observable that does not separate the families, $0.422\pm0.021$ against $0.442\pm0.023$ with $p\approx0.11$: conditioned on being in the transition window, coalitions address a comparable fraction of the logical space in both ensembles.

Two caveats are essential and we state them as limitations of the
experiment rather than as interpretation. First, this comparison controls physical block length and check-weight distributions but not the encoded dimension. In fact, the two families are
\emph{not} matched in $k_L$: the quasi-cyclic instances have
$k_L\in\{29,65\}$ and the random ones $k_L\in\{9,17\}$, so the quasi-cyclic family also has the higher secret rate, and the larger $\Aramp$ is at least partly a rate effect of the kind already visible in
Section~\ref{subsec:access-results}. Imposing circulant structure at fixed seed dimensions changes the seed ranks, so $k_L$ cannot be held fixed while varying only regularity; this is intrinsic to the comparison and not a defect of the sampling. Second, a single lifting size and a single seed shape are used, so no scaling claim in $n$ or $L$ is available. What the ensemble does establish is narrower than the loose reading of Section~\ref{subsec:access-results} and more than the four headline codes support on their own: at fixed block length and check-weight profile, the quasi-cyclic instances studied exhibit sharper authorization transitions and wider ramp regions than the random biregular instances. However, because the two ensembles differ substantially in encoded dimension and secret rate, these differences cannot be attributed exclusively to quasi-cyclic structure.

\begin{figure}
    \centering
    \includegraphics[width=0.98\linewidth]{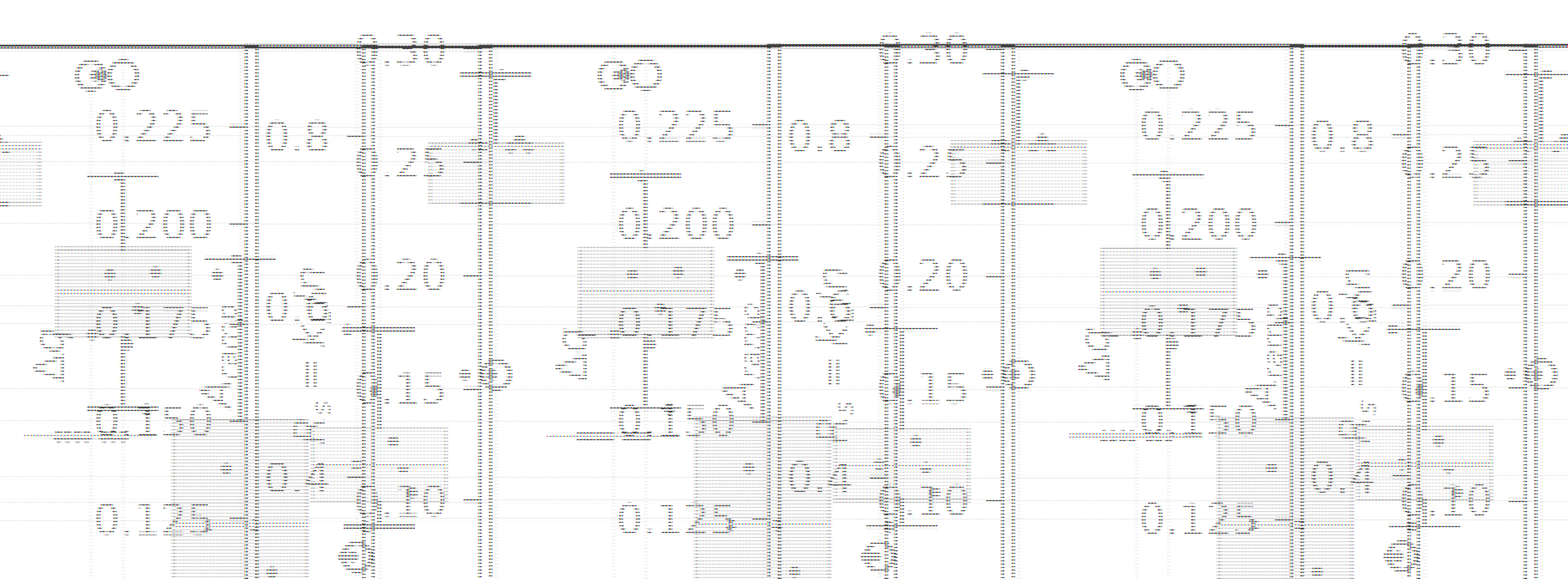}
    \caption{Controlled ensemble at $n=225$ and $\Wtot=1512$: transition
    width $\Delta q$, ramp area $\Aramp$ and secret rate $R_s$ for $8$
    quasi-cyclic and $8$ random biregular hypergraph-product instances with
    identical seed dimensions and weight profiles. Boxes show quartiles,
    dots individual instances. The rightmost panel makes the residual
    confounder explicit: the families are not matched in $k_L$.}
    \label{fig:ensemble-regularity}
\end{figure}

These results highlight differences in authorization behavior between the sampled quasi-cyclic and random biregular constructions, while the independent effects of algebraic regularity and secret rate remain to be established.
\subsection{Topology-aware embedding results}
\label{subsec:embedding-results}

\begin{figure*}[t]
    \centering
    \includegraphics[width=0.92\linewidth]{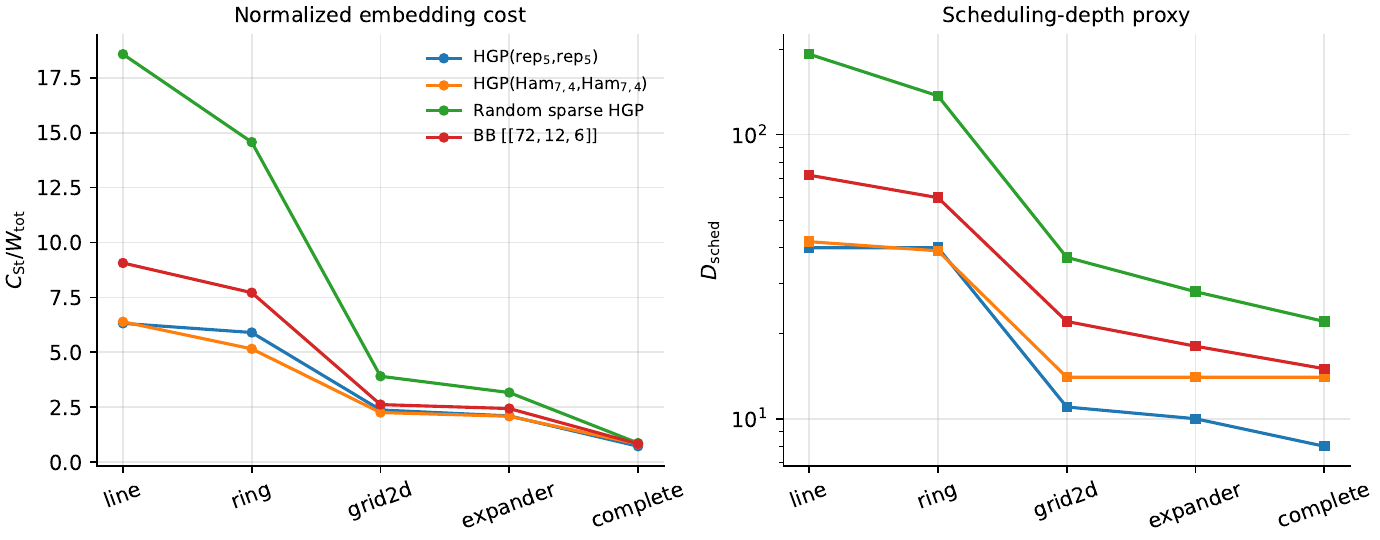}
    \caption{Normalized topology-aware stabilizer cost $\CSt/\Wtot$ for the
    four codes on five physical topologies. Sparsity of the Tanner graph
    does not by itself imply cheap physical realization: the same code
    spans more than an order of magnitude in normalized cost depending on
    the network graph.}
    \label{fig:embedding-cost}
\end{figure*}

We embed each code into five physical network graphs on
$N=n+m_X+m_Z$ nodes: a path (line), a ring, a two-dimensional square
grid, a bounded-degree random expander of degree four, and the complete
graph. Node placement is the identity map $\pi=\mathrm{id}$, so the
reported costs measure the intrinsic difficulty of the code--topology
pair rather than the quality of a heuristic optimizer; the improvements
achievable by Algorithm~\ref{alg:topology-placement} are quantified
separately in Section~\ref{subsec:placement-results}, where they range
from $9.7\,\%$ to $61.4\,\%$ depending on the code and the graph.
For each pair we compute the Steiner routing cost $\CSt$ of
Definition~\ref{def:steiner-cost}, using the classical
$2-$approximation of Kou, Markowsky and Berman for the underlying Steiner
tree problem \cite{KouMarkowskyBerman1981}, its normalization
$\CSt/\Wtot$, and the scheduling-depth proxy $\Dsched$.
Table~\ref{tab:embedding-summary} reports the full data and
Figure~\ref{fig:embedding-cost} the normalized costs.

\begin{table*}[t]
    \centering
    \caption{Topology-aware embedding results. $C_{\mathrm{St}}$ is the total Steiner routing cost, $W_{\mathrm{tot}}$ the Tanner incidence count, and $D_{\mathrm{sched}}$ the scheduling-depth proxy.}
    \label{tab:embedding-summary}
    \begin{tabular}{llrrrrr}
        \toprule
        Code & Topology & Placement & $W_{\mathrm{tot}}$ & $C_{\mathrm{St}}$ & $C_{\mathrm{St}}/W_{\mathrm{tot}}$ & $D_{\mathrm{sched}}$ \\
        \midrule
        HGP(rep$_5$,rep$_5$) & line & identity & 144 & 910 & 6.319 & 40 \\
        HGP(rep$_5$,rep$_5$) & ring & identity & 144 & 850 & 5.903 & 40 \\
        HGP(rep$_5$,rep$_5$) & grid2d & identity & 144 & 342 & 2.375 & 11 \\
        HGP(rep$_5$,rep$_5$) & expander & identity & 144 & 303 & 2.104 & 10 \\
        HGP(rep$_5$,rep$_5$) & complete & identity & 144 & 104 & 0.722 & 8 \\
        HGP(Ham$_{7,4}$,Ham$_{7,4}$) & line & identity & 240 & 1534 & 6.392 & 42 \\
        HGP(Ham$_{7,4}$,Ham$_{7,4}$) & ring & identity & 240 & 1238 & 5.158 & 39 \\
        HGP(Ham$_{7,4}$,Ham$_{7,4}$) & grid2d & identity & 240 & 542 & 2.258 & 14 \\
        HGP(Ham$_{7,4}$,Ham$_{7,4}$) & expander & identity & 240 & 503 & 2.096 & 14 \\
        HGP(Ham$_{7,4}$,Ham$_{7,4}$) & complete & identity & 240 & 198 & 0.825 & 14 \\
        Random sparse HGP & line & identity & 1440 & 26752 & 18.578 & 192 \\
        Random sparse HGP & ring & identity & 1440 & 20985 & 14.573 & 137 \\
        Random sparse HGP & grid2d & identity & 1440 & 5630 & 3.910 & 37 \\
        Random sparse HGP & expander & identity & 1440 & 4556 & 3.164 & 28 \\
        Random sparse HGP & complete & identity & 1440 & 1248 & 0.867 & 22 \\
        BB $[[72,12,6]]$ & line & identity & 432 & 3918 & 9.069 & 72 \\
        BB $[[72,12,6]]$ & ring & identity & 432 & 3334 & 7.718 & 60 \\
        BB $[[72,12,6]]$ & grid2d & identity & 432 & 1132 & 2.620 & 22 \\
        BB $[[72,12,6]]$ & expander & identity & 432 & 1056 & 2.444 & 18 \\
        BB $[[72,12,6]]$ & complete & identity & 432 & 360 & 0.833 & 15 \\
        \bottomrule
    \end{tabular}
\end{table*}

The data support four conclusions.

First, on the complete graph the normalized cost is below unity for every
code, between $0.722$ and $0.867$. This is the expected calibration: a
weight-$w$ check can be connected by a star with $w-1$ edges, so
$\CSt\approx\Wtot-(m_X+m_Z)$, and the routing cost carries no locality
penalty.

Second, the penalty of one-dimensional connectivity is severe and grows
with block length. For the $[[41,1]]$ code the line gives
$\CSt/\Wtot=6.32$; for the $[[208,20]]$ random code the same topology
gives $18.58$, a factor $21.4$ above its complete-graph value of $0.867$.
This is a concrete instance of the general obstruction that geometrically
local implementations of qLDPC codes must pay in routing or in ancilla
overhead \cite{DelfosseBeverlandTremblay2021}.

Third, moderate connectivity recovers most of the loss. On the
two-dimensional grid the normalized costs fall to $2.38$, $2.26$, $3.91$
and $2.62$ for the four codes, and the degree-four expander does slightly
better still, $1.94$, $2.18$, $3.08$ and $2.32$. The expander improves on
the grid despite having the same average degree, because its logarithmic
diameter shortens the long-range paths that dominate $\CSt$; for the
$[[58,16]]$ code the two topologies are nearly tied ($2.26$ versus
$2.18$), showing that the advantage depends on how well the network graph
matches the code interaction pattern rather than on expansion alone.

Fourth, the scheduling-depth proxy $\Dsched$ behaves differently from
$\CSt$ and must be reported separately. Going from line to grid, the
random HGP code improves $\CSt$ by a factor $4.75$ but $\Dsched$ by a
factor $5.19$ ($192\to37$); going from grid to complete, $\CSt$ improves
by $4.51$ while $\Dsched$ improves only by $1.68$ ($37\to22$). In the
dense regime the depth is limited by the edge-colouring term rather than
by per-link congestion, so adding connectivity buys total cost without
buying latency. For the BB code specifically, $\Dsched$ on the line equals
$72=n$, the worst-case serialization of a $(6,3)$-regular code on a path,
and drops to $15$ on the complete graph. Any deployment estimate that
reports only $\CSt$ therefore overstates the benefit of dense
connectivity.

\subsection{Recovery-support and recovery-routing costs}
\label{subsec:recovery-results}

For sampled authorized coalitions we construct cleaned logical
representatives by solving the linear systems of
Eqs.~\eqref{eq:x-cleaning-linear-system}
and~\eqref{eq:z-cleaning-linear-system} for all $2k_L$ generators of a
symplectic logical basis, followed by a stabilizer-assisted weight
reduction restricted to $A$. We report the aggregate recovery weight
$\Wrec(A)$, i.e.\ the total support weight of the $2k_L$ cleaned
representatives, and the recovery-routing cost $\Crec(A;\pi,\Gphys)$,
averaged over sampled authorized coalitions of a given size,
\begin{align}
    \overline{W}_{\mathrm{rec}}(a)
    &:=
    \mathbb{E}\bigl[\Wrec(A)\,\big|\,|A|=a,\;A\in\mathcal{A}_{\mathrm{auth}}\bigr],
    \label{eq:mean-recovery-weight}\\
    \overline{C}_{\mathrm{rec}}(a)
    &:=
    \mathbb{E}\bigl[\Crec(A;\pi,\Gphys)\,\big|\,|A|=a,\;A\in\mathcal{A}_{\mathrm{auth}}\bigr].
    \label{eq:mean-recovery-routing-cost}
\end{align}

\begin{figure*}[t]
    \centering
    \includegraphics[width=0.92\linewidth]{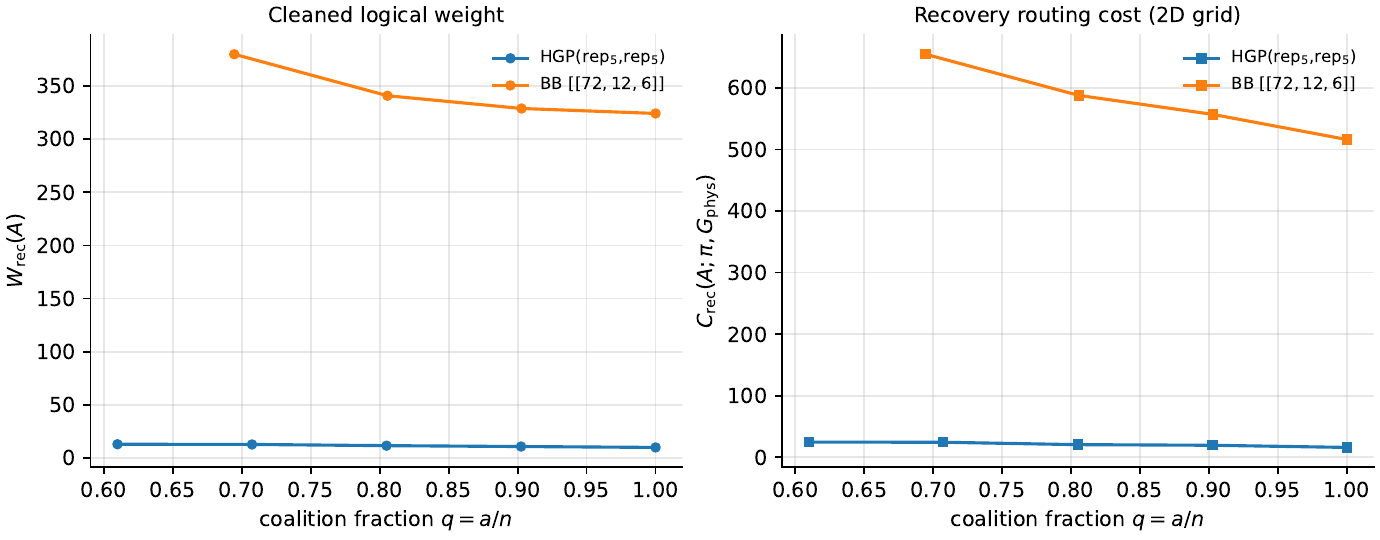}
    \caption{Mean aggregate recovery weight
    $\overline{W}_{\mathrm{rec}}(a)$ (left) and mean topology-aware
    recovery cost $\overline{C}_{\mathrm{rec}}(a)$ (right), conditioned on
    authorization, for the $[[41,1]]$ baseline and the BB $[[72,12,6]]$
    code on the two-dimensional grid. Both decrease with coalition size:
    larger coalitions admit lighter and more local recovery operators.}
    \label{fig:recovery-cost}
\end{figure*}

Figure~\ref{fig:recovery-cost} shows that both quantities decrease
monotonically with $a$. For the $[[41,1]]$ code,
$\overline{W}_{\mathrm{rec}}$ falls from $13.0$ at $a=25$ to $10.0$ at
$a=41$, while $\overline{C}_{\mathrm{rec}}$ falls from $24.7$ to $16.0$.
For the BB code, where the basis contains $2k_L=24$ representatives, the
aggregate weight falls from $379.7$ at $a=50$ to $324.0$ at $a=72$ and the
routing cost from $654.5$ to $516.0$. Two points deserve emphasis. First,
recovery cost is not a function of authorization status: all sampled
coalitions here are authorized, yet their recovery cost varies by up to
$27\%$ with size, and coalitions of equal size show a spread that the
mean alone does not capture. Second, the ratio
$\overline{C}_{\mathrm{rec}}/\overline{W}_{\mathrm{rec}}$ stays close to
$1.6$--$1.9$ on the grid, so the routing overhead of recovery is a
roughly multiplicative topology factor on top of the algebraic support
weight. Information-theoretic accessibility and physical
reconstructability are thus distinct design criteria: a coalition can be
authorized in the exact QSS sense and still require a high-weight,
spatially spread recovery operator.


\subsection{Optimized versus identity qubit placement}
\label{subsec:placement-results}

The embedding costs of Section~\ref{subsec:embedding-results} use the identity placement $\pi=\mathrm{id}$, which assigns code qubit $i$ to physical node $i$. That choice is arbitrary and, on structured graphs, bad. To separate the intrinsic routing cost of a code from the cost of an arbitrary labelling, we run a local search over placements, proposing transpositions and accepting those that decrease $\CSt$, for $800$ iterations from the identity.

\begin{figure}
    \centering
    \includegraphics[width=0.98\linewidth]{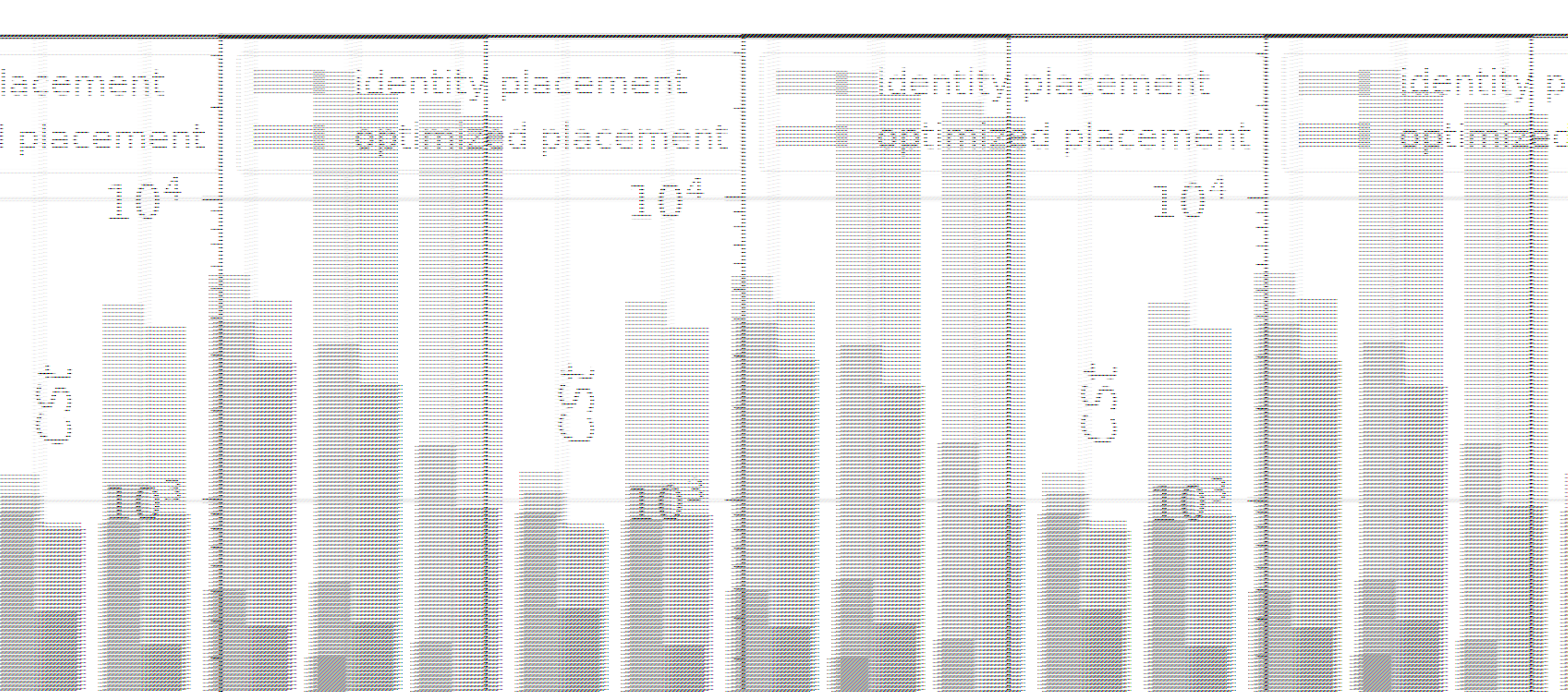}
    \caption{Steiner-routing cost $\CSt$ at the identity placement and
    after $800$ accepted-swap local-search iterations, per code and
    physical topology (logarithmic vertical axis). Optimized values are
    upper bounds on the optimum of an NP-hard assignment problem.}
    \label{fig:placement-optimization}
\end{figure}

Figure~\ref{fig:placement-optimization} and Table~\ref{tab:placement-optimization} report identity and optimized costs side by side. The reductions are substantial and systematically ordered. They are largest for the smallest code on the most constrained graphs, reaching $61.4\,\%$ for the $[[41,1]]$ code on a ring and $52.5\,\%$ on a line, and smallest for the largest code, $16.7\,\%$ on a grid and $17.3\,\%$ on an expander for the $[[208,20]]$ instance; the BB code sits in between, at $20.8$--$27.7\,\%$. Two readings are consistent with this pattern: on graphs of low degree there is more room to shorten Steiner trees by relabelling, and at larger $n$ a fixed budget of $800$ accepted-swap explores a proportionally smaller part of the assignment space, so the reported gains at $n=208$ should be read as a lower bound on what placement optimization can achieve.

The practical implication is that the identity-placement numbers of
Section~\ref{subsec:embedding-results} overstate the network cost of every
code, and by an amount that depends on both the code and the graph. Because
the reduction factor is not uniform across codes, comparisons of $\CSt$
between codes at identity placement are not reliable rankings; the
qualitative conclusions of Section~\ref{subsec:embedding-results} that
survive optimization are the ordering of topologies for a fixed code and
the outlier status of the random instance. The reported improvements represent achievable reductions obtained by the heuristic. Further optimization may yield lower routing costs, but global optimality is not established

\subsection{Comparative summary}
\label{subsec:numerical-summary}
The four instances occupy distinct regions of the joint access--network design space. The $[[41,1]]$ baseline is closest to threshold behavior but has negligible secret rate, whereas the random HGP instance exhibits both the largest ramp region and the highest embedding overhead. Among the instances studied, the BB $[[72,12,6]]$ code provides the most balanced combination of secret rate, authorization-transition width, block-level individual security, and routing cost. These results reinforce that $R_s$, $\Delta_q$, $A_{\rm ramp}$, $C_{\rm St}/W_{\rm tot}$, and $D_{\rm sched}$ should be considered jointly rather than optimizing code rate alone.

\section{Block-Level Quantum Secret Sharing and Comparative Discussion}
\label{sec:block-level-qss}

Sections~\ref{sec:access-structure}--\ref{sec:numerical-results} treat
every physical qubit as belonging to a distinct participant. This
one-qubit-per-party convention is convenient analytically, but it is not
the operationally relevant regime for high-rate qLDPC codes. A code such
as the bivariate bicycle code $[[72,12,6]]$ encodes $k_L=12$ logical
qubits into $n=72$ shares, and a network of $72$ independent parties is
neither typical nor necessary. In practice, a moderate number of
participants each hold a \emph{block} of physical qubits.

In this section we formalize block-level QSS, show that the induced
coarse-grained access structure inherits the exact rank criterion of
Theorem~\ref{thm:exact-css-qss-access}, and derive the resulting
threshold guarantees. We then quantify the partial information available
to intermediate coalitions, and close with a comparison to GHZ- and
graph-state-based QSS protocols together with an explicit statement of
the limitations of the present framework.

\subsection{Participant partitions and induced access structures}
\label{subsec:block-partitions}

\begin{definition}[Share partition]
\label{def:share-partition}
Let $m\leq n$. A \emph{share partition} of a CSS code
$\mathcal{C}=[[n,k_L,d]]$ into $m$ participants is a partition
\begin{equation}
    \mathcal{P}
    =
    \{B_1,B_2,\ldots,B_m\},
    \qquad
    \bigsqcup_{j=1}^{m}B_j=[n],
    \qquad
    B_j\neq\varnothing,
    \label{eq:share-partition}
\end{equation}
where participant $j$ holds the physical qubits indexed by $B_j$. We
write
\begin{equation}
    b_j:=|B_j|,
    \qquad
    b_{\max}:=\max_{j\in[m]}b_j,
    \qquad
    b_{\min}:=\min_{j\in[m]}b_j.
    \label{eq:block-sizes}
\end{equation}
The partition is \emph{balanced} when $b_j=n/m$ for every $j$.
\end{definition}

A coalition of participants is a subset $J\subseteq[m]$. Its physical
support is the union of the corresponding blocks,
\begin{equation}
    B(J)
    :=
    \bigcup_{j\in J}B_j
    \subseteq[n],
    \label{eq:coalition-support}
\end{equation}
and its complement satisfies $B([m]\setminus J)=B(J)^c$.

\begin{definition}[Block-level access structure]
\label{def:block-access-structure}
The block-level access structure induced by $(\mathcal{C},\mathcal{P})$
consists of the three families
\begin{align}
    \mathcal{B}_{\mathrm{auth}}
    &:=
    \left\{J\subseteq[m]:B(J)\in\mathcal{A}_{\mathrm{auth}}\right\},
    \label{eq:block-auth}\\
    \mathcal{B}_{\mathrm{forb}}
    &:=
    \left\{J\subseteq[m]:B(J)\in\mathcal{A}_{\mathrm{forb}}\right\},
    \label{eq:block-forb}\\
    \mathcal{B}_{\mathrm{int}}
    &:=
    \left\{J\subseteq[m]:B(J)\in\mathcal{A}_{\mathrm{int}}\right\}.
    \label{eq:block-int}
\end{align}
\end{definition}

Because Definition~\ref{def:block-access-structure} is stated entirely in
terms of the qubit-level classification, the exact criterion of
Section~\ref{sec:access-structure} applies verbatim after substituting
$A\mapsto B(J)$.

\begin{proposition}[Exact block-level criterion]
\label{prop:block-level-criterion}
Let $J\subseteq[m]$ and set $E:=B(J)$. Then $J\in\mathcal{B}_{\mathrm{auth}}$
if and only if
\begin{equation}
    \ell_X(E^c)=\ell_Z(E^c)=0,
    \label{eq:block-auth-condition}
\end{equation}
and $J\in\mathcal{B}_{\mathrm{forb}}$ if and only if
\begin{equation}
    \ell_X(E)=\ell_Z(E)=0,
    \label{eq:block-forb-condition}
\end{equation}
where $\ell_X$ and $\ell_Z$ are given by the rank formulas of
Proposition~\ref{prop:rank-formula-supported-logicals}.
\end{proposition}

The result follows immediately from Definition~\ref{def:block-access-structure} and Theorem~\ref{thm:exact-css-qss-access} applied to $B(J)\subseteq[n]$.

An important computational consequence is that block-level enumeration involves $2^m$ participant coalitions rather than $2^n$ qubit coalitions. For the moderate participant counts considered below, the complete block-level access structure can therefore be enumerated exactly. We denote the corresponding exact size-conditioned probabilities by $P_{\rm auth}^{\mathcal P}(t)$, $P_{\rm forb}^{\mathcal P}(t)$, and $P_{\rm int}^{\mathcal P}(t)$.

\subsection{Distance guarantees and thresholds}
\label{subsec:coarse-graining}

Grouping shares can only make the access structure coarser: aggregating
qubits into blocks removes the ability to distinguish coalitions that
differ inside a block, but it never invalidates the distance-based
guarantees. The following two statements make this precise.

\begin{proposition}[Block-level distance guarantees]
\label{prop:block-distance-guarantees}
Let $\mathcal{C}=[[n,k_L,d]]$ with share partition $\mathcal{P}$. For
every $J\subseteq[m]$,
\begin{equation}
    \sum_{j\in J}b_j\leq d-1
    \quad\Longrightarrow\quad
    J\in\mathcal{B}_{\mathrm{forb}},
    \label{eq:block-forbidden-guarantee}
\end{equation}
and
\begin{equation}
    \sum_{j\notin J}b_j\leq d-1
    \quad\Longrightarrow\quad
    J\in\mathcal{B}_{\mathrm{auth}}.
    \label{eq:block-authorized-guarantee}
\end{equation}
In particular, for a balanced partition with $b_j=n/m$,
\begin{align}
    |J|
    \leq
    \Bigl\lfloor\frac{m(d-1)}{n}\Bigr\rfloor
    &\Longrightarrow
    J\in\mathcal{B}_{\mathrm{forb}},
    \label{eq:balanced-block-forb}\\
    |J|
    \geq
    m-\Bigl\lfloor\frac{m(d-1)}{n}\Bigr\rfloor
    &\Longrightarrow
    J\in\mathcal{B}_{\mathrm{auth}}.
    \label{eq:balanced-block-guarantees}
\end{align}
\end{proposition}
\emph{Proof.} See Appendix~\ref{app:block-proofs-subsection-B}.

Equation~\eqref{eq:balanced-block-guarantees} exposes a genuine tension.
Reducing the number of participants $m$ enlarges each block, so a single
participant may control more qubits than the code distance permits. Once
$b_{\max}\geq d$, the guaranteed-forbidden region can become empty and a
single participant may already be an intermediate coalition. This yields
a design constraint that is absent from the qubit-level analysis.

\begin{corollary}[Individual-share security condition]
\label{cor:individual-security}
If $b_{\max}\leq d-1$, then every singleton $\{j\}$ satisfies
$\{j\}\in\mathcal{B}_{\mathrm{forb}}$: no individual participant obtains
any information about the secret. For a balanced partition, this holds
whenever
\begin{equation}
    m\geq\left\lceil\frac{n}{d-1}\right\rceil.
    \label{eq:min-participants}
\end{equation}
\end{corollary}

The proof follows directly from Proposition~\ref{eq:block-forbidden-guarantee} applied to singleton participant coalitions.


\begin{remark}[The partition is a design variable]
\label{rem:block-placement-matters}
Balanced partitions with identical block sizes need not induce identical access structures, because the result depends on how logical-operator supports intersect the blocks. Hence $\mathcal P$ controls the induced access structure, whereas the placement $\pi$ of Section~IV controls the network-routing cost. The two variables may therefore be optimized jointly.
\end{remark}

\subsection{Joint optimization of partition and placement}
\label{subsec:joint-optimization}

Multiple-share assignment in QSS is established; in particular, Matsumoto~\cite{matsumoto2017quantum} optimizes share assignment for a prescribed access structure with the objective of reducing average share size. Our optimization problem is different: the CSS qLDPC code is fixed, while the participant partition $\mathcal P$ changes the induced access structure and the placement $\pi$ changes its network realization. We therefore optimize access and routing objectives jointly.
Remark~\ref{rem:block-placement-matters} suggests a multiobjective design
problem. Given a code $\mathcal{C}$ and a physical network
$G_{\mathrm{phys}}$, we seek a partition $\mathcal{P}$ and a placement
$\pi$ that simultaneously
\begin{enumerate}[label=\textup{(\roman*)}]
    \item maximize the security margin, measured as the largest $t$ such
    that every coalition of at most $t$ participants is forbidden,
    \begin{equation}
        t_{\mathrm{sec}}(\mathcal{P})
        :=
        \max\left\{
            t:\;
            \forall J\subseteq[m],\,|J|\leq t
            \;\Longrightarrow\;
            J\in\mathcal{B}_{\mathrm{forb}}
        \right\},
        \label{eq:security-margin}
    \end{equation}
    \item minimize the reconstruction threshold, measured as the smallest
    $t$ such that every coalition of at least $t$ participants is
    authorized,
    \begin{equation}
        t_{\mathrm{rec}}(\mathcal{P})
        :=
        \min\left\{
            t:\;
            \forall J\subseteq[m],\,|J|\geq t
            \;\Longrightarrow\;
            J\in\mathcal{B}_{\mathrm{auth}}
        \right\},
        \label{eq:reconstruction-threshold}
    \end{equation}
    \item minimize the topology-aware embedding cost
    $C_{\mathrm{St}}(\pi;G_{\mathrm{phys}})$ of
    Eq.~\eqref{eq:total-stabilizer-cost}.
\end{enumerate}

The gap
\begin{equation}
    \Delta t(\mathcal{P})
    :=
    t_{\mathrm{rec}}(\mathcal{P})-t_{\mathrm{sec}}(\mathcal{P})-1
    \label{eq:block-ramp-width}
\end{equation}
is the block-level ramp width. A perfect threshold scheme at the
participant level corresponds to $\Delta t(\mathcal{P})=0$, in which case
$\mathcal{B}_{\mathrm{int}}=\varnothing$ and the construction realizes a
genuine $((t_{\mathrm{rec}},m))$ threshold QSS scheme on $m$
participants. Whether a given high-rate qLDPC code admits a partition
achieving $\Delta t=0$, and at what cost in $m$, is a natural
combinatorial question raised by this framework. The no-cloning
constraint of Section~\ref{subsec:qss-access} imposes
\begin{equation}
    t_{\mathrm{rec}}(\mathcal{P})
    >
    \frac{m}{2}
    \label{eq:block-no-cloning}
\end{equation}
for any partition, since two disjoint participant coalitions cannot both
be authorized.

\begin{algorithm}[t]
\caption{Joint partition and placement search}
\label{alg:joint-partition-placement}
\begin{algorithmic}[1]
\Require CSS matrices $H_X,H_Z$, network $G_{\mathrm{phys}}$, target
         participant count $m$
\Ensure Partition $\mathcal{P}$ and placement $\pi$
\State Initialize a balanced partition $\mathcal{P}$ of $[n]$ into $m$ blocks
\State Compute $t_{\mathrm{sec}}(\mathcal{P})$, $t_{\mathrm{rec}}(\mathcal{P})$
       by exact enumeration over $2^m$ coalitions using
       Proposition~\ref{prop:block-level-criterion}
\State Initialize $\pi$ mapping each block to a network node
\For{$r=1$ to $N_{\mathrm{iter}}$}
    \State Propose a swap of two qubits between distinct blocks, or a
           relocation of one block
    \State Recompute $t_{\mathrm{sec}}$, $t_{\mathrm{rec}}$, and
           $C_{\mathrm{St}}$
    \State Accept the move if it improves the scalarized objective
           $-\lambda_1 t_{\mathrm{sec}}
            +\lambda_2 t_{\mathrm{rec}}
            +\lambda_3 C_{\mathrm{St}}$
\EndFor
\State \Return $\mathcal{P},\pi$
\end{algorithmic}
\end{algorithm}
Algorithm \ref{alg:joint-partition-placement} describes a multiobjective optimization problem defined as $$\min_{P, \pi}\left(-t_{sec}, t_{rec}, C_{st}\right).$$
The scalarization by parameters $\lambda_1 > 0,~\lambda_2 > 0$ and $\lambda_3 > 0$ is used only for the numerical demonstration.
\subsection{Partial information at intermediate coalitions}
\label{subsec:partial-information}

Intermediate coalitions are the defining feature of ramp schemes, and the
classification of Theorem~\ref{thm:exact-css-qss-access} tells us that
they exist but not how much of the logical register they can address. We
now make that count exact.
The object quantified below has a known ancestor in the
information-group formalism for stabilizer QSS. Gheorghiu showed
that the information available to an intermediate set can be
described by a subgroup of the logical Pauli group, referred to as
the information group~\cite{Gheorghiu2012}. In canonical form, this
group separates into a symplectic part, generated by conjugate
logical Pauli pairs, and an isotropic part containing mutually
commuting logical information. Matsumoto subsequently gave
linear-algebraic conditions, in terms of stabilizer check matrices,
for qualified and forbidden sets and quantitative descriptions of
the information available to intermediate sets~\cite{Matsumoto2019}.

The contribution below is therefore not the existence of partial
logical information at intermediate coalitions. Rather, for CSS
codes we identify a directly computable invariant of its
noncommuting quantum part: the rank of the symplectic pairing
restricted to logical classes admitting representatives on the
coalition. This quantity can be obtained directly from restricted
parity-check linear algebra, without explicit enumeration of the
information group. Its precise relation to the information-group
formalism is stated in Remark~\ref{rem:kappa-vs-information-group}.

For a coalition $A$, let
\begin{equation}
    \mathcal{L}^X_A
    =
    \frac{\ker H_Z\cap\Ftwo^{A}}{\row H_X\cap\Ftwo^{A}},
    \qquad
    \mathcal{L}^Z_A
    =
    \frac{\ker H_X\cap\Ftwo^{A}}{\row H_Z\cap\Ftwo^{A}},
    \label{eq:local-logical-quotients}
\end{equation}
so that $\dim\mathcal{L}^X_A=\ell_X(A)$ and $\dim\mathcal{L}^Z_A=\ell_Z(A)$
in the notation of Definition~\ref{def:supported-logical-dimensions}.

\begin{definition}[Accessible logical dimension]
\label{def:accessible-logical-dimension}
Fix bases $\{x_1,\dots,x_{\ell_X(A)}\}$ of $\mathcal{L}^X_A$ and
$\{z_1,\dots,z_{\ell_Z(A)}\}$ of $\mathcal{L}^Z_A$, and let
$M(A)\in\Ftwo^{\ell_X(A)\times\ell_Z(A)}$ be the symplectic pairing matrix
\begin{equation}
    M(A)_{ij}
    =
    \langle x_i,z_j\rangle
    =
    \bigl|\supp(x_i)\cap\supp(z_j)\bigr|
    \bmod 2 .
    \label{eq:local-pairing-matrix}
\end{equation}
The \emph{accessible logical dimension} of $A$ is
\begin{equation}
    \kappa(A)
    :=
    \rank_{\Ftwo} M(A).
    \label{eq:accessible-logical-dimension}
\end{equation}
\end{definition}

\begin{proposition}[Properties of $\kappa$]
\label{prop:kappa-properties}
The quantity $\kappa(A)$ does not depend on the chosen bases, and
\begin{enumerate}[label=\textup{(\roman*)}]
    \item $0\le\kappa(A)\le\min\{\ell_X(A),\ell_Z(A)\}\le k_L$;
    \item $\kappa$ is monotone: $A\subseteq B$ implies
    $\kappa(A)\le\kappa(B)$;
    \item $\kappa(A)=k_L$ if and only if $A$ is authorized;
    \item $\kappa(A)=0$ if and only if the algebra generated by locally supported logical Pauli operators on $A$ is abelian; in particular $\kappa(A)=0$ whenever $A$ is forbidden.
\end{enumerate}
\end{proposition}

\emph{Proof.} See Appendix~\ref{app:block-proofs-subsection-D}.

\begin{remark}
Equation~\eqref{eq:accessible-logical-dimension} refines the ternary classification of Theorem~\ref{thm:exact-css-qss-access}: intermediate coalitions are graded by
\[
0\leq\kappa(A)<k_L.
\]
The normalized quantity $\kappa(A)/k_L$ measures the fraction of independent logical Pauli pairs that can be represented jointly on $A$. Its operational interpretation is discussed below.
\end{remark}

To turn the rank $\kappa(A)$ into an operational statement we first
identify the group whose symplectic form $M(A)$ represents.

\begin{lemma}[Local logical group and its symplectic form]
\label{lem:local-logical-group}
Let $\mathcal{P}_A$ denote the Pauli group acting trivially outside $A$,
let $\mathcal{S}$ be the stabilizer group and $\mathcal{N}(\mathcal{S})$
its normalizer. Define the \emph{local logical group}
\begin{equation}
    \mathcal{L}_A
    :=
    \frac{\mathcal{N}(\mathcal{S})\cap\mathcal{P}_A}
         {\mathcal{S}\cap\mathcal{P}_A},
    \label{eq:local-logical-group}
\end{equation}
which for a CSS code decomposes as
$\mathcal{L}_A\cong\mathcal{L}^X_A\oplus\mathcal{L}^Z_A$ with the spaces of
Eq.~\eqref{eq:local-logical-quotients}. The commutator induces a
well-defined alternating bilinear form
$\omega_A:\mathcal{L}_A\times\mathcal{L}_A\to\Ftwo$, whose Gram matrix in
any basis adapted to the CSS splitting is
\begin{equation}
    \Omega(A)
    =
    \begin{pmatrix}
        0 & M(A) \\
        M(A)^{T} & 0
    \end{pmatrix},
    \qquad
    \rank_{\Ftwo}\Omega(A)=2\kappa(A).
    \label{eq:local-gram-matrix}
\end{equation}
\end{lemma}

\emph{Proof.} See Appendix~\ref{app:block-proofs-subsection-E}.

\begin{proposition}[Accessible logical subsystem at intermediate coalitions]
\label{prop:accessible-subsystem}
Let $A$ be any coalition and let $\kappa=\kappa(A)$. There exist logical Pauli operators
\[
\overline X'_1,\overline Z'_1,\ldots,
\overline X'_\kappa,\overline Z'_\kappa
\in \mathcal N(\mathcal S)\cap\mathcal P_A
\]
whose logical classes satisfy the canonical commutation relations
\[
\omega_A(\overline X'_i,\overline Z'_j)=\delta_{ij},
\qquad
\omega_A(\overline X'_i,\overline X'_j) =
\omega_A(\overline Z'_i,\overline Z'_j)=0.
\]

The $C^*$-algebra generated by these operators on the code space is isomorphic to the full matrix algebra
\[
M_{2^\kappa}(\mathbb C).
\]
Consequently, the logical Hilbert space admits a decomposition
\[
\mathcal H_L
\simeq
\mathcal H_{\mathrm{acc}}\otimes\mathcal H_{\mathrm{res}},
\qquad
\dim\mathcal H_{\mathrm{acc}}=2^\kappa,
\]
such that this matrix algebra acts as
\[
\mathcal B(\mathcal H_{\mathrm{acc}})\otimes I_{\mathrm{res}}.
\]

Moreover, this logical factor is exactly correctable from the shares held by $A$ in the operator-algebra quantum-error-correction sense: there exists a recovery channel $\mathcal R_A$ acting on $A$ such that the observables of
$\mathcal B(\mathcal H_{\mathrm{acc}})\otimes I_{\mathrm{res}}$ are perfectly reconstructed.

For $\kappa(A)=k_L$, this reduces to full authorization. For $\kappa(A)=0$, no nontrivial quantum logical subsystem is recoverable from $A$ alone, although commuting logical information may still be accessible.
\end{proposition}

\emph{Proof sketch.}
By Lemma~\ref{lem:local-logical-group}, the local logical group carries an alternating form of rank $2\kappa(A)$. Symplectic Gram--Schmidt therefore produces $\kappa(A)$ hyperbolic pairs of logical Pauli operators supported on $A$. These pairs generate a full matrix algebra $M_{2^{\kappa(A)}}(\mathbb C)$ on the logical code space. The finite-dimensional matrix-factor decomposition then yields $\mathcal H_L\simeq\mathcal H_{\rm acc}\otimes \mathcal H_{\rm res}$ with $\dim\mathcal H_{\rm acc}=2^{\kappa(A)}$. The complete operator-algebraic argument is given in Appendix~\ref{app:block-proofs-subsection-F}.

Proposition~\ref{prop:accessible-subsystem} refines the tripartite
classification into a graded one. An intermediate coalition is not simply
``partially informed'': it can reconstruct a tensor factor consisting of
$\kappa(A)$ logical qubits, and retains at most partial correlations with
the residual factor. This graded structure is characteristic of ramp
secret-sharing schemes built from stabilizer codes, where intermediate sets
acquire a controlled amount of information about the secret
\cite{OgawaEtAl2005,ZhangMatsumoto2014,Matsumoto2019}. Reporting the
empirical distribution of $\kappa(A)$ conditioned on $|A|$ therefore
provides a strictly finer diagnostic than $P_{\mathrm{int}}(a)$ alone, and
we include it among the observables of
Section~\ref{sec:numerical-results}.

\begin{remark}[Interpretation of $\kappa$]
\label{rem:kappa-vs-information-group}
The existence of partial logical information for intermediate sets is established in the stabilizer-QSS information-group formalism. Here $\kappa(A)$ provides a CSS specific directly computable characterization of the noncommuting part of that local logical algebra.

Importantly, $\kappa(A)$ is not an information-theoretic leakage measure. The radical of the local symplectic form may contain commuting logical information, so
$\kappa(A)=0$ does not in general imply that $A$ is forbidden. Accordingly, $\kappa(A)/k_L$ is referred to as the normalized accessible logical dimension, not as a fraction of leaked information.
\end{remark}


\subsection{Block-Level Numerical Evaluation}
\label{subsec:Numerical evaluation}

\paragraph{Baseline partitions}
\label{subsec:block-results}

We next coarse-grain each code into $m$ participant blocks according to Definition~\ref{def:share-partition} and enumerate all $2^m$ participant coalitions exactly, using the block-level criterion of Proposition~\ref{prop:block-level-criterion}. Contiguous balanced partitions are used throughout, with $m$ chosen so that $2^m$ enumeration remains exact: $m=8$ for $[[41,1]]$, $m=10$ for $[[58,16]]$, $m=13$ for the random $[[208,20]]$ instance, and $m=12$ for the BB code. Table~\ref{tab:block-summary} reports the resulting thresholds and Figure~\ref{fig:block-access} (in appendix~\ref{app:numerical-details-subsection-7}) the exact block-level profiles.

\begin{table}
    \centering
    \caption{Block-level access structure. $t_{\mathrm{sec}}$ and $t_{\mathrm{rec}}$ are obtained by exact enumeration of all $2^m$ participant coalitions.}
    \label{tab:block-summary}
    \begin{tabular}{lrrrrrc}
        \toprule
        Code & $m$ & $b$ & $t_{\mathrm{sec}}$ & $t_{\mathrm{rec}}$ & $\Delta t$ & indiv.\ secure \\
        \midrule
        HGP(rep$_5$,rep$_5$) & 8 & 6 & 0 & 8 & 7 & no \\
        HGP(Ham$_{7,4}$,Ham$_{7,4}$) & 10 & 6 & 0 & 10 & 9 & no \\
        Random sparse HGP & 13 & 16 & 0 & 13 & 12 & no \\
        BB $[[72,12,6]]$ & 12 & 6 & 1 & 11 & 9 & yes \\
        \bottomrule
    \end{tabular}
\end{table}

The BB $[[72,12,6]]$ instance is the only one with $t_{\rm sec}\ge1$. In particular, its balanced $m=12$ partition has block size $b=6>d-1=5$ and therefore lies outside the sufficient distance guarantee of Corollary~\ref{cor:individual-security}. Nevertheless, the exact rank criterion certifies every singleton participant as forbidden. This illustrates that Corollary~\ref{cor:individual-security} is sufficient but not necessary.

\paragraph{Sensitivity to participant partition}
\label{subsec:partition-sweep}

The previous paragraph evaluated one partition per code, namely
the contiguous one. To measure how much of the block-level access structure
is a property of the partition rather than of the code, we sweep random
balanced partitions of the same code into the same number of participants
and extract $\tsec$, $\trec$ and $\Delta_t$ exactly for each, by complete
enumeration of the $2^m$ block coalitions. Two structured references are
included for comparison: the contiguous partition, and a Tanner-ordered
partition that groups qubits according to a breadth-first traversal of the
Tanner graph.

Table~\ref{tab:partition-sweep} and Figure~\ref{fig:partition-sweep} (in appendix~\ref{app:numerical-details-subsection-8}) show
that the spread is large and that the contiguous choice is a poor default.
For the $[[41,1]]$ code at $m=8$, random partitions realize
$\tsec\in[0,2]$, $\trec\in[6,8]$ and $\Delta_t\in[3,7]$, so the best
sampled partition reduces the ramp width from $7$ to $3$ relative to the
contiguous partition, which attains $(\tsec,\trec)=(0,8)$ and hence the worst value
$\Delta_t=7$ in the whole sample; moreover only $2\,\%$ of random
partitions leave individual shares insecure, against the certainty of doing
so for the contiguous one. For the $[[58,16]]$ code at $m=10$ the picture is
the opposite in a revealing way: exactly half of the random partitions have
$\tsec=0$, and both structured references also do, so here insecure
individual shares are a genuine feature of the code at this coarse-graining
and not an artefact of one ordering. For the BB $[[72,12,6]]$ code at
$m=12$, every sampled partition has $\tsec\ge1$, the Tanner-ordered
partition reaches $(\tsec,\trec)=(2,10)$, and $\Delta_t$ never falls below
$7$.

Two consequences follow. First, statements of the form "code $\mathcal{C}$ induces such-and-such threshold" are ill-posed without naming the partition. Second, the Tanner-ordered heuristic is a useful but unreliable
proxy: it strictly improves on the contiguous partition for two of the three codes and ties on the third, yet it is beaten by random partitions in
the first case, which is what motivates the explicit search in the bellow paragraph. All values are achievable, hence upper bounds on the optimum of
Eq.~\eqref{eq:optimal-ramp-width}.

\begin{table}
    \centering
    \caption{
    Distribution of the block-level thresholds over random balanced
    partitions of the same code into $m$ participants, compared
    with the contiguous and Tanner-ordered reference partitions.
    The access structure is a property of the pair
    (code, partition), not of the code alone.}
    \label{tab:partition-sweep}
    \begin{tabular}{p{0.11\textwidth}p{0.02\textwidth}p{0.03\textwidth}p{0.05\textwidth}p{0.05\textwidth}p{0.04\textwidth}p{0.04\textwidth}p{0.06\textwidth}}
        \toprule
        Code & $m$ & $N_{\mathcal{P}}$ & $\tsec$ range & $\trec$ range & $\Delta_t$ range & $\overline{\Delta_t}$ & $\Pr[\tsec=0]$ \\
        \midrule
        HGP(rep$_5$,
        rep$_5$) & 8 & 60 & $[0,2]$ & $[6,8]$ & $[3,7]$ & 3.48 & 0.02 \\
        HGP(Ham$_{7,4}$,
        Ham$_{7,4}$) & 10 & 40 & $[0,1]$ & $[9,10]$ & $[7,9]$ & 8.00 & 0.50 \\
        BB $[[72,12,6]]$ & 12 & 15 & $[1,2]$ & $[10,11]$ & $[7,9]$ & 7.35 & 0.00 \\
        \bottomrule
    \end{tabular}
\end{table}

The complete distributions of $\Delta_t$ over sampled partitions are shown in Appendix~\ref{app:numerical-details}.

\paragraph{Effect of participant-partition optimization}
\label{subsec:partition-optimization-results}

The block-level construction of Section~\ref{sec:block-level-qss} is sensitive to the choice of partition. Running the local-search heuristic of Algorithm~\ref{alg:joint-partition-placement} for $40$ iterations on the two structured baselines improves the block-level thresholds substantially. For $[[41,1]]$ with $m=8$ participants, the contiguous partition gives $(\tsec,\trec)=(0,8)$, i.e.\ no nontrivial security guarantee and unanimity for recovery, whereas the optimized partition gives $(\tsec,\trec)=(2,6)$: any two participants are provably forbidden and any six can reconstruct, reducing the block-level ramp from
$\Delta t=7$ to $\Delta t=3$. For $[[58,16]]$ with $m=10$, the improvement is from $(0,10)$ to $(1,9)$, i.e.\ $\Delta t=9\to7$. In both cases the optimizer attains the best value observed anywhere in the random sweep of Section~\ref{subsec:partition-sweep}, namely $\Delta_t=3$ at $m=8$ and $\Delta_t=7$ at $m=10$, which is consistent with those values being at or near the optimum of Eq.~\eqref{eq:optimal-ramp-width} for these instances, though the search certifies only achievability.

This is the strongest design-level finding of the numerical study: the block-level access structure of a fixed code is not an invariant of the code but a property of the code together with the participant partition, and a few tens of local moves can turn a zero-security-guarantee configuration into a nontrivial ramp scheme with $\tsec=2$, $\trec=6$ and $\Delta_t=3$. We stress that the optimized configuration remains a ramp scheme rather than a perfect $((t,m))$ threshold scheme: coalitions with cardinalities strictly between $t_{\rm sec}$ and $t_{\rm rec}$ can be intermediate, and their accessible quantum logical subsystem is quantified by $\kappa(A)$ according to Proposition~\ref{prop:accessible-subsystem}. Since the underlying optimization is NP-hard, the reported values are upper bounds on the achievable $\Delta t$.

\subsection{Security notions and their relationship}
\label{subsec:security-taxonomy}

Several distinct guarantees are in play, and the literature uses
overlapping words for them. Table~\ref{tab:security-notions} separates
them, states which are addressed here, and fixes the vocabulary used in
the rest of the paper.

\begin{table*}[t]
    \centering
    \caption{Security notions relevant to code-induced QSS, the criterion
    that defines each, and whether the present framework establishes it.
    The exact criteria in the first three rows are those of
    Theorem~\ref{thm:exact-css-qss-access}; the last three are stated for
    contrast and are not established here.}
    \label{tab:security-notions}
    \setlength{\tabcolsep}{4pt}
    \small
    \begin{tabular}{p{0.20\textwidth}p{0.08\textwidth}p{0.49\textwidth}p{0.15\textwidth}}
        \toprule
        Notion & Secret & Defining criterion & Established here \\
        \midrule
        Correctness (authorized) & quantum &
        $\ell_X(A^c)=\ell_Z(A^c)=0$: exact recovery on $A$ & established exactly \\
        Privacy (forbidden) & quantum &
        $\ell_X(A)=\ell_Z(A)=0$: reduced state independent of secret &
        established exactly \\
        Graded accessibility (intermediate) & quantum &
        $\kappa(A)$ exactly recoverable quantum logical qubits; the radical of $\omega_A$ may additionally contain commuting logical information & established exactly \\
        \midrule
        Approximate QSS & quantum &
        $(\varepsilon,\delta)$ trace-distance recovery and privacy &
        defined, not evaluated \\
        Strong security & classical &
        no information on any individual secret symbol &
        not established; requires additional message randomization~\cite{ZhangMatsumoto2014,Matsumoto2019} \\
        Active (verifiable) QSS & either &
        robustness against corrupted shares & not established; outside the erasure-only model \\
        \bottomrule
    \end{tabular}
\end{table*}

\begin{remark}[Strong security is an additional requirement]
\label{rem:strong-security}
Proposition~\ref{prop:accessible-subsystem} concerns recoverable quantum logical subsystems, not strong security for classical secrets. Additional message randomization is required for the latter~\cite{ZhangMatsumoto2014,Matsumoto2019}.
\end{remark}

\subsection{Comparison with GHZ- and graph-state QSS}
\label{subsec:comparison-ghz}

Early QSS protocols were formulated using specific entangled resource states. The scheme of Hillery, Bu\v{z}ek, and Berthiaume uses GHZ correlations~\cite{HilleryBuzekBerthiaume1999}, while graph-state constructions provide a broader framework supporting both threshold and non-threshold access structures~\cite{MarkhamSanders2008,Gheorghiu2012}. Table~\ref{tab:qss-comparison} places representative instances of these approaches alongside code-based threshold QSS and the CSS qLDPC framework considered here.

\begin{table*}[t]
    \centering
    \caption{
    Comparison of QSS resource paradigms. The rate column refers to the number of quantum secrets shared per physical share; the locality column refers to the weight of the operators that must be measured jointly.}
    \label{tab:qss-comparison}
    \begin{tabular}{llccl}
        \toprule
        Paradigm & Resource & Secrets per share & Check weight & Access structure \\
        \midrule
        Representative GHZ-based scheme $((n,n))$ &
        $n$-party GHZ & $1/n$ & $n$ (global) & Unanimity only \\
        Representative single-secret graph-state & Graph state & $1/n$ & $\deg(v)+1$ & Threshold / non-threshold \\
        CGL threshold $((t,n))$ & $[[2t-1,1,t]]$ code & $1/n$ & $O(t)$ & Perfect threshold \\
        This work & CSS qLDPC $[[n,k_L,d]]$ & $k_L/n$ & $O(1)$ & Ramp, exactly computable \\
        \bottomrule
    \end{tabular}
\end{table*}

The comparison is deliberately limited to representative constructions and should not be interpreted as a general limitation of GHZ- or graph-state QSS. In particular, the graph-state formalism encompasses substantially more general stabilizer-based schemes, and communication-efficient variants of threshold QSS are also known.

The main distinction of the present framework is that a CSS qLDPC code can encode $k_L$ logical qubits into $n$ physical shares while retaining bounded-weight stabilizer checks. The resulting secret rate is therefore $R_s=k_L/n$, rather than the $1/n$ rate of representative single-secret constructions. Moreover, the access structure is not prescribed independently of the code: it is induced by the pair $(\mathcal C,\mathcal P)$ and can be computed exactly using the rank criterion of Proposition~\ref{prop:block-level-criterion}. Different participant partitions can consequently realize different ramp access structures from the same underlying code.

These features come with a tradeoff. Code-induced access structures generally contain intermediate coalitions, so they need not realize perfect threshold schemes. The participant partition therefore becomes a design variable, motivating the optimization of the block-level ramp width introduced below. The price is what the partition search of Section~\ref{subsec:joint-optimization} tries to reduce, and it is natural to ask how far it can be reduced. Define the \emph{optimal block-level ramp width} of a code $\mathcal{C}$ at $m$ participants as
\begin{equation}
    \Delta_t^{\star}(\mathcal{C},m)
    :=
    \min_{\mathcal{P}}
    \bigl[\,\trec(\mathcal{C},\mathcal{P})
    -\tsec(\mathcal{C},\mathcal{P})-1\,\bigr],
    \label{eq:optimal-ramp-width}
\end{equation}
the minimum over all balanced partitions $\mathcal{P}$ of $[n]$ into $m$
blocks.


\subsection{Limitations and scope}
\label{subsec:limitations}

We state explicitly what the present framework does and does not
establish.

\paragraph{Passive share loss, not active cheating.}
The QEC--QSS correspondence used here models unavailable shares as erasures at known locations. It does not address participants who actively corrupt their shares; verifiable QSS against malicious participants requires additional mechanisms and is outside our scope.

\paragraph{Noiseless shares.}
Our exact access-structure results assume ideal encoding, storage, and transmission. In a realistic network, noise and imperfect entanglement distribution must be incorporated into the recovery analysis. Accordingly, $C_{\rm St}$ should be interpreted as a routing proxy rather than a fidelity-aware resource budget.

\paragraph{Finite-size, instance-level conclusions.}
The numerical results concern specific finite-length code instances and should not be interpreted as asymptotic threshold guarantees. In particular, the observed values of $q_{50}$ and $\Delta_q$ do not establish asymptotic scaling laws.

\paragraph{Heuristic optimization.}
The partition and placement searches are heuristic and do not certify global optimality. The reported optimized ramp widths and routing costs therefore establish achievable values rather than exact optima.

\paragraph{Quantum secrets.}
Our privacy and graded-accessibility analysis concerns quantum secrets. Strong security for classical secrets, which requires protecting individual secret symbols from intermediate coalitions, generally requires additional message randomization and is not established here.

These limitations delimit rather than weaken the contribution. The exact rank criterion, the block-level coarse-graining, the graded leakage count $\kappa(A)$, and the topology-aware cost decomposition are all independent of the noise model, and each becomes a component of a fault-tolerant analysis once noise is introduced.

\section{Noisy Channels, Approximate Recovery, and Network Deployment}
\label{sec:discussion}

The framework developed so far is exact but noiseless: coalitions are
classified by whether logical operators can be supported on them, and
recovery is treated as an algebraic cleaning problem. A deployed networked
QSS scheme operates instead over noisy channels, with imperfect
entanglement distribution and finite-fidelity local operations. This
section states precisely how the exact quantities degrade in that setting,
which of them survive unchanged, and what the model does not capture.

\subsection{From exact to approximate access structures}
\label{subsec:approximate-qss}

The exact classification of Theorem~\ref{thm:exact-css-qss-access} is the
$\varepsilon\to0$ limit of an approximate notion. Let $\rho^{S}$ denote the
encoded state of a secret system $S$, let $\mathcal{N}$ be the channel
describing share distribution and storage, and let
$\rho_A=\Tr_{A^c}\bigl[(\mathcal{I}_S\otimes\mathcal{N})(\rho)\bigr]$ be
the state held by coalition $A$.

\begin{definition}[$(\varepsilon,\delta)$-approximate access structure]
\label{def:approximate-access}
A coalition $A$ is \emph{$\delta$-authorized} if there exists a recovery
map $\mathcal{R}_A$ acting only on the shares of $A$ with
\begin{equation}
    \bigl\|
        \mathcal{R}_A(\rho_A)-\rho^{S}
    \bigr\|_1
    \le\delta ,
    \label{eq:delta-authorized}
\end{equation}
and \emph{$\varepsilon$-forbidden} if
\begin{equation}
    \max_{\rho^{S}_1,\rho^{S}_2}
    \bigl\|
        \rho_A(\rho^{S}_1)-\rho_A(\rho^{S}_2)
    \bigr\|_1
    \le\varepsilon .
    \label{eq:eps-forbidden}
\end{equation}
\end{definition}

The exact access structure of Theorem~\ref{thm:exact-css-qss-access} is recovered in the limit $\varepsilon,\delta\rightarrow0$. Under noise, the exact authorization--privacy duality is replaced by the corresponding approximate quantum-error-correction relation. The algebraic quantities $\ell_X(A)$, $\ell_Z(A)$, and $\kappa(A)$ remain properties of the underlying code and therefore continue to characterize its noiseless logical structure.

Importantly, $\kappa(A)$ and $\varepsilon$ quantify different effects. The former measures the dimension of the noncommuting logical subsystem algebraically accessible from $A$, whereas the latter quantifies distinguishability arising from the noisy channel. A complete numerical study of the approximate access structure is left for future work.

\subsection{Cost Model under Noisy Entanglement Distribution}
\label{subsec:noisy-cost}

The unit-weight Steiner proxy can be extended to heterogeneous noisy links by assigning an effective cost to each physical edge. If link $e$ succeeds with
heralding probability $p_e$, requires an effective purification overhead factor $r_e$, and carries load $L_e(\pi)$, then a simple reweighted proxy is

\begin{equation}
\label{eq:noisy-steiner-cost}
C_{\rm St}^{\rm noisy}(\pi;G_{\rm phys}) = \sum_{e\in E_{\rm phys}}\frac{r_e}{p_e}L_e(\pi).
\end{equation}

This preserves the structure of the placement problem of Section~\ref{sec:topology-cost} while replacing unit edge weights by channel-dependent costs. It remains a resource proxy: a complete deployment model would additionally require a physical error model, purification protocol, memory noise, and operation fidelities. We leave such a fault-tolerant evaluation for future work.

\subsection{Deployment Implications}
\label{subsec:deployment}

The numerical results suggest three practical design principles. First, secret rate should not be optimized in isolation: the authorization-transition width $\Delta_q$ and ramp area $A_{\rm ramp}$ provide complementary information about coalition access. Second, the participant partition $\mathcal P$ is itself
a design variable, since different partitions of the same code can produce substantially different block-level thresholds. Third, Tanner sparsity does not guarantee network locality: the physical placement and network topology can substantially change $C_{\rm St}$ and $D_{\rm sched}$.

Consequently, a networked QSS design should be evaluated jointly in terms of code rate, access-structure sharpness, participant partition, and topology-aware implementation cost.

The extensions above are intended as a bridge from the exact algebraic framework to a physical network model, rather than as a fault-tolerance analysis. The finite-size, heuristic-optimization, proxy-cost, and security-model limitations stated in
Section~\ref{subsec:limitations} therefore continue to apply.

\section{Conclusion}
\label{sec:conclusion}

We have developed an exact framework for quantum secret sharing induced by
CSS qLDPC codes, and applied it to four finite-length code instances.

On the analytical side, for CSS codes, the logical-operator characterization of coalition access reduces to an explicit rank criterion over $\Ftwo$
(Theorem~\ref{thm:exact-css-qss-access}): a coalition is authorized when
$\ell_X(A)=\ell_Z(A)=k_L$, forbidden when
$\ell_X(A)=\ell_Z(A)=0$, and intermediate otherwise. The CSS complement
duality $\ell_X(A)=k_L-\ell_Z(A^c)$
(Lemma~\ref{lem:css-complement-duality}) converts the information-theoretic
no-cloning duality into an algebraic identity and yields the exact profile
symmetries $\Pauth(a)=\Pforb(n-a)$ and $\Pint(a)=\Pint(n-a)$. The
accessible logical dimension $\kappa(A)$, defined as the $\Ftwo$-rank of
the symplectic pairing matrix between locally supported logical classes,
grades the intermediate family that the ternary classification leaves
undifferentiated, and is basis-independent, monotone, and maximal exactly
on authorized coalitions (Proposition~\ref{prop:kappa-properties}).

On the network side, we separated the combinatorial cost of a code from
the cost of realizing it: the Tanner incidence count $\Wtot$ is an
intrinsic property of the parity-check matrices, while the Steiner routing
cost $\CSt$ and the scheduling depth $\Dsched$ depend on the physical
graph. Numerically, $\CSt/\Wtot$ spans from $0.83$ to $9.07$ for the
bivariate bicycle code and from $0.87$ to $18.58$ for the random
hypergraph-product code, according to topology alone. Sparsity and
locality are therefore independent resources, and a qLDPC code is not
automatically network-friendly.

On the design side, three findings stand out. The bivariate bicycle
$[[72,12,6]]$ code is the only instance among those studied that combines
a competitive rate, the sharpest access transition ($\Delta q=0.101$),
individually secure shares at the block level, and moderate embedding
cost. The sufficient condition $m\ge\lceil n/(d-1)\rceil$ for individual
share security (Corollary~\ref{cor:individual-security}) is not necessary,
as the bivariate bicycle instance with block size $b=6>d-1=5$
demonstrates. And block-level thresholds are properties of the code
together with the participant partition: local search improves $[[41,1]]$
from $(\tsec,\trec)=(0,8)$ to $(2,6)$, i.e.\ from no security guarantee to
a nontrivial ramp scheme with $\Delta_t=3$, still not a perfect
$((t,m))$ threshold scheme.

Several directions follow naturally. The exact criterion and the graded measure $\kappa$ apply verbatim to lifted-product and quantum Tanner codes
\cite{PanteleevKalachev2022,LeverrierZemor2022}, where the asymptotic behaviour of $\Delta q$ and $\Aramp$ as functions of the rate is open; a linear-distance family would be expected to sharpen the transition further, and testing this is the most direct extension of the present work. On the network side, replacing unit link weights by heralding-rate weights as in Eq.~\eqref{eq:noisy-steiner-cost} and coupling the placement heuristic to a concrete modular architecture would turn the cost model into a deployment planning tool. Finally, evaluating the approximate access structure of Definition~\ref{def:approximate-access} numerically under a realistic noise model, and connecting $\kappa(A)$ to the coherent information of the restricted channel, would bridge the exact algebraic picture developed here and the fault-tolerant analysis that a physical implementation requires.

\bibliographystyle{apsrev4-2}
\bibliography{refs}

\appendices

\renewcommand{\thesection}{\Alph{section}}
\renewcommand{\thesubsection}{\arabic{subsection}}

\section{QEC--QSS Preliminaries: Supplementary Proof}
\label{app:secII-proofs}

\subsection{Proof of Proposition~\ref{prop:logical-erasure-criterion}}

For an erasure on $E$, the relevant error set is the Pauli group $\mathcal{P}(E)$ supported entirely on $E$. By the Knill--Laflamme condition, the erasure is correctable if and only if, for every $P,Q\in\mathcal{P}(E)$,
\begin{equation}
    \Pi P^\dagger Q\Pi=c_{P,Q}\Pi,
\end{equation}
where $\Pi$ is the projector onto the code space and $c_{P,Q}$ is a scalar.

Since $P^\dagger Q$ is also supported on $E$, there are two possibilities. If $P^\dagger Q\notin N(\mathcal{S})$, then it anticommutes with at least one stabilizer generator and therefore maps the code space to an orthogonal syndrome subspace, so that
\begin{equation}
    \Pi P^\dagger Q\Pi=0.
\end{equation}

If $P^\dagger Q\in N(\mathcal{S})$, then the Knill--Laflamme condition requires it to act trivially on the code space. For a stabilizer code, this occurs precisely when $P^\dagger Q\in\mathcal{S}$, modulo an irrelevant global phase.

Consequently, the erasure of $E$ is correctable if and only if every Pauli operator supported on $E$ that belongs to $N(\mathcal{S})$ also belongs to $\mathcal{S}$. Equivalently,
\begin{equation}
N(\mathcal{S})\cap\mathcal{P}(E) = \mathcal{S}\cap\mathcal{P}(E).
\end{equation}
Thus, $E$ is correctable if and only if it supports no nontrivial logical Pauli operator.

\section{Proofs for the Exact Algebraic Characterization}
\label{app:secIII-proofs}

\subsection{Proof of Lemma~\ref{lem:local-normalizer-dimensions}}
\label{app:secIII-proofs-subsection-1}
Every vector $x\in\mathbb{F}_2^E$ is uniquely identified with its restriction $x_E\in\mathbb{F}_2^{|E|}$. The condition $H_Zx^T=0$ becomes
\[
    (H_Z|_E)x_E^T=0.
\]
Thus,
\[
    \ker(H_Z)\cap\mathbb{F}_2^E \cong \ker(H_Z|_E).
\]
The rank-nullity theorem gives
\[
    \dim\ker(H_Z|_E) = |E|-\rank(H_Z|_E).
\]
The proof for $H_X$ is identical.

\subsection{Proof of Lemma~\ref{lem:local-stabilizer-dimensions}}
\label{app:secIII-proofs-subsection-2}
Consider the linear restriction map
\begin{equation}
    \pi_{E^c}: \operatorname{row}(H_X) \longrightarrow \mathbb{F}_2^{E^c}, \qquad x\longmapsto x|_{E^c}. \label{eq:restriction-map-x-stabilizers}
\end{equation}
Its kernel consists precisely of stabilizer vectors whose support is contained in $E$:
\[
    \ker(\pi_{E^c}) = \operatorname{row}(H_X)\cap\mathbb{F}_2^E.
\]
The image of $\pi_{E^c}$ is the row space of $H_X|_{E^c}$ and has dimension $\rank(H_X|_{E^c})$. Applying rank-nullity to Eq.~\eqref{eq:restriction-map-x-stabilizers} yields
\[
    \dim\ker(\pi_{E^c}) = \rank(H_X)-\rank(H_X|_{E^c}).
\]
The statement for $H_Z$ follows analogously.

\subsection{Proof of Proposition~\ref{prop:rank-formula-supported-logicals}}
\label{app:secIII-proofs-subsection-3}

By Definition~\ref{def:supported-logical-dimensions},
\[
    \ell_X(E) = \dim\!\left(\ker(H_Z)\cap\mathbb{F}_2^E\right) - \dim\!\left(\operatorname{row}(H_X)\cap\mathbb{F}_2^E\right).
\]
Substituting Eqs.~\eqref{eq:kernel-hz-restricted} and \eqref{eq:local-x-stabilizer-dim} gives Eq.~\eqref{eq:ellx-rank-formula}. The derivation of Eq.~\eqref{eq:ellz-rank-formula} is symmetric under the exchange
\[
    X\longleftrightarrow Z.
\]

\subsection{Proof of Theorem~\ref{thm:exact-css-qss-access}}
\label{app:secIII-proofs-subsection-4}

By Proposition~\ref{prop:logical-erasure-criterion}, an erasure set $E$ is correctable if and only if it supports no nontrivial logical operator. For a CSS code, this is equivalent to
\[
    \ell_X(E)=\ell_Z(E)=0.
\]

A coalition $A$ is authorized if and only if the erasure of its complement $A^c$ is correctable; this establishes Eq.~\eqref{eq:authorized-ell-condition}. Likewise, $A$ is forbidden if and only if the erasure of $A$ itself is correctable, yielding Eq.~\eqref{eq:forbidden-ell-condition}. Finally, $A$ is intermediate when it is neither authorized nor forbidden, which means that neither $A$ nor $A^c$ is a correctable erasure set. This gives Eq.~\eqref{eq:intermediate-ell-condition}.

\subsection{Proof of Corollary~\ref{cor:distance-based-regions}}
\label{app:secIII-proofs-subsection-5}

If $|A|\leq d-1$, then $A$ cannot support a nontrivial logical operator by the definition of the distance. Hence $\ell_X(A)=\ell_Z(A)=0$ and $A$ is forbidden. If $|A|\geq n-d+1$, then $|A^c|\leq d-1$, so $A^c$ is correctable and $A$ is authorized.

\subsection{Proofs of Propositions~\ref{prop:authorization-monotonicity} and~\ref{prop:forbidden-monotonicity}}
\label{app:secIII-proofs-subsection-6}
If $A$ is authorized, then $A^c$ is a correctable erasure set. Since $A\subseteq B$, one has
\[
    B^c\subseteq A^c.
\]
Every subset of a correctable erasure set is itself correctable. Therefore $B^c$ is correctable and $B$ is authorized.

In the same way, if $A$ is forbidden, then $A$ is a correctable erasure set. Since $B\subseteq A$, the erasure of $B$ is also correctable. Hence $B$ is forbidden.

\subsection{Proofs of Corollary~\ref{cor:pure-qss-complementarity}}
\label{app:secIII-proofs-subsection-7}

Let $V:\mathcal{H}_S\rightarrow \mathcal{H}_A\otimes\mathcal{H}_{A^c}$ be the encoding isometry. If $A$ is authorized, then there exists a recovery channel $\mathcal{R}_A$ such that
\[
\mathcal{R}_A\circ\mathcal{N}_A=\operatorname{id}_S.
\]
Perfect reversibility of $\mathcal{N}_A$ implies that its complementary channel $\mathcal{N}_{A^c}$ is constant on the input state. Hence there exists a density operator $\sigma_{A^c}$ such that
\[
\mathcal{N}_{A^c}(\rho)=\sigma_{A^c}
\]
for every logical state $\rho$. Therefore $A^c$ is forbidden.

Conversely, if $A^c$ is forbidden, then the complementary channel $\mathcal{N}_{A^c}$ is constant on the logical input. By the information--disturbance theorem for complementary channels, $\mathcal{N}_A$ is perfectly reversible on the code space. Hence there exists a recovery channel $\mathcal{R}_A$ satisfying
\[
\mathcal{R}_A\circ\mathcal{N}_A=\operatorname{id}_S,
\]
and $A$ is authorized.

Therefore,
\[
A\in\mathcal{A}_{\mathrm{auth}} \quad\Longleftrightarrow\quad A^c\in\mathcal{A}_{\mathrm{forb}}.
\]
Exchanging $A$ and $A^c$ gives the converse relation.

\subsection{Proofs of Lemma~\ref{lem:css-complement-duality}}
\label{app:secIII-proofs-subsection-8} 
The identities follow directly from the restricted-rank expressions of Proposition~\ref{prop:rank-formula-supported-logicals}.

For any coalition $A\subseteq[n]$, Proposition~\ref{prop:rank-formula-supported-logicals}
gives
\[
\ell_X(A) = |A| - \operatorname{rank}(H_Z|_A) - \operatorname{rank}(H_X) + \operatorname{rank}(H_X|_{A^c}),
\]
whereas, applied to the complementary coalition $A^c$,
\[
\ell_Z(A^c)
=
|A^c| - \operatorname{rank}(H_X|_{A^c}) - \operatorname{rank}(H_Z) +\operatorname{rank}(H_Z|_A).
\]
Adding the two expressions, the restricted-rank terms cancel, yielding
\[
\ell_X(A)+\ell_Z(A^c)
=
|A|+|A^c|
-\operatorname{rank}(H_X) - \operatorname{rank}(H_Z).
\]
Since
\[
|A|+|A^c|=n
\]
and, for a CSS code,
\[
k_L
=
n-\operatorname{rank}(H_X)-\operatorname{rank}(H_Z),
\]
we obtain
\[
\ell_X(A)+\ell_Z(A^c)=k_L,
\]
or equivalently,
\[
\boxed{
\ell_X(A)=k_L-\ell_Z(A^c).
}
\]

Exchanging $X$ and $Z$ gives
\[
\ell_Z(A)+\ell_X(A^c)=k_L,
\]
and therefore
\[
\boxed{
\ell_Z(A)=k_L-\ell_X(A^c).
}
\]
This proves both complement-duality identities.

\subsection{Proofs of Corollary~\ref{cor:access-profile-symmetry}}
\label{app:secIII-proofs-subsection-9}

By Theorem~\ref{thm:exact-css-qss-access}, $A$ is authorized if and only if $\ell_X(A)=\ell_Z(A)=k_L$, which by Lemma~\ref{lem:css-complement-duality} is equivalent to
$\ell_X(A^c)=\ell_Z(A^c)=0$, i.e.\ to $A^c$ being forbidden. Averaging the indicator functions over the uniform distribution on subsets of size $a$, and using the size-preserving bijection $A\mapsto A^{c}$ between subsets of size $a$ and subsets of size $n-a$, gives the first identity; the second follows because the intermediate family is the complement of the union of
the two extreme families and is itself closed under complementation.

\section{Technical Details for Topology-Aware Embedding and Constructive Recovery}
\label{app:secIV-details}

\subsection{Scheduling-Depth Computation}
\label{app:secIV-details-subsection-1}

\begin{remark}[Computability and tightness of $\Dsched$]
\label{rem:dsched-computation}
Since $T_X$ and $T_Z$ are bipartite, K\"onig's edge-coloring theorem gives $\chi'(T)=\Delta(T)$ exactly, where $\Delta(T)=\max\{\Delta^{\mathrm{check}},\Delta^{\mathrm{qubit}}\}$ is the maximum of the largest check weight and the largest number of checks incident on a data qubit. Our implementation computes $\chi'(T)$ by largest-first greedy edge coloring, which for the instances reported in Section~\ref{sec:numerical-results} attains the K\"onig bound and is therefore optimal; in general greedy edge coloring on a bipartite graph uses at most $2\Delta(T)-1$ colors, so the reported values are upper bounds within a factor two of optimal. The link loads are read off the same approximate Steiner trees used for $\CSt$, so $\Dsched$ inherits the approximation factor of that construction. Both terms are computable in time linear in the number of Tanner edges and Steiner-tree edges, respectively.
\end{remark}

\subsection{Physical Interpretation of the Cost Proxies}
\label{app:secIV-details-subsection-2}

We consider a modular architecture in which nonlocal stabilizer checks are measured using distributed entanglement, such as GHZ or cat states. In this model, $W_{\mathrm{tot}}$ counts local data--ancilla interactions, $\CSt$ captures the entanglement-routing cost along Steiner trees, and $\Dsched$ captures scheduling depth. Related distributed parity-measurement and modular implementations have been studied in Refs.~\cite{ChandraKaurSeshadreesan2025, ChandraEtAl2026,ShawRengaswamy2026}, with alternative realizations based on surgery or connectivity-reduction techniques~\cite{BaspinBerentCohen2025, ZhaoYanNi2025,BerthusenEtAl2025}. The precise resource constants are architecture dependent, while the three proxies retain distinct dependence on the physical embedding.

\subsection{Steiner-Tree Approximation and Placement Search}
\label{app:secIV-details-subsection-3}

Depending on the
architecture, $\mathfrak{P}$ may enforce:
\begin{itemize}
    \item one data qubit per participant node;
    \item a maximum number of data qubits per node;
    \item preassigned subsets of nodes;
    \item geographical or modular constraints;
    \item co-location restrictions for security reasons.
\end{itemize}

The exact optimization in Eq.~\eqref{eq:placement-optimization} is computationally difficult for two independent reasons: the Steiner-tree problem is NP-hard in general, and the placement space can grow factorially with $n$. We therefore use a heuristic two-level procedure.

First, for a fixed placement $\pi$, we approximate the Steiner cost of each stabilizer support using a metric-closure minimum spanning tree (MST). Let $d_G(u,v)$ be the shortest-path distance between two physical vertices in $G_{\mathrm{phys}}$. For a terminal set $U=\{u_1,\ldots,u_t\}$, construct the complete metric graph with edge weights $d_G(u_i,u_j)$. The weight of an MST in this metric closure is a polynomial-time upper bound on the Steiner-tree cost. The corresponding shortest paths are then expanded in $G_{\mathrm{phys}}$ and merged to obtain a routing tree.

Second, we optimize the placement using local search:
\begin{enumerate}
    \item initialize $\pi$ using either a random placement or a Tanner-graph-informed placement;
    \item evaluate $C_{\mathrm{St}}(\pi;G_{\mathrm{phys}})$;
    \item propose a local modification, such as swapping two qubit locations or moving one qubit to an unused node;
    \item accept improvements greedily, or according to a simulated annealing rule;
    \item repeat until convergence or a fixed computational budget is reached.
\end{enumerate}

\subsection{Proofs of Proposition~\ref{prop:linear-steiner-cost}}
\label{app:secIV-details-subsection-5}

By Eq.~\eqref{eq:total-stabilizer-cost} and the bounded-distortion assumption,
\[
\begin{aligned}
    C_{\mathrm{St}}(\pi_n;G_{\mathrm{phys},n})
    \leq
    c_{\mathrm{emb}}
    \Bigl(
        &\sum_{s\in\mathcal{S}_X}|\supp(s)|\\
        &+
        \sum_{s\in\mathcal{S}_Z}|\supp(s)|
    \Bigr).
\end{aligned}
\]
The expression in parentheses is $W_{\mathrm{tot}}$. Since the family is qLDPC, the stabilizer weights and qubit degrees are bounded, implying $W_{\mathrm{tot}}=O(n)$. Therefore,
\[
    C_{\mathrm{St}}(\pi_n;G_{\mathrm{phys},n})=O(n).
\]

\subsection{Proofs of Proposition~\ref{prop:logical-cleaning}}
\label{app:secIV-details-subsection-6}

Since $A$ is authorized, the erasure of $A^c$ is correctable. By the logical-operator criterion, $A^c$ supports no nontrivial logical operator. Therefore, the restriction of any logical operator to $A^c$ must be removable by multiplication with a stabilizer. In the binary CSS representation, multiplication by an $X$-type or $Z$-type stabilizer is addition of a row-space vector, yielding the linear systems in Eqs.~\eqref{eq:x-cleaning-linear-system} and \eqref{eq:z-cleaning-linear-system}. Their solutions produce the desired cleaned representatives.

\subsection{Dependence of Recovery Cost on Basis and Heuristic}
\label{app:secIV-details-subsection-7}

Equation~\eqref{eq:recovery-weight} is \emph{not} an invariant of the pair
$(\mathcal{C},A)$. It depends on two arbitrary choices, and we write
$\Wrec(A;\mathcal{B},\mathcal{H})$ when the dependence matters.

First, the choice of logical basis $\mathcal{B}=\{\bar X_j,\bar Z_j\}$: a
symplectic change of basis $\mathcal{B}\mapsto\mathcal{B}'$ leaves the
encoded subsystem untouched but replaces each generator by a product of
generators, whose cleaned support can be larger or smaller. Second, the
choice of solver $\mathcal{H}$ for the underdetermined systems
\eqref{eq:x-cleaning-linear-system}--\eqref{eq:z-cleaning-linear-system}:
any particular solution $v$ may be shifted by an element of
$\ker(H_X|_{A^c})^{T}$, again changing the weight. Minimizing
$\mathrm{wt}(x_A)$ over that coset is the problem of finding a minimum-weight
coset representative of a binary code, which is NP-hard in general, so no
polynomial-time procedure is expected to return the minimum.

What we report is therefore an achievable upper bound produced by a fixed,
documented pipeline: the logical basis is the one returned by the
Gaussian-elimination construction of
Section~\ref{subsec:stabilizer-logicals}, and $\mathcal{H}$ is the
least-index particular solution of the row-reduced system, followed by a
greedy pass that adds any stabilizer row strictly decreasing the weight.
Comparisons of $\Wrec$ across coalition sizes, as in
Section~\ref{sec:numerical-results}, are meaningful because the basis and
solver are held fixed; comparisons against a hypothetical optimum are not
claimed.

\subsection{Constructive Cleaning Algorithm}
\label{app:secIV-details-subsection-8}

For completeness we state the cleaning procedure and its cost.

\begin{algorithm}[Cleaning a logical basis onto $A$]
\label{alg:cleaning}
\textbf{Input:} $H_X,H_Z\in\Ftwo^{\ast\times n}$; a logical basis
$\mathcal{B}=\{\bar X_j,\bar Z_j\}_{j=1}^{k_L}$; a coalition
$A\subseteq[n]$.\\
\textbf{Output:} cleaned representatives supported on $A$, or
\textsc{fail}.
\begin{enumerate}[label=\textup{\arabic*.}, leftmargin=2em]
    \item Form the restricted matrices $H_X|_{A^c}$ and $H_Z|_{A^c}$ and
    row-reduce them once, caching the transforms.
    \item For each $j$: solve $u(H_X|_{A^c})=\bar X_j|_{A^c}$ by back
    substitution. If the system is inconsistent, return \textsc{fail}.
    Otherwise set $\bar X_{j,A}\leftarrow\bar X_j+uH_X$.
    \item Greedily add rows of $H_X$ that strictly reduce
    $\mathrm{wt}(\bar X_{j,A})$ while preserving
    $\supp(\bar X_{j,A})\subseteq A$.
    \item Repeat steps 2--3 with $(H_Z,\bar Z_j)$.
    \item Return $\{\bar X_{j,A},\bar Z_{j,A}\}$.
\end{enumerate}
\end{algorithm}

The row reduction in step~1 costs $O(m\,|A^c|\min\{m,|A^c|\})$ binary
operations for $m$ checks and is shared by all $2k_L$ systems; each
subsequent solve costs $O(m\,|A^c|)$, and the greedy pass costs
$O(\|H_X\|_0)$ per generator. The total is therefore
$O(m|A^c|\min\{m,|A^c|\}+k_L(m|A^c|+\|H_X\|_0+\|H_Z\|_0))$, polynomial in
the block length. The procedure returns \textsc{fail} exactly when some
system in step~2 is inconsistent, that is, when the restriction
$\bar X_j|_{A^c}$ lies outside the row space of $H_X|_{A^c}$; by
Proposition~\ref{prop:logical-cleaning} this happens if and only if $A$ is
not authorized, so \textsc{fail} is itself a correct authorization test
and is used as a consistency check in Section~\ref{sec:numerical-results}.

\section{Additional Numerical and Validation Results}
\label{app:numerical-details}

\subsection{Code Construction and Sampling Details}
\label{app:numerical-details-subsection-1}

\paragraph{Repetition-derived HGP baseline.}
The hypergraph product of two length-five repetition codes gives $[[41,1]]$ with $\rank H_X=\rank H_Z=20$. Because $k_L=1$, this instance reproduces the classical single-secret situation and is the sharpest available test of the implementation: authorized, forbidden and intermediate families must satisfy the exact complementarity of Corollary~\ref{cor:access-profile-symmetry} with no intermediate coalition carrying partial logical information beyond $\kappa\in\{0,1\}$.

\paragraph{Hamming-derived HGP baseline.}
The hypergraph product of two copies of the $[7,4,3]$ Hamming code gives $[[58,16]]$ with $\rank H_X=\rank H_Z=21$. This code has the highest secret rate of the four instances, $R_s=16/58\approx0.276$, and therefore the widest ramp region.

\paragraph{Hypergraph-product construction.}
For classical parity-check matrices
$H_1\in\Ftwo^{m_1\times n_1}$ and $H_2\in\Ftwo^{m_2\times n_2}$ we use the
convention
\begin{align}
    H_X &= \bigl[\,H_1\otimes I_{n_2}\;\big|\;I_{m_1}\otimes H_2^{T}\,\bigr],
    \label{eq:hgp-hx-used}\\
    H_Z &= \bigl[\,I_{n_1}\otimes H_2\;\big|\;H_1^{T}\otimes I_{m_2}\,\bigr],
    \label{eq:hgp-hz-used}
\end{align}
which satisfies $H_XH_Z^{T}=H_1\otimes H_2^{T}+H_1\otimes H_2^{T}=0$ over $\Ftwo$ and yields $n=n_1n_2+m_1m_2$. HGP codes were introduced by Tillich and Z\'emor as a CSS qLDPC family with nonzero rate and distance scaling as the square root of the block length for suitable classical seeds~\cite{TillichZemor2014,BreuckmannEberhardt2021}.

\paragraph{Random sparse HGP code.}
To separate the effect of sparsity from that of algebraic regularity, we build an HGP code from two independent random sparse matrices with target column weight three. The instance used throughout has $n=208$, $\rank H_X=96$, $\rank H_Z=92$, hence
\begin{equation}
    k_L=n-\rank H_X-\rank H_Z=208-96-92=20,
    \label{eq:random-hgp-parameters}
\end{equation}
and the finite-size secret rate representing the logical qubits per physical share qubit
\begin{equation}
    R_s:=\frac{k_L}{n}=\frac{20}{208}\approx 0.0962 .
    \label{eq:random-hgp-secret-rate}
\end{equation}
The asymmetry $\rank H_X\neq\rank H_Z$ is characteristic of random seeds and is a useful stress test, since it breaks the $X\leftrightarrow Z$ symmetry of the two structured baselines while leaving Lemma~\ref{lem:css-complement-duality} intact.

\paragraph{Bivariate bicycle code.}
We use the $[[72,12,6]]$ bivariate bicycle (BB) code defined on $\Ftwo[x,y]/(x^{6}-1,y^{6}-1)$ by the sparse polynomials
\begin{equation}
    A(x,y)=x^3+y+y^2,
    \qquad
    B(x,y)=y^3+x+x^2 ,
    \label{eq:bb72-polynomials}
\end{equation}
with $\rank H_X=\rank H_Z=30$, rate $R_s=12/72=1/6$, distance $d=6$ and weight-six stabilizer generators \cite{BravyiEtAl2024,BB72Code}. BB codes are of particular practical interest because of their low-overhead fault-tolerant-memory performance and the existence of specialized syndrome-extraction and decoding constructions~\cite{BravyiEtAl2024,MorphingBBCodes2024,BerthusenEtAl2025}.

For each instance we record the code parameters, the CSS ranks, the row- and column-weight statistics of $H_X$ and $H_Z$, the total incidence count $\Wtot$, the access-transition observables $q_{10},q_{50},q_{90},\Delta q,\Aramp$ of Section~\ref{subsec:access-transition-observables}, and the topology-aware costs $\CSt(\pi;\Gphys)$ and $\Crec(A;\pi,\Gphys)$.

We sample $21$ coalition sizes per code, uniformly spaced in $\{0,1,\dots,n\}$. When $\binom{n}{a}\le N_{\mathrm{exact}}$ we enumerate every coalition of size $a$; otherwise we draw $N_{\mathrm{MC}}$ independent uniform coalitions of that cardinality, with $N_{\mathrm{MC}}$ between $60$ and $200$ depending on $n$. Statistical uncertainties on sampled probabilities are binomial standard errors,
\begin{equation}
    \sigma_{\widehat{P}} = \sqrt{\frac{\widehat{P}\,(1-\widehat{P})}{N_{\mathrm{MC}}}} ,
    \label{eq:binomial-standard-error}
\end{equation}
so that a probability estimated at $\widehat{P}=0.5$ with $N_{\mathrm{MC}}=150$ carries $\sigma_{\widehat P}\approx0.041$. Sparsity statistics for the random HGP and BB instances appear in Tables~\ref{tab:random-hgp-sparsity} and~\ref{tab:bb72-sparsity}: the random instance has mean check weight $7.50$ and mean qubit degree $3.46$ with a broad spread, whereas the BB code is exactly $(6,3)$-regular, every check acting on six qubits and every qubit on three checks of each type.

\begin{table}
    \centering
    \caption{Sparsity statistics for the representative random sparse HGP code.}
    \label{tab:random-hgp-sparsity}
    \begin{tabular}{lccc}
        \toprule
        Matrix quantity & Minimum & Maximum & Mean \\
        \midrule
        $H_X$ row weight & $4$ & $10$ & $7.50$ \\
        $H_Z$ row weight & $4$ & $11$ & $7.50$ \\
        $H_X$ column weight & $1$ & $8$ & $3.46$ \\
        $H_Z$ column weight & $1$ & $7$ & $3.46$ \\
        \bottomrule
    \end{tabular}
\end{table}

\begin{table}[t]
    \centering
    \caption{Sparsity statistics for the bivariate bicycle $[[72,12,6]]$ code.}
    \label{tab:bb72-sparsity}
    \begin{tabular}{lccc}
        \toprule
        Matrix quantity & Minimum & Maximum & Mean \\
        \midrule
        $H_X$ row weight & $6$ & $6$ & $6.00$ \\
        $H_Z$ row weight & $6$ & $6$ & $6.00$ \\
        $H_X$ column weight & $3$ & $3$ & $3.00$ \\
        $H_Z$ column weight & $3$ & $3$ & $3.00$ \\
        \bottomrule
    \end{tabular}
\end{table}

\subsection{Implementation Consistency Checks}
\label{app:numerical-details-subsection-2}

Three structural predictions of Section~\ref{sec:access-structure} are
tested on randomly drawn coalitions:
\begin{enumerate}[label=\textup{(\roman*)}]
    \item the duality identities
    $\ell_X(A)=k_L-\ell_Z(A^c)$ and $\ell_Z(A)=k_L-\ell_X(A^c)$ of
    Lemma~\ref{lem:css-complement-duality} ($20$ random coalitions per
    code);
    \item the complementarity of
    Corollary~\ref{cor:pure-qss-complementarity}, i.e.\ $A$ authorized
    $\iff$ $A^c$ forbidden ($20$ coalitions per code);
    \item the monotonicity of
    Proposition~\ref{prop:authorization-monotonicity}, tested on $30$
    nested pairs $A\subseteq B$ per code.
\end{enumerate}
Across the four codes, all $280$ tests pass with zero violations. Since
these identities are exact statements over $\Ftwo$ rather than statistical
ones, a single violation would invalidate the implementation; their
systematic satisfaction on codes with $n$ ranging from $41$ to $208$ and
with both symmetric and asymmetric rank profiles is therefore a
meaningful validation.

\subsection{Statistical Robustness and Bootstrap Analysis}
\label{app:numerical-details-subsection-3}

\begin{table*}[t]
    \centering
    \caption{
    Nonparametric bootstrap confidence intervals for the
    finite-size transition observables. Each interval is the
    central 95\,\% interval over 6000 resamples of the sampled
    coalitions at fixed code instance, so it quantifies
    coalition-sampling uncertainty only. The asymptotic erasure
    benchmark is $q_{\mathrm{cap}}=(1+R_s)/2$.}
    \label{tab:bootstrap-ci}
    \begin{tabular}{lcccc}
        \toprule
        Code & $R_s$ & $q_{\mathrm{cap}}$ & $q_{50}$ [95\,\% CI] & $\Delta q$ [95\,\% CI] \\
        \midrule
        HGP(rep$_5$,rep$_5$) & 0.0244 & 0.5122 & 0.5573 [0.5469, 0.5684] & 0.1978 [0.1826, 0.2176] \\
        HGP(Ham$_{7,4}$,Ham$_{7,4}$) & 0.2759 & 0.6379 & 0.8022 [0.7890, 0.8149] & 0.1816 [0.1601, 0.2011] \\
        Random sparse HGP & 0.0962 & 0.5481 & 0.8688 [0.8535, 0.8805] & 0.1934 [0.1392, 0.2372] \\
        BB $[[72,12,6]]$ & 0.1667 & 0.5833 & 0.6507 [0.6432, 0.6585] & 0.1003 [0.0853, 0.1158] \\
        \bottomrule
    \end{tabular}
\end{table*}

\begin{figure}[h]
    \centering
    \includegraphics[width=0.98\linewidth]{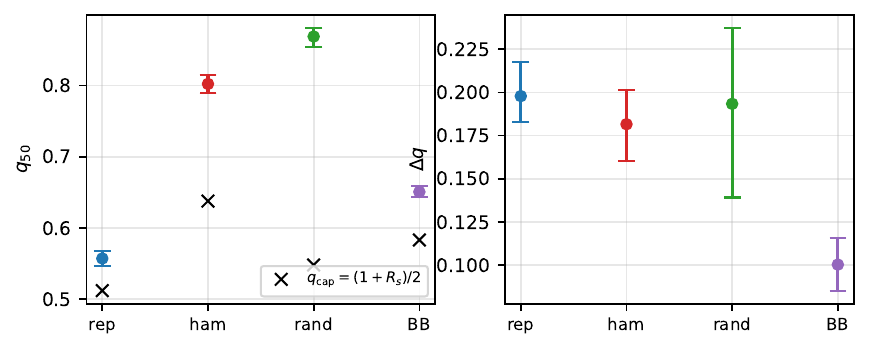}
    \caption{Bootstrap point estimates and $95\,\%$ intervals for $q_{50}$
    (left) and $\Delta q$ (right), over $6000$ resamples of the sampled
    coalitions at fixed code instance. Crosses in the left panel mark
    $q_{\mathrm{cap}}=(1+R_s)/2$. The intervals quantify coalition-sampling
    uncertainty only, not instance-to-instance variability.}
    \label{fig:bootstrap-intervals}
\end{figure}

The sampling uncertainty of the observables in Section~\ref{subsec:access-results} is assessed by a nonparametric bootstrap. For each code, coalition outcomes are resampled independently at each size $a$, and $q_{10}$, $q_{50}$, $q_{90}$, and $\Delta q$ are recomputed over $6000$ replicates. We report the central $95\,\%$ intervals in Table~\ref{tab:bootstrap-ci} and Figure~\ref{fig:bootstrap-intervals}.
The BB $[[72,12,6]]$ code has a $\Delta q$ interval disjoint from those of the three HGP instances, supporting a significantly sharper transition among the codes studied. The HGP intervals overlap substantially, so no ordering among them is claimed. The $q_{50}$ intervals are narrow and remain above the corresponding $q_{\mathrm{cap}}=(1+R_s)/2$, indicating that this finite-size observation is robust to coalition-sampling uncertainty.

\subsection{Supplementary Controlled-Ensemble Data}
\label{app:numerical-details-subsection-4}

Table~\ref{tab:ensemble-regularity} reports the complete statistics for the controlled comparison of the quasi-cyclic and random HGP ensembles discussed in Section~\ref{subsec:ensemble-regularity}. Both ensembles have the same block length and check-weight profile, while the quasi-cyclic family additionally imposes a circulant structure. The table reports the mean, standard deviation, and observed range of the access-structure metrics, together with the corresponding permutation-test $p$-values. As noted in the main text, the two ensembles are not matched in $k_L$, so the comparison does not isolate quasi-cyclic structure as the sole cause of the observed differences.

\begin{table*}[t]
    \centering
    \caption{
    Controlled ensemble at fixed block length and fixed
    check-weight profile. Both families are hypergraph products of
    $9\times12$ seed
    matrices of column weight $3$ and row
    weight $4$, giving $n=225$ and identical Tanner incidence
    count $\Wtot=1512$ for every instance;
    the quasi-cyclic family additionally imposes a circulant block
    structure of lifting size $L=3$. Entries are
    mean $\pm$ standard deviation over instances, with the range in
    brackets, and $p$ is a two-sided permutation test on the
    difference of means. The two families are \emph{not} matched in
    $k_L$ (see text), which is a limitation of the comparison.}
    \label{tab:ensemble-regularity}
    \begin{tabular}{lccc}
        \toprule
        Observable & Quasi-cyclic ($N=8$) & Random ($N=8$) & $p$ \\
        \midrule
        $\Delta q$ & $0.1379\pm0.0232$ [0.0954, 0.1733] & $0.1874\pm0.0268$ [0.1530, 0.2382] & 0.001 \\
        $\Aramp$ & $0.7329\pm0.1342$ [0.5205, 0.8725] & $0.3895\pm0.1727$ [0.2144, 0.6571] & 0.001 \\
        $q_{50}$ & $0.8708\pm0.0699$ [0.7588, 0.9379] & $0.6884\pm0.0887$ [0.5954, 0.8267] & 0.001 \\
        $\overline{\kappa}/k_L$ & $0.4225\pm0.0209$ [0.4079, 0.4564] & $0.4420\pm0.0228$ [0.3936, 0.4605] & 0.107 \\
        \midrule
        $k_L$ observed & $29,\,65$ & $9,\,17$ & \\
        \bottomrule
    \end{tabular}
\end{table*}



\subsection{Complete Placement-Optimization Results}
\label{app:numerical-details-subsection-9}

Table~\ref{tab:placement-optimization} reports the complete
placement-optimization results underlying the comparison
presented in the main text. For each code and physical
topology, we compare the identity placement with the
placement obtained by local search using the
Steiner-routing proxy $\CSt$. The table also reports the
relative improvement, the normalized optimized cost
$\CSt/W_{\mathrm{tot}}$, and the corresponding scheduling
depth $\Dsched$. Since the placement problem is solved
heuristically, the optimized values are achievable upper
bounds rather than certified global optima.

\begin{table*}[t]
    \centering
    \caption{
    Optimized qubit placement versus the identity placement. The
    optimizer performs $800$ accepted-swap iterations of local search on the
    Steiner-routing proxy $\CSt$; since the underlying assignment
    problem is NP-hard, the optimized values are upper bounds.
    $\Dsched$ is reported at the optimized placement.}
    \label{tab:placement-optimization}
    \begin{tabular}{llccccc}
        \toprule
        Code & $\Gphys$ & $\CSt$ (identity) & $\CSt$ (optimized) & Gain & $\CSt/\Wtot$ (opt.) & $\Dsched$ (opt.) \\
        \midrule
        HGP(rep$_5$,rep$_5$) & expander & 303 & 177 & 41.6\,\% & 1.229 & 8 \\
         & grid2d & 342 & 157 & 54.1\,\% & 1.090 & 8 \\
         & line & 910 & 432 & 52.5\,\% & 3.000 & 17 \\
         & ring & 850 & 328 & 61.4\,\% & 2.278 & 12 \\
        HGP(Ham$_{7,4}$,Ham$_{7,4}$) & expander & 503 & 380 & 24.5\,\% & 1.583 & 14 \\
         & grid2d & 542 & 392 & 27.7\,\% & 1.633 & 14 \\
         & line & 1534 & 953 & 37.9\,\% & 3.971 & 25 \\
         & ring & 1238 & 847 & 31.6\,\% & 3.529 & 21 \\
        Random sparse HGP & expander & 4556 & 3746 & 17.8\,\% & 2.601 & 22 \\
         & grid2d & 5630 & 4645 & 17.5\,\% & 3.226 & 29 \\
         & line & 26752 & 22521 & 15.8\,\% & 15.640 & 154 \\
         & ring & 20985 & 18940 & 9.7\,\% & 13.153 & 111 \\
        BB $[[72,12,6]]$ & expander & 1056 & 806 & 23.7\,\% & 1.866 & 15 \\
         & grid2d & 1132 & 897 & 20.8\,\% & 2.076 & 15 \\
         & line & 3918 & 2921 & 25.4\,\% & 6.762 & 57 \\
         & ring & 3334 & 2410 & 27.7\,\% & 5.579 & 39 \\
        \bottomrule
    \end{tabular}
\end{table*}

\section{Proofs and Technical Details for Block-Level QSS}
\label{app:block-proofs}

\subsection{Proof of Proposition~\ref{prop:block-distance-guarantees}}
\label{app:block-proofs-subsection-B}
The physical support $B(J)$ has cardinality $\sum_{j\in J}b_j$. Applying Corollary~\ref{cor:distance-based-regions} to $B(J)$ gives Eq.~\eqref{eq:block-forbidden-guarantee}, and applying it to $B(J)^c$, whose cardinality is $\sum_{j\notin J}b_j$, gives Eq.~\eqref{eq:block-authorized-guarantee}. For the balanced case, $\sum_{j\in J}b_j=|J|n/m\leq d-1$ is equivalent to $|J|\leq m(d-1)/n$, and $|J|$ is an integer.


\subsection{Proof of Proposition~\ref{prop:kappa-properties}}
\label{app:block-proofs-subsection-D}

A change of bases replaces $M(A)$ by $PM(A)Q^{T}$ with $P$ and $Q$ invertible over $\Ftwo$, which preserves the rank; hence $\kappa$ is well defined. Item~(i) follows from $\rank M(A)\le\min\{\ell_X(A),\ell_Z(A)\}$ together with $\ell_X(A),\ell_Z(A)\le k_L$. 
For item~(ii), suppose that $A\subseteq B$. Then every logical class admitting a representative supported in $A$ also admits a representative supported in $B$, and therefore
\[
L_A^X\subseteq L_B^X,
\qquad
L_A^Z\subseteq L_B^Z.
\]
Choose bases of $L_B^X$ and $L_B^Z$ whose first vectors form bases of $L_A^X$ and $L_A^Z$, respectively. In these bases, the pairing matrix $M(A)$ appears as a submatrix of $M(B)$. Hence
\[
\operatorname{rank} M(A) \leq \operatorname{rank} M(B),
\]
and consequently
\[
\kappa(A)\leq\kappa(B).
\].
For item~(iii), if $\kappa(A)=k_L$, then item (i) implies
\[
\ell_X(A)=\ell_Z(A)=k_L.
\]
By the complement-duality identities of Lemma~\ref{lem:css-complement-duality}, this is equivalent to
\[
\ell_X(A^c)=\ell_Z(A^c)=0.
\]
Theorem~\ref{thm:exact-css-qss-access} therefore implies that $A$ is authorized. Conversely, if $A$ is authorized, Theorem~\ref{thm:exact-css-qss-access} gives
\[
\ell_X(A^c)=\ell_Z(A^c)=0.
\]
Lemma~\ref{lem:css-complement-duality} then yields
\[
\ell_X(A)=\ell_Z(A)=k_L.
\]
Thus the full logical $X$- and $Z$-spaces admit representatives supported in $A$. Since their logical symplectic pairing is nondegenerate, the corresponding pairing matrix has rank $k_L$. Hence
\[
\kappa(A)=k_L.
\]
Item~(iv) restates $M(A)=0$, i.e.\ the absence of an encoded conjugate pair jointly accessible on $A$.

\subsection{Proof of Lemma~\ref{lem:local-logical-group}}
\label{app:block-proofs-subsection-E}

The quotient in Eq.~\eqref{eq:local-logical-group} is well defined because $\mathcal{S}\cap\mathcal{P}_A$ is normal in $\mathcal{N}(\mathcal{S})\cap\mathcal{P}_A$, and elements of $\mathcal{S}\cap\mathcal{P}_A$ commute with every element of $\mathcal{N}(\mathcal{S})$, so the commutator descends to the quotient and $\omega_A$ is well defined. For a CSS code an element of $\mathcal{N}(\mathcal{S})\cap\mathcal{P}_A$ splits uniquely, modulo $\mathcal{S}\cap\mathcal{P}_A$, into an $X$-type and a $Z$-type part supported on $A$, which gives the direct-sum decomposition and $\dim\mathcal{L}_A=\ell_X(A)+\ell_Z(A)$. Two $X$-type elements commute, as do two $Z$-type elements, so the diagonal blocks of the Gram matrix vanish; the off-diagonal block is exactly the pairing matrix \eqref{eq:local-pairing-matrix}. Finally, over $\Ftwo$, $\rank\begin{psmallmatrix}0&M\\M^{T}&0\end{psmallmatrix}=2\rank M$.

\subsection{Proof of Proposition~\ref{prop:accessible-subsystem}}
\label{app:block-proofs-subsection-F}

By Lemma~\ref{lem:local-logical-group}, the local logical group $L_A$ carries an alternating bilinear form $\omega_A$ whose Gram matrix satisfies
\[
\operatorname{rank}_{\mathbb F_2}\Omega(A)=2\kappa(A).
\]
Applying symplectic Gram--Schmidt to $L_A$ therefore yields a decomposition of its nondegenerate part into $\kappa(A)$ hyperbolic planes, together with the radical
\[
\operatorname{rad}(\omega_A) = \{v\in L_A:\omega_A(v,w)=0\ \text{for all }w\in L_A\}.
\]
Hence there exist logical classes
\[
\overline X'_1,\overline Z'_1,\ldots, \overline X'_\kappa,\overline Z'_\kappa
\]
satisfying
\[
\omega_A(\overline X'_i,\overline Z'_j)=\delta_{ij},
\qquad
\omega_A(\overline X'_i,\overline X'_j) = \omega_A(\overline Z'_i,\overline Z'_j)=0.
\]
By the definition of $L_A$, each of these classes admits a representative in $\mathcal N(\mathcal S)\cap\mathcal P_A$ and is therefore physically representable on the shares held by $A$. Restricted to the code space, these $2\kappa$ logical Pauli operators satisfy the canonical Pauli commutation relations for $\kappa$ logical qubits. The $C^*$-algebra they generate is consequently a full matrix factor
\[
\mathcal A_{\mathrm{acc}} \simeq M_{2^\kappa}(\mathbb C)
\]
inside the full logical algebra $\mathcal B(\mathcal H_L)\simeq M_{2^{k_L}}(\mathbb C)$.

By the structure theorem for finite-dimensional matrix factors, there exists a decomposition
\[
\mathcal H_L \simeq \mathcal H_{\mathrm{acc}}\otimes\mathcal H_{\mathrm{res}},
\qquad
\dim\mathcal H_{\mathrm{acc}}=2^\kappa,
\]
under which
\[
\mathcal A_{\mathrm{acc}} = \mathcal B(\mathcal H_{\mathrm{acc}}) \otimes I_{\mathrm{res}}.
\]

The algebra $\mathcal A_{\mathrm{acc}}$ is represented by logical observables supported entirely on $A$. In the operator-algebra quantum-error-correction framework, such a matrix factor is a correctable logical algebra for the channel obtained by restricting the encoded state to $A$. Equivalently, there exists a recovery channel $\mathcal R_A$ such that, in the Heisenberg picture\cite{Beny2007Observables},
\[
(\mathcal R_A\circ\mathcal N_A)^\dagger(O) = O
\]
for every
\[
O\in \mathcal B(\mathcal H_{\mathrm{acc}}) \otimes I_{\mathrm{res}},
\]
where the equality is understood on the logical code space. Thus the $\mathcal H_{\mathrm{acc}}$ factor is exactly recoverable from the shares in $A$.
When $\kappa(A)=k_L$, Proposition~\ref{prop:kappa-properties} implies that $A$ is authorized and $\mathcal H_{\mathrm{acc}}=\mathcal H_L$. When $\kappa(A)=0$, the local logical algebra contains no noncommuting logical Pauli pair and hence no nontrivial quantum logical subsystem, although its radical may still contain commuting logical information.

\subsection{Complete Block-Level Access Profiles}
\label{app:numerical-details-subsection-7}

\begin{figure}[h]
    \centering
    \includegraphics[width=0.92\linewidth]{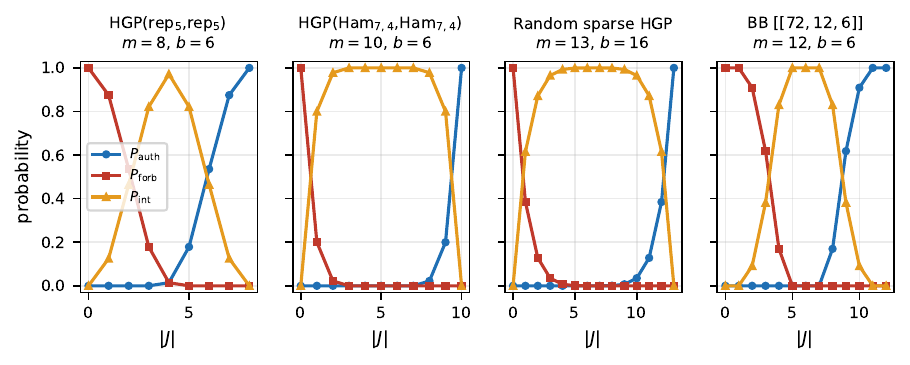}
    \caption{Exact block-level access profiles obtained by enumerating all $2^m$ participant coalitions for each code and contiguous balanced partition. Only the BB $[[72,12,6]]$ code has $P_{\mathrm{forb}}(|J|=1)=1$, i.e.\ individually secure shares.}
    \label{fig:block-access}
\end{figure}

Figure~\ref{fig:block-access} reports the complete block-level access profiles for the representative participant partitions discussed in the main text. For each participant-coalition size, it shows the exact fractions of forbidden, intermediate, and authorized coalitions obtained from the rank-based classification.
These profiles provide the full access-structure picture underlying the summary thresholds $t_{\rm sec}$ and $t_{\rm rec}$ reported in the main text.

\subsection{Partition-Sweep Distributions}
\label{app:numerical-details-subsection-8}

\begin{figure}[h]
    \centering
    \includegraphics[width=0.98\linewidth]{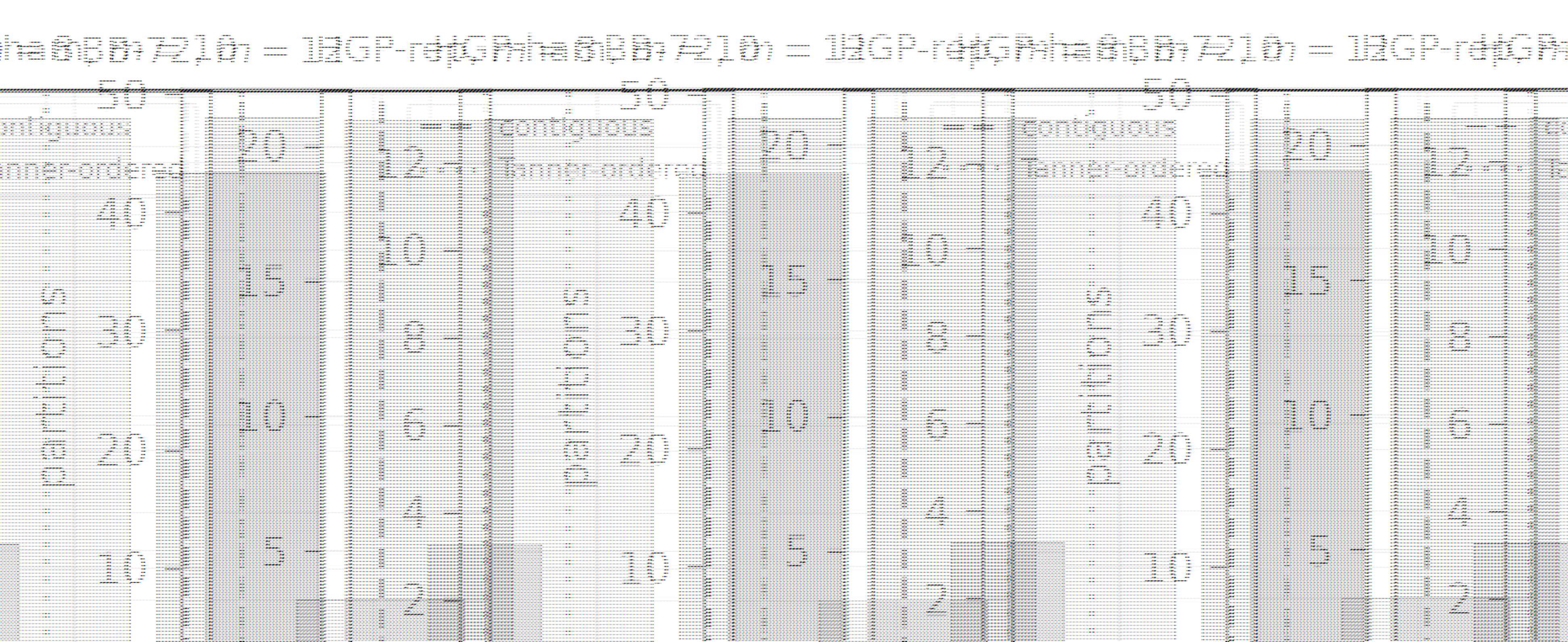}
    \caption{Distribution of the block-level ramp width $\Delta_t$ over random balanced partitions of the same code into $m$ participants, with the contiguous (dashed) and Tanner-ordered (dotted) references marked. Thresholds are extracted by exact enumeration of all $2^m$ block coalitions.}
    \label{fig:partition-sweep}
\end{figure}

Figure~\ref{fig:partition-sweep} shows the distributions of $t_{\rm sec}$, $t_{\rm rec}$, and the ramp width $\Delta_t$ obtained by sweeping over the sampled participant partitions discussed in the main text. The distributions illustrate the sensitivity of the block-level access structure to the choice of partition and complement the aggregate statistics reported in the main text.

\end{document}